\documentclass[fleqn,usenatbib,useAMS]{mnras}

\usepackage{graphicx}	
\usepackage{amsmath}	
\usepackage{amssymb}	
\usepackage{multicol}        
\usepackage{bm}		
\usepackage{pdflscape}	
\usepackage{ mathrsfs }

\usepackage[T1]{fontenc}
\usepackage{ae,aecompl}

\usepackage{mathptmx}
\usepackage[dvipsnames]{xcolor}

\title[TOI-220\,$b$: a transiting warm sub-Neptune]{TOI-220\,$b$: a warm sub-Neptune discovered by \textit{TESS}\thanks{Based on observations made with ESO Telescopes at the La Silla Observatory under programs ID 1102.C-0923, 1102.C-0249, 0102.C-0584, 60.A.9700, and 60.A-9709.}}
\author[Hoyer, S. et al.]{\parbox{\textwidth}{\Large S. Hoyer,$^{1}$\thanks{Contact e-mail: \href{mailto:sergio.hoyer@lam.fr}{sergio.hoyer@lam.fr}}
D.~Gandolfi,$^{2}$ 
D.\,J.~Armstrong,$^{3,4}$ 
M.~Deleuil,$^{1}$ 
L.~Acu{\~n}a,$^{1}$ 
J.\,R.~de~Medeiros,$^{5}$ 
E.~Goffo,$^{2}$ 
J.~Lillo-Box,$^{6}$ 
E.~Delgado~Mena,$^{7,8}$ 
T.\,A.~Lopez,$^{1}$,   
A.~Santerne,$^{1}$   
S.~Sousa,$^{7,8}$ 
M.~Fridlund,$^{9,10}$  
V.~Adibekyan,$^{7,8}$   
K.\,A.~Collins,$^{11}$ 
L.\,M.~Serrano,$^{2}$ 
P.~Cort{\'e}s-Zuleta,$^{1}$  
S.\,B. Howell,$^{12}$ 
H.~Deeg,$^{13,14}$ 
A.~Aguichine,$^{1}$ 
O.~Barrag{\'a}n,$^{15}$ 
E.\,M.~Bryant,$^{3,4}$ 
B.\,L.~Canto~Martins,$^{5}$ 
K.\,I.~Collins,$^{16}$ 
B.\,F.~Cooke,$^{3,4}$  
R.\,F.~D{\'i}az,$^{17}$ 
M.~Esposito,$^{18}$   
E.~Furlan,$^{18}$  
S.~Hojjatpanah,$^{7,8}$ 
J.~Jackman,$^{3,4}$ 
J.\,M.~Jenkins,$^{12}$ 
E.\,L.\,N.~Jensen,$^{20}$ 
D.\,W.~Latham,$^{9}$ 
I.\,C.~Le\~ao,$^{5}$ 
R.\,A.~Matson,$^{21}$ 
L.\,D.~Nielsen,$^{22}$ 
A.~Osborn,$^{3,4}$
J.\,F.~Otegi,$^{22,23}$ 
F.~Rodler,$^{24}$ 
S.~Sabotta,$^{18}$  
N.\,J.~Scott,$^{12}$  
S.~Seager,$^{25,26,27}$ 
C.~Stockdale,$^{28}$ 
P.\,A.~Str{\o}m,$^{3,4}$ 
R. Vanderspek,$^{25}$ 
V.~Van~Eylen,$^{29}$ 
P.\,J.~Wheatley,$^{3,4}$ 
J.\,N.~Winn,$^{30}$ 
J.\,M.~Almenara,$^{31}$ 
D.~Barrado,$^{6}$ 
S.\,C.\,C.~Barros,$^{7,8}$ 
D.~Bayliss,$^{3,4}$   
F.~Bouchy,$^{22}$ 
P.\,T.~Boyd,$^{32}$ 
J.~Cabrera,$^{33}$ 
W.\,D.~Cochran,$^{34}$ 
O.\,Demangeon,$^{7,8}$ 
J.\,P.~Doty,$^{35}$  
X.~Dumusque,$^{22}$ 
P.~Figueira,$^{7,24}$ 
W.~Fong,$^{25}$  
S.~Grziwa,$^{36}$ 
A.\,P.~Hatzes$^{18}$ 
P.~Kab{\'a}th,$^{37}$ 
E.~Knudstrup,$^{38}$ 
J.~Korth,$^{36}$ 
J.\,H.~Livingston,$^{39}$ 
R.~Luque,$^{13,14}$   
O.~Mousis,$^{1}$ 
S.\,E.~Mullally,$^{40}$ 
H.\,P.~Osborn,$^{25,41}$ 
E.~Pall\'e,$^{13,14}$  
C.\,M.~Persson,$^{10}$  
S.~Redfield,$^{42}$  
N.\,C.~Santos,$^{7,8}$ 
J.~Smith,$^{12,43}$ 
J.~\v{S}ubjak,$^{37}$ 
J.\,D.~Twicken,$^{12,43}$ 
S.~Udry$^{22}$ 
D.\,A.~Yahalomi$^{11,44}$ 
}
\vspace{0.5cm}
\\
\parbox{\textwidth}{
The authors' affiliations are shown in Appendix \ref{sec:affiliations}}
\vspace{-0.5cm}}
\date{Last updated XX; in original form XX}

\pubyear{2020}

\begin{document}
\label{firstpage}
\pagerange{\pageref{firstpage}--\pageref{lastpage}}
\maketitle

\begin{abstract}
In this paper we report the discovery of TOI-220\,$b$, a new sub-Neptune detected by the Transiting Exoplanet Survey Satellite (\textit{TESS}) and confirmed by radial velocity follow-up observations with the HARPS spectrograph. Based on the combined analysis of \textit{TESS} transit photometry and high precision radial velocity measurements we estimate a planetary mass of 13.8\,$\pm$\,1.0\,M$_{\earth}$ and radius of 3.03\,$\pm$\,0.15\,R$_{\earth}$, implying a bulk density of 2.73\,$\pm$\,0.47\,$\textrm{g\,cm}^{-3}$. TOI-220\,$b$ orbits a relative bright (V=10.4) and old (10.1$\pm$1.4\,Gyr) K dwarf star with a period of $\sim$10.69\,d. Thus, TOI-220\,$b$ is a new warm sub-Neptunes with very precise mass and radius determinations. A Bayesian analysis of the TOI-220\,$b$ internal structure indicates that due to the strong irradiation it receives, the low density of this planet could be explained with a steam atmosphere in radiative-convective equilibrium and a supercritical water layer on top of a differentiated interior made of a silicate mantle and a small iron core.     
\end{abstract}
\begin{keywords}
 Planetary systems -- Planets and satellites: fundamental parameters -- Planets and satellites: individual: TYC 8897-01263-1 -- Stars: fundamental parameters -- Techniques: photometric -- Techniques: radial velocities
\end{keywords}



\section{Introduction}
\label{sec:intro}
Launched in April 2018, the Transiting Exoplanet Survey Satellite (\textit{TESS}) space mission \citep{Ricker2015} has discovered about 98 confirmed new exoplanets and more than 2450 candidates\footnote{Source: NASA Exoplanet Archive \citep{Akeson2013}, as of 20 January 2021.}. Among all the confirmed exoplanets, one class of particular interest are the small-sized planets (R\,$<$\,5\,R$_{\earth}$) orbiting bright stars (V\,$<$\,11 mag) with periods shorter than 10 days. Within this parameter space there is the so-called Neptunian desert \citep{SzaboKiss2011,Mazeh2016}, which is a region that presents a significant paucity of hot/highly irradiated planets with respect to the overall planet population. This deficit could be seen as evidence of photoevaporation and/or tidal disruptions \citep{OwenLai2018} or of core-powered atmospheric mass loss mechanism \citep{Ginzburg2018,Gupta2019}. Currently, of all the known planets in the Neptunian desert only a very small number have precisely measured masses \citep[e.g.][]{West2019NGTS4b,Armstrong2020TOI849b, Jenkins2020LTT9779b}, preventing a comprehensive understanding of the formation history of these objects. On the other side, there is a population of sub-Neptune size planets exposed to a milder stellar irradiation that usually present a gas-rich envelope with equilibrium temperatures below 1000\,K \citep[e.g.][]{Morley2015,Crossfield2017,Gao2020Nat}.  In this paper we present the discovery and confirmation of TOI-220\,$b$, a sub-Neptune orbiting TYC\,8897-01263-1, an old (10.1\,Gyr) K0\,V high proper motion southern star with a magnitude of 9.69 in the \textit{TESS} bandpass (V=10.47). The brightness of the star, therefore, enabled a precise high SNR radial velocity (RV) follow-up campaign with the High Accuracy Radial velocity Planet Searcher \citep[HARPS,][]{Pepe2002} spectrograph mounted at the ESO La Silla 3.6\,m telescope, in the framework of the NCORES \citep[e.g.,][]{Nielsen2020,Armstrong2020TOI849b} and KESPRINT \citep[e.g.,][]{Gandolfi2018,Carleo2020} collaborations. The confirmation of the planetary nature of TOI-220\,$b$ is based on the simultaneous analysis of the \textit{TESS} photometry and HARPS RV measurements. 

The rest of the paper is organized as follows. The gathered observations of the TOI-220 system are described in Sect.~\ref{sec:obs}. The stellar analysis, the combined modelling and the timing analysis of the transits are presented in Sect.~\ref{sec:analysis}. In Sect.~\ref{sec:composition} we describe the Bayesian analysis of the internal structure of TOI-220\,$b$ and the final discussion is presented in Sect.~\ref{sec:discussion}. 

\section{Observations}
\label{sec:obs}

\subsection{\textit{TESS} observations}
\label{subsec:tess}

\begin{figure*}
    \centering
    \includegraphics[scale=0.6]{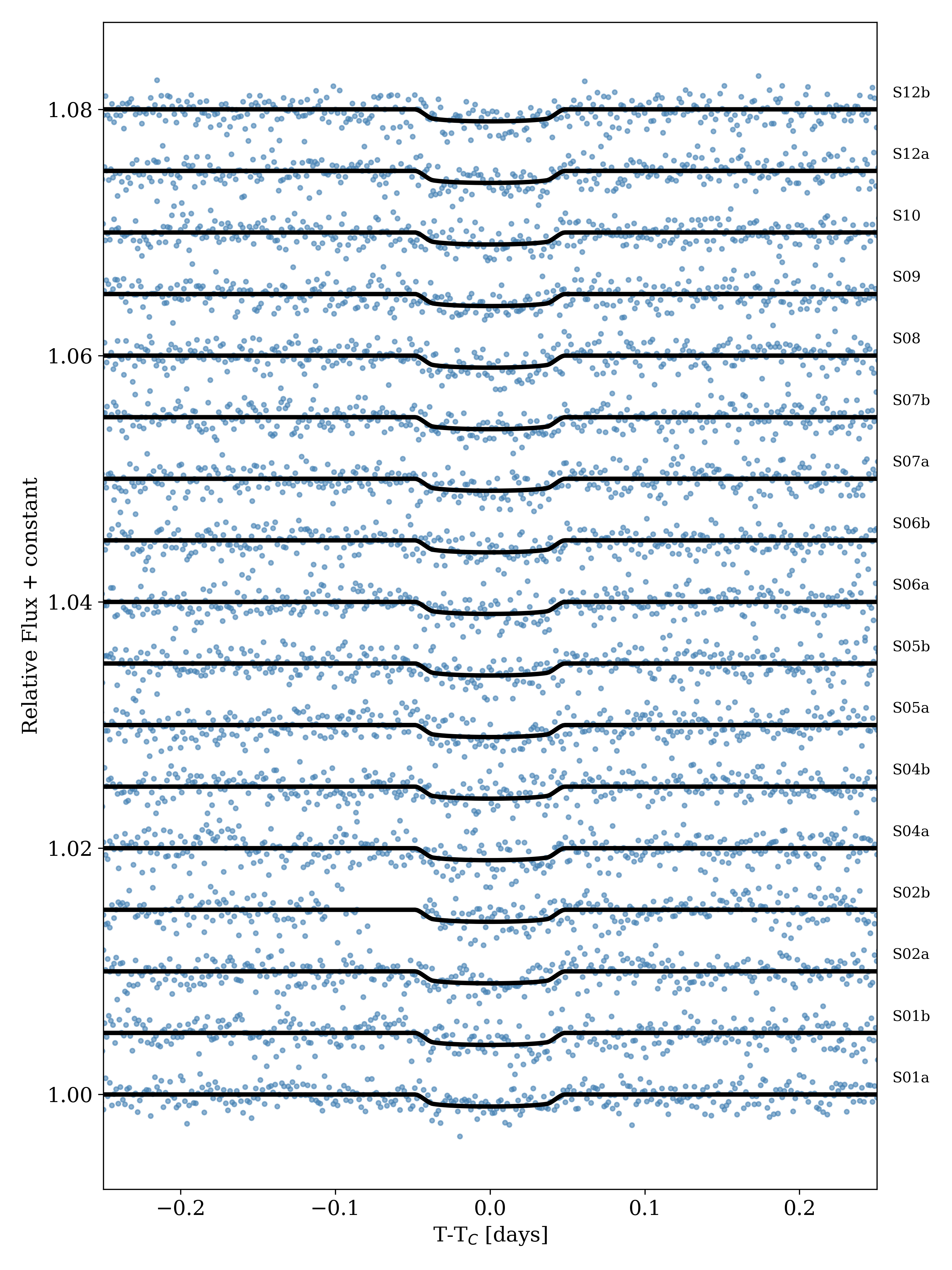}
    \caption{TOI-220\,$b$ transit light curves extracted from the \textit{TESS} data. The black solid line represents the best transit model (Sect.~\ref{sec:pastis}). The labels indicate the corresponding \textit{TESS} sector. }
    \label{fig:lcs_all_vertical}
\end{figure*}

During {\it TESS} Southern ecliptic hemisphere observations, TOI-220 was observed through sectors 1--12 with the exception of sector~3. By the end of the writing of this work, the sectors 27 and 28 light curves were released, as part of the {\it TESS} extended survey in the Southern hemisphere. These data were included in the timing analysis (Sect.~\ref{sec:ttv}) and in the search for transits of the hypothetical planet $c$ (Appendix~\ref{sec:bls_residuals}). The transit signals of TOI-220.01 (aka TOI-220\,$b$) were detected by the \textit{TESS} Science Processing Operations Center \citep[SPOC,][]{Jenkins2016} transit search pipeline \citep{Jenkins2002,Jenkins2010, Li2019}. The transit detection threshold crossing event, with a periodicity of 10.695\,d, a depth of $\sim$900\,ppm, and a SNR$\sim$28, subsequently passed all SPOC data validation diagnostic tests \citep{Twicken2018} and was promoted to {\it TESS} Object of Interest (TOI) status as a planet candidate (Guerrero et al., submitted). 
For each {\it TESS} sector, the light curve of the target was retrieved from the Mikulski Archive for Space Telescopes (MAST) archive. In this work, the two-minute cadence Presearch Data Conditioning Simple Aperture Photometry (PDCSAP) light curves \citep{Stumpe2012, Stumpe2014, Smith2012} detrended with the SPOC algorithms were used. Two or three transits of TOI-220\,$b$ were identified using a simple Box Least Squares (BLS) analysis on per-Sector light curves. After a first inspection of the flux time series of each sector some transits of TOI-220\,$b$ were discarded due to SNR considerations and/or due to large systematics in the flux time series. This was the case for the full light curve of Sector 11. A few transits with poor coverage (incomplete transits) were also discarded. Therefore, only one or two transits per \textit{TESS} Sector were finally extracted. A total of 17 out of the 24 transits of TOI-220\,$b$ observed by \textit{TESS} were considered for further analysis. After a first estimation of the individual time of transits using the period obtained from the BLS analysis, we extracted the photometric light curves around 0.4 days before/after the transit central time, when possible. The {\it TESS} sector, central time, and the 2\,min RMS of each transit used in this work are reported in Table~\ref{tab:tess_transits} and the light curves are shown in Fig.~\ref{fig:lcs_all_vertical}. 

    \begin{table}
        \centering
        \caption{Summary of TOI-220\,$b$ transits extracted from the \textit{TESS} data.}
        \label{tab:tess_transits}
        \begin{tabular}{cccl}
            \hline
            \textit{TESS} & transit mid-time & RMS(2\,min) \\
            Sector & [BJD\_TDB] & [ppm]  \\
            \hline
     
            S01 & 2458335.9024 & 890 \\ 
            S01 & 2458346.5976 & 941 \\ 
            S02 & 2458357.2928 & 897 \\ 
            S02 & 2458378.6832 & 903 \\ 
            S04 & 2458421.4640 & 979 \\ 
            S04 & 2458432.1592 & 909 \\ 
            S05 & 2458442.8544 & 942 \\ 
            S05 & 2458453.5496 & 949 \\ 
            S06 & 2458474.9400 & 952 \\ 
            S06 & 2458485.6352 & 882 \\ 
            S07 & 2458496.3304 & 908 \\ 
            S07 & 2458507.0257 & 929 \\ 
            S08 & 2458528.4161 & 924 \\ 
            S09 & 2458549.8065 & 911 \\ 
            S10 & 2458592.5873 & 833 \\ 
            S12 & 2458635.3681 & 961 \\ 
            S12 & 2458646.0633 & 938 \\ 

            \hline
        \end{tabular}
    \end{table}
\normalsize

\subsection{HARPS follow-up}
\label{subsec:harps}

We obtained radial velocity measurements of TOI-220 with the HARPS spectrograph \citep{Mayor2003} mounted on the 3.6\,m telescope at ESO's La Silla Observatory. HARPS is a stabilized high-resolution spectrograph with resolving power of R\,$\approx$\,115000, capable of sub m\,s$^{-1}$ RV precision. In total 101 observations were taken between 18 November 2018 and 11 October 2019 in high accuracy mode, of which 91 were used in the combined fit. Observations were performed under programmes 1102.C-0923 (PI: Gandolfi), 1102.C-0249 (PI: Armstrong), 0102.C-0584 (PI: De Medeiros), 60.A-9700 and 60.A-9709 (technical time). A typical exposure time of 1800\,s was used, on occasion adjusted between 1500\,s and 2100\,s depending on sky condition and observing schedule. We achieved a typical signal-to-noise per pixel of 60 and an RV precision of 1.0 m\,s$^{-1}$.

Eight observations acquired on the nights between 25 and 27 November 2018 were excluded. On these dates the ThAr lamp used for wavelength calibration of HARPS was deteriorating and was subsequently exchanged on 28 November 2018. The changing flux ratio between the thorium and argon emission lines of the dying ThAr lamp induced a 2\,m\,s$^{-1}\,$d$^{-1}$ drift in the wavelength solution of the instrument over 5\,d. The problematic data were confirmed by comparing unpublished data from the HARPS-N solar telescope \citep{Dumusque2015, CollierCameron2019} and those of the Helios program on HARPS, which also observes the Sun daily. Another observation done on the night of the 19 January 2019 was excluded due to an earthquake on that night\footnote{ \url{https://www.volcanodiscovery.com/earthquakes/2019/01/20/01h32/magnitude7-NearCoastofCentralChile-quake.html}.}. Finally, one observation on the night of the 17 April 2019 was excluded due to its very low signal-to-noise ratio. This gives 91 useful spectra. 

The data were reduced with the HARPS data reduction software \citep[{\sc DRS};][]{Lovis2007}. Radial velocity measurements were derived using the weighted cross-correlation function (CCF) method using a K0 numerical mask \citep{Pepe2002}. For each spectrum, the {\sc DRS} provides also the contrast, the full width at half maximum (FWHM), and the bisector inverse slope (BIS) of the cross-correlation function, as well as the Ca\,{\sc ii} H\,\&\,K lines activity indicator ($\log$\,R$^{\prime}_\mathrm{HK}$). We also extracted additional activity indexes and spectral diagnostics, namely, H$\alpha$, Na D, chromaticity (CRX), and differential line width (dLW), using the code {\sc SERVAL} \citep{Zechmeister2018}. The 91 {\sc DRS} and {\sc SERVAL} RV measurements and activity indicators are listed in Tables~\ref{tab:rv_data} and \ref{tab:serval_data}. Time stamps are given in Barycentric Julian Dates in Barycentric Dynamical Time (BJD\_TDB). 
The RV time series along with the best fitting model (Sect.~\ref{sec:pastis}) are shown in Fig.~\ref{fig:rvs_models}. We found an RV jitter below 2\,m\,s$^{-1}$, consistent with the low activity level of the star (see Sect.~\ref{sec:spectra_analysis} and \ref{sec:pastis}).

\subsection{High resolution imaging}
\label{subsec:hr}

\begin{figure}
    \centering
    \includegraphics[scale=0.5]{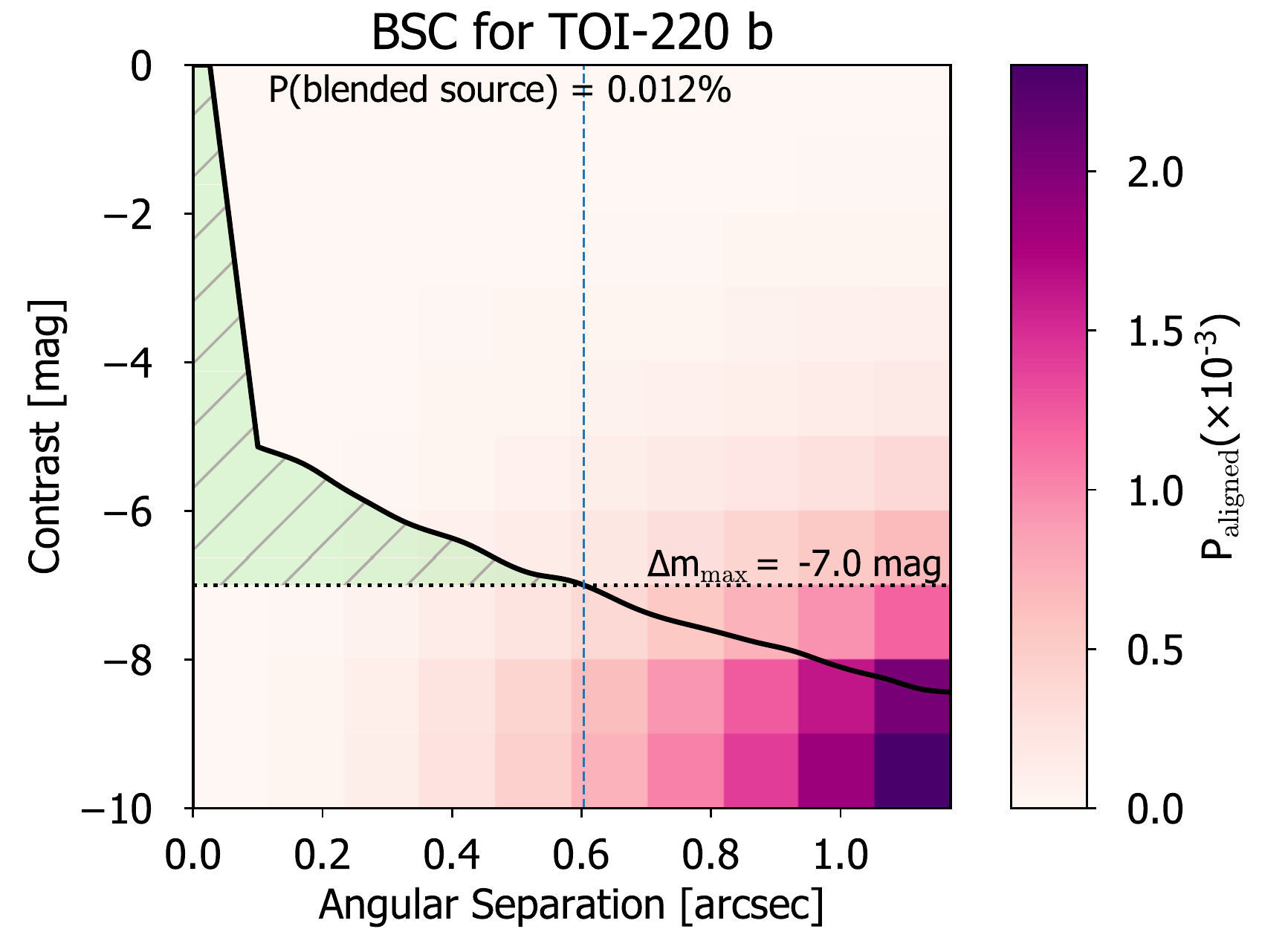} 
    \caption{Contrast curve from the Gemini/Zorro instrument for the 832\,nm filter (solid black line). The color on each angular separation and contrast bin represents the probability of a chance-aligned source with these properties at the location of the target, based on TRILEGAL model (see Sect.~\ref{subsec:hr} within the main text). The maximum contrast of a blended binary capable of mimicking the planet transit depth is shown as a dotted horizontal line. The green-shaded region represents the non-explored regime by the high-spatial resolution image. The BSC corresponds to the integration of $P_{\rm aligned}$ over this shaded region.}
    \label{fig:bsc_zorro832}
\end{figure}

To search for contaminating stars in the \textit{TESS} photometric aperture, TOI-220 was observed on 9 January 2020 with the Zorro speckle instrument at Gemini South telescope\footnote{\url{https://www.gemini.edu/sciops/instruments/alopeke-zorro/}.}. Zorro provides simultaneous speckle imaging in two bands, 562\,nm and 832\,nm, with output data products including a reconstructed image, and robust limits on companion detections \citep{howell2011}. The contrast curve for the 832\,nm is shown in Fig.~\ref{fig:bsc_zorro832}. We also calculated from our high-resolution images the probability of contamination from blended and undetectable sources in the \textit{TESS} aperture. We call this the blended source confidence (BSC) and the steps for estimating this probability are fully described in \cite{lillo-box14b}. We used a {\sc python} implementation of this technique ({\sc bsc}, by J. Lillo-Box) which uses the {\sc trilegal}\footnote{\url{http://stev.oapd.inaf.it/cgi-bin/trilegal}.} galactic model \citep[version 1.6;][]{girardi12} to retrieve a simulated source population of the region around the corresponding target\footnote{This is done in {\sc python} by using the {\sc astrobase} implementation by \cite{astrobase}.}. From this simulated population, the density of stars around the target position (radius r=1$^{\circ}$) was derived with the associated probability of chance-alignment at a given contrast magnitude and separation. We used the default parameters for the bulge, halo, thin/thick disks, and the lognormal initial mass function from \cite{chabrier01}. The Zorro contrast curve was then used to constrain this parameter space and estimate the final probability of undetected potentially contaminant sources. We considered as potentially contaminant sources those with a maximum contrast magnitude of $\Delta m_{\rm max} = -2.5\log{\delta}$, with $\delta$ being the transit depth of the candidate planet in the \textit{TESS} band. This represents the maximum magnitude that a blended star can have to mimic this transit depth. We converted the depth in the \textit{TESS} passband to the Zorro filters by using simple conversions using the TIC catalogue magnitudes and linking the 562\,nm filter to the SDSSr band and the 832\,nm filter to the SDSSz band. The corresponding conversions imply $\Delta m_{\rm 562\,nm} = 0.954\Delta m_{\rm TESS}$ and $\Delta m_{\rm 832\,nm} = 0.920\Delta m_{\rm TESS}$. 

\begin{figure}
    \centering
    \includegraphics[scale=0.5]{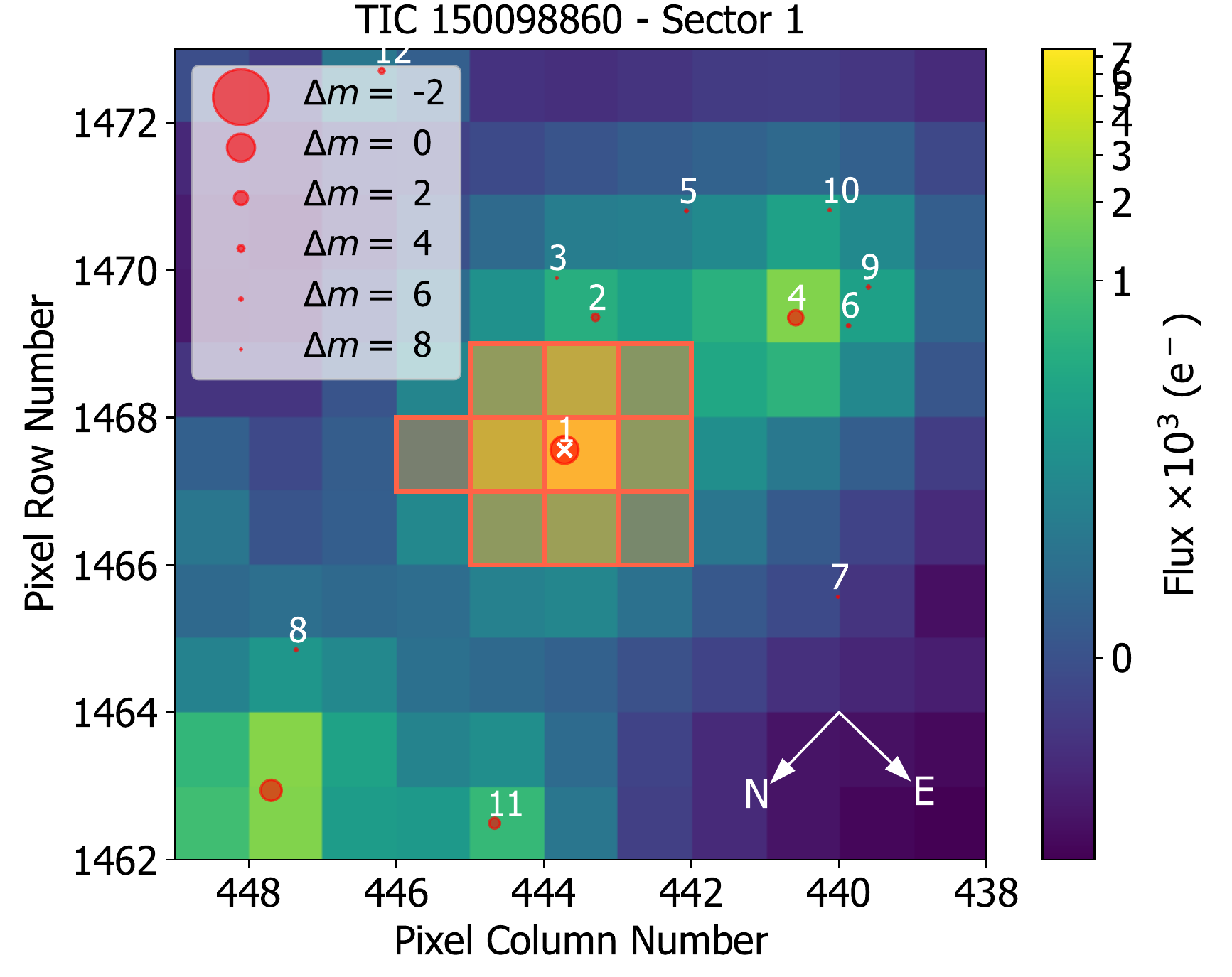} 
    \caption{Target Pixel File of the \textit{TESS} frame for TOI-220 corresponding to Sector 1 and computed with {\sc tpfplotter} \citep{aller20}. The SPOC pipeline aperture is highlighted as red squares and the Gaia DR2 catalogue is over plotted with symbol sizes proportional to the magnitude contrast with the target.}
    \label{fig:tpfplot}
\end{figure}

We applied this technique to TOI-220 using the two contrast curves available from the Gemini/Zorro instrument. The result provides a very low probability for an undetected source capable of mimicking the transit signal of 0.055\,\% for the 562\,nm image and 0.015\,\% for the 832\,nm image. We underline that using high-resolution images allowed us to significaly improve these probabilities by 26\,\% and 81\,\% for the 562\,nm and 832\,nm images, respectively, when comparing to the raw probabilities (i.e, without high-resolution images). As an example, we show in Fig.~\ref{fig:bsc_zorro832} the contrast curve and the results of the BSC analysis. In addition, by using the {\sc tpfplotter} code \citep{aller20} we ruled out the presence of contaminant sources within the \textit{TESS} photometric aperture down to a $\Delta {\rm mag} =8$\,mag in the Gaia DR2 catalogue (see Fig. \ref{fig:tpfplot}).

\subsection{Ground based photometric follow-up}

We observed transit ingresses of TOI-220\,$b$ in Pan-STARRS $z$-short band on 13 February 2019 and 21 December 2019 from the LCOGT \citep{Brown:2013} 1.0\,m network node at South Africa Astronomical Observatory. We used the {\tt TESS Transit Finder}, which is a customized version of the {\tt Tapir} software package \citep{Jensen:2013}, to schedule our transit observations. The $4096\times4096$ LCOGT SINISTRO cameras have an image scale of 0.389$\arcsec$ per pixel, resulting in a $26\arcmin\times26\arcmin$ field of view. The images were calibrated by the standard LCOGT {\sc banzai} pipeline, and photometric data were extracted with {\tt AstroImageJ} \citep{Collins:2017}. The 13 February 2019 images were focused and have typical stellar point-spread-functions with a FWHM of $1\farcs9$, and circular apertures with radius $4\farcs3$ were used to extract the differential photometry. The 21 December 2019 images were defocused and have typical stellar point-spread-functions with a FWHM of $5\farcs2$, and circular apertures with radius $7\farcs8$ were used to extract the differential photometry. The light curves are presented in Appendix \ref{fig:lco_transits}.

\section{Data analysis}
\label{sec:analysis}

\subsection{Stellar parameters}
\label{sec:spectra_analysis}

We derived the spectroscopic parameters and chemical abundance of TOI-220 from the co-added spectrum, which has an S/N per pixel at 5500\,\AA{} of $\sim$600.  We used {\sc ares+moog}, and followed the methodology described in \citet{Santos-13} and \citet{Sousa-14} to derive the stellar atmospheric parameters (T$_{\mathrm{eff}}$, $\log g$, microturbulence, [Fe/H]), and their respective uncertainties. We first measured equivalent widths (EW) of iron lines on the combined HARPS spectrum of TOI-220 using the {\sc ares} v2 code\footnote{The {\sc ares} v2 code can be downloaded at \url{http://www.astro.up.pt/~sousasag/ares}.} \citep{Sousa-15}. Then we used a minimization process where we assume ionization and excitation equilibrium to converge in the best set of spectroscopic parameters. This process makes use of a grid of Kurucz model atmospheres \citep{Kurucz-93} and the radiative transfer code MOOG \citep{Sneden-73}. Stellar abundances of the elements were derived using the classical curve-of-growth analysis method assuming local thermodynamic equilibrium \citep[e.g.][]{Adibekyan-12, Adibekyan-15, Delgado-14, Delgado-17}. Abundances of the volatile elements, C and O, were derived following the method of \cite{Delgado-10, Bertrandelis-15}. Since the two analysed spectral lines of oxygen (6158.17\AA{} and 6300.3\AA{}) are usually weak and the 6300.3\AA{} line is blended with a Ni line \citep[see e.g.][]{Bertrandelis-15}, the EWs of these lines were manually measured with the task {\sc splot} in {\sc IRAF}. All the [X/H] ratios are obtained by doing a differential analysis with respect to a high S/N solar spectrum from HARPS, obtained from archival observations of the asteroid Vesta. 

The stellar parameters and abundances of the elements are presented in Table \ref{tab:stellar_parameters}. These results are in agreement with the values obtained from the Bayesian analysis described in Sect.~\ref{sec:pastis}. We found that the [X/Fe] ratios of $\alpha$ elements are slightly above solar. This fact together with a metallicity of -0.22\,dex indicate that TOI-220 is probably an old star from the thin disk \citep[e.g.][]{Delgado-19}. Moreover, we used the chemical abundances of some elements to derive ages through the so-called chemical clocks (i.e. certain chemical abundance ratios which have a strong correlation for age). We applied the 3D formulas described in Table~10 of \citet{Delgado-19}, which also consider the variation in age produced by the effective temperature and iron abundance. We used the chemical clocks [Y/Mg], [Y/Zn], [Y/Ti], [Y/Si], [Y/Al], [Sr/Ti], [Sr/Mg] and [Sr/Si] from which we obtain a weighted average age of 10.1\,$\pm$\,1.4 Gyr. This age is in agreement (within their uncertainties) with the age obtained from the Bayesian analysis described in the Sect.~\ref{sec:pastis}.

Following \citet{Johnson-87} we calculated the space velocity components (UVW) with respect to the local standard of rest, adopting the standard solar motion (U$_{\odot}$, V$_{\odot}$, W$_{\odot}$) = (11.1, 12.24, 7.25) km\,s$^{-1}$ of \citet{Schonrich-10}. Then, following \citet{Bensby-03} we calculated the probability of the star belonging to different stellar populations. For more details about the calculations of the kinematic properties of the stars see \citet{Adibekyan-12}. Our calculations suggest that TOI-220 belongs to the galactic thin disk with a 98\,\% probability, in agreement with the chemical classification.

Finally, \citet{Martins2020a} studied TOI-220 in their recent study of 1000 TOIs and found no evidence of a coherent rotation period in the photometry, indicative of a low activity level. With a mean value of the Ca\,{\sc ii} H\,\&\,K lines activity index of $\log$\,R$^{\prime}_\mathrm{HK}$\,=\,$-5.07\pm0.05$, TOI-220 is magnetically less active than the Sun whose $\log$\,R$^{\prime}_\mathrm{HK}$ ranges from $-$4.86 to $-$4.95 from solar maximum to minimum \citep{Hall2007}.


\subsection{Periodogram analysis of the HARPS measurements}
\label{sec:periodograms}

We performed a frequency analysis of the HARPS RVs and spectral diagnostics to search for the Doppler reflex motion induced by TOI-220\,$b$ and unveil the presence of additional signals that might be associated to stellar activity and/or arise from the orbital motion of other planets in the system.

\begin{figure}
\centering
    \includegraphics[width=\linewidth, trim={0 1.9cm 0 0cm}, clip]{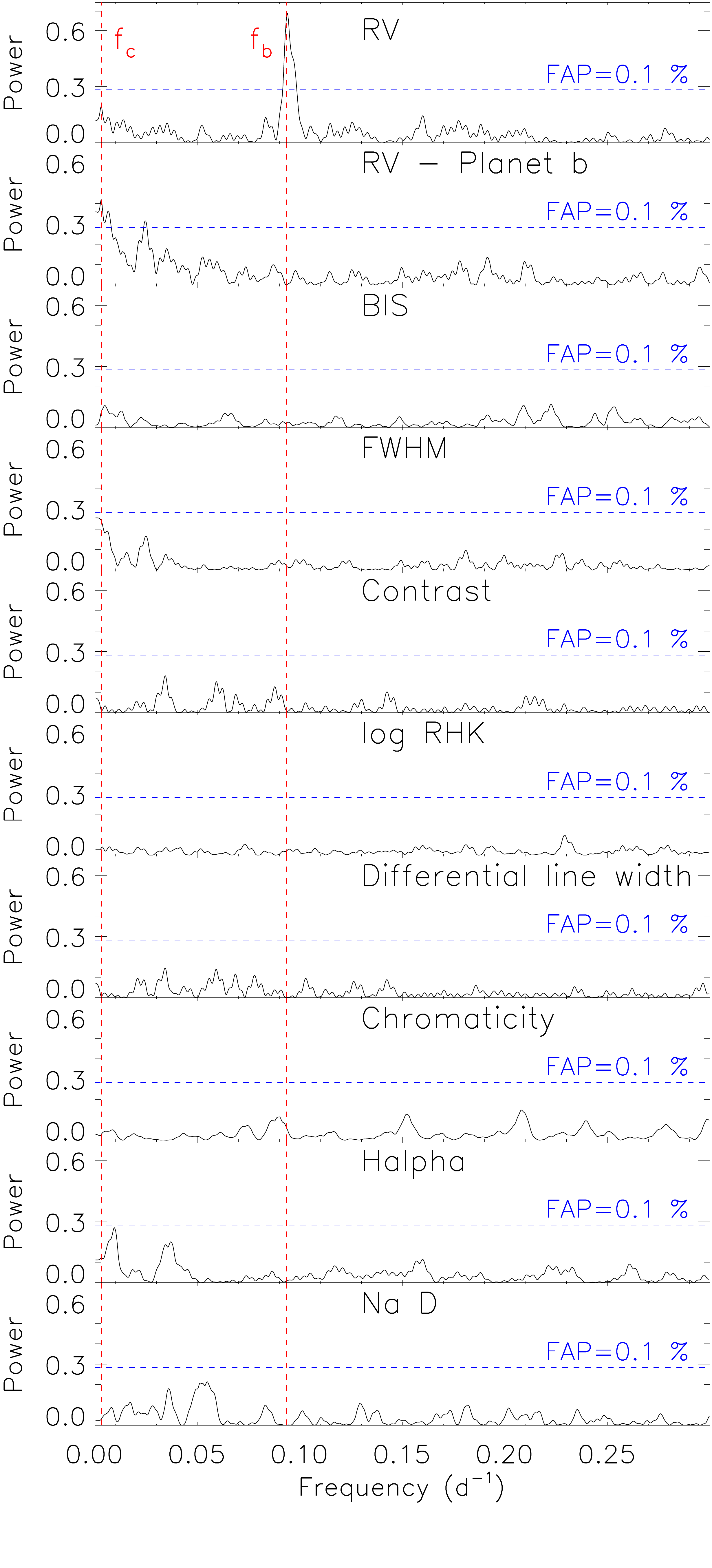}
    \caption{From top to bottom: Generalized Lomb-Scargle periodograms of the HARPS RV measurements (upper panel); the RV residuals following the subtraction of the signal of the transiting planet (second panel); the activity indicators of TOI-220 (remaining panels). The 0.1\,\% false alarm probability (FAP) estimated using the bootstrap method is shown with horizontal dashed lines. The vertical red lines mark the orbital frequencies the transiting planet TOI-220\,$b$ (f$_\mathrm{b}$\,=\,0.093\,d$^{-1}$) and of the additional Doppler signal we found in the HARPS RVs ($f_\mathrm{c}$\,=\,0.0032\,d$^{-1}$). }
    \label{fig:periodograms_activity}
\end{figure}

The generalized Lomb-Scargle (GLS) periodogram \citep{Zechmeister2009} of the HARPS DRS RVs (Fig.~\ref{fig:periodograms_activity}, upper panel) shows a significant peak at f$_\mathrm{b}$\,=\,0.093\,d$^{-1}$, i.e., the orbital frequency of the transiting planet detected in the \textit{TESS} light curve. We estimated its false-alarm probability (FAP) using the bootstrap method described in \cite{Murdoch1993}. Briefly, we computed the periodogram of 10$^6$ mock time-series obtained by shuffling the RV measurements and their uncertainties, while keeping the time-stamps fixed. We defined the FAP as the fraction of those periodograms whose highest power exceeds the observed power of f$_\mathrm{b}$ in the periodogram of the original HARPS data at any frequency. We found no false positives out of our 10$^6$ trials, implying that f$_\mathrm{b}$ has a FAP\,$<$\,10$^{-6}$. Moreover, the peak at f$_\mathrm{b}$ does not appear in any of the periodograms of the activity indicators\footnote{For the CCF's FWHM and BIS we arbitarily adopted uncertanties twice as large as those of the RV measurements.} (Fig.~\ref{fig:periodograms_activity}), confirming that the Doppler signal is induced by the presence of a \emph{bona fide} planet, namely, TOI-220\,$b$.

We subtracted the Doppler signal of the transiting planet TOI-220\,b from the HARPS RVs by fitting a circular model,fixing the period and T$_0$ to the values derived from the modelling of the \textit{TESS} light curves (with T$_0$ being equal to the mid-time of the first \textit{TESS} transit), while allowing for the systemic velocity and RV semi-amplitude to vary. The GLS periodogram of the RV residuals displays significant (FAP\,$\,<\,$\,0.1\,\%) power at frequencies lower than $\sim$0.008\,d$^{-1}$ (125 d), with a peak at about 0.0032\,d$^{-1}$  (309 d) (Fig.~\ref{fig:periodograms_activity}, second panel). Despite this peak being undetected in the periodograms of the activity indicators and line asymmetry diagnostics (Fig.~\ref{fig:periodograms_activity}) -- suggesting the presence of an additional outer companion -- we note that its period is poorly constrained, being comparable to the temporal baseline of our observations (327 d). Moreover, the position of this peak depends on the last poorly-sampled measurements. Therefore, as discussed in Sect.~\ref{sec:rv_1pl}, the exact nature of this long period signal can not be firmly constrained with the currently available RV data.

\subsection{PASTIS analysis}
\label{sec:pastis}
The \textit{Planet Analysis and Small Transit Investigation Software} \citep{Diaz2014, Santerne2015}, {\sc pastis}, was used for the joint analysis of the transit light curves, the radial velocities and the spectral energy distribution of TOI-220. The {\sc pastis} package implements a fully Bayesian analysis of the data and uses the Markov Chain Monte Carlo (MCMC) method to sample the posterior distribution of the resulting parameters \citep[see e.g.][and references therein]{Santerne2019, Lopez2019}.  We draw the posterior distributions of all free parameters after merging the best MCMC chains, i.e. those with the largest model likelihood ($\log$\,{\it L}) from a sample of 30 chains with $10^6$ steps each. The convergence of these chains is checked by a Kologorov-Smirnov (K-S) test after the \textit{burn-in} period (BI) of each chain has been removed. Here, the BI phase corresponds to the initial portion of each chain with a $\log$\,{\it L} mean and variance 2\,$\sigma$ away from the corresponding values of the last 10\,\% of the chain. Then, after removing the BI phase, the selected chains correspond to those with a median $\log$\,{\it L} within 2\,$\sigma$ of the {\it best} chain (i.e. the chain with the largest median $\log$\,{\it L}) and with a K-S test p-value above 10$^{-30}$.

As mentioned in Sect.~\ref{sec:obs}, 17 individual transits of TOI-220\,$b$ were extracted from the \textit{TESS} light curves and input to {\sc pastis}.  
Additionally, a nightly binning was applied to the RV data in order to minimize the jitter induced by stellar activity and/or granulation, ending up thus with 69 points out of the total 91 RV measurements. The transits were modelled using {\sc jktebop} software \citep{Southworth2008} and the RV data using keplerian orbits. 

\subsubsection{Stellar SED analysis}
\label{sec:pastis_sed}

{\sc pastis} also models the host star by fitting the spectral energy distribution (SED) to the BT-Settl atmosphere models \citep{Allard2012}, Dartmouth evolutionary tracks \citep{Dotter2008} and the quadratic limb darkening coefficients for \textit{TESS} bandpass based on the tables from \cite{Claret2011}. 

\begin{figure}
    \centering
    \includegraphics[width=\linewidth]{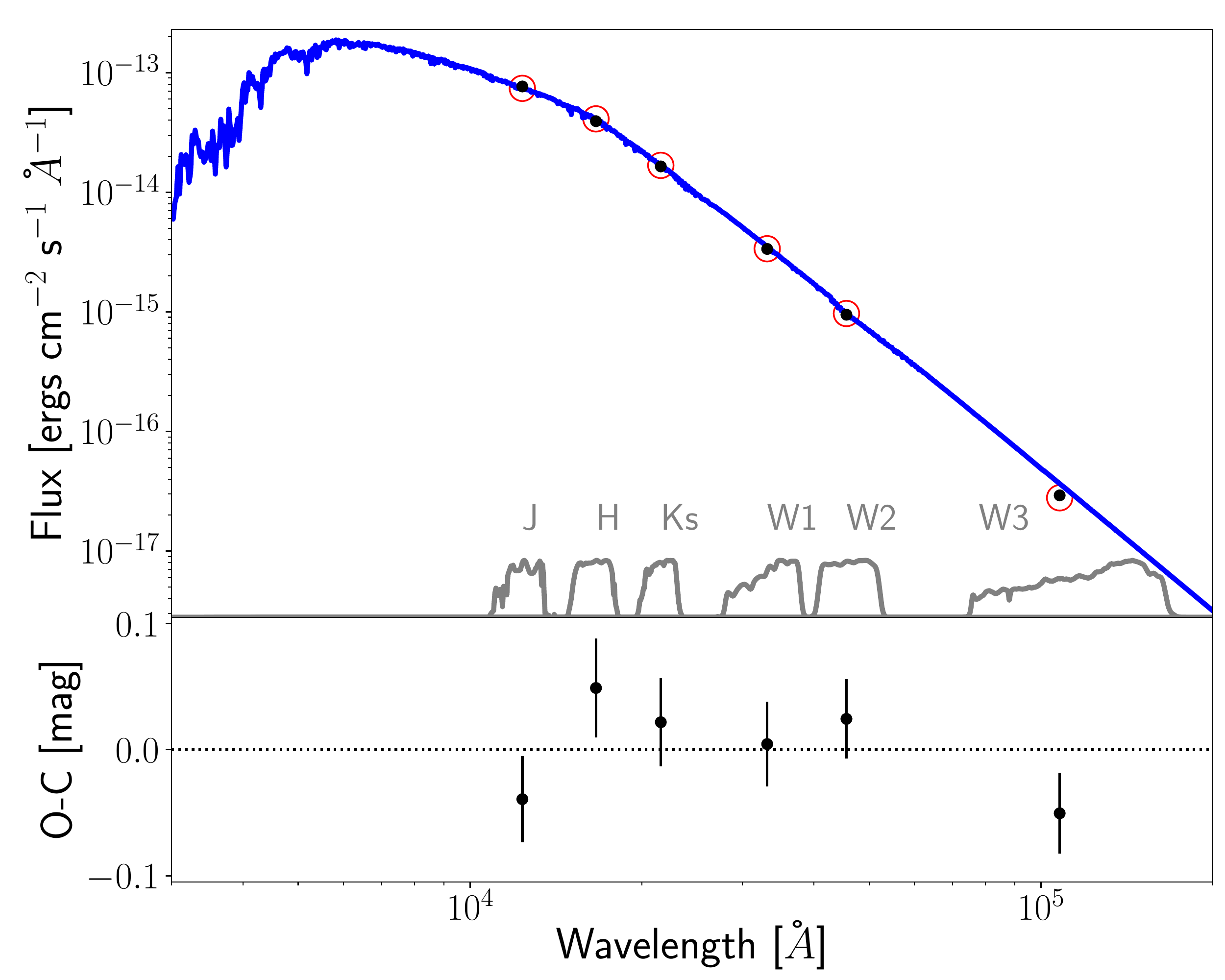} 
    
    \caption{Top panel: spectral energy distribution of TOI-220 with the best stellar atmosphere model from the BT-SETTL library obtained with { \sc pastis} (blue solid line). Residuals are shown in the bottom panel (error bars include the fitted jitter from the analysis).  }
    \label{fig:sed}
\end{figure}

After a preliminary analysis using all the available magnitudes of TOI-220 (Table \ref{tab:star-ids}) we obtained a very poor fit of the visual magnitudes, in particular in the V band. We attributed this result to inconsistencies with the magnitude zero point used or with unreliable V magnitude estimations due to saturation of stars brighter than V=10\,mag in the case of the APASS catalogue \citep{Henden2015}. As {\sc pastis} does not implement the use of Gaia magnitudes by the time of this analysis, we decided to use only 2MASS \citep{2MASS} and AllWISE \citep{AllWISE2013} magnitudes for the bayesian analysis. For this, we took TOI-220's NIR magnitudes as reported in the \textit{TESS} Input Catalogue \citep[TICv8,][]{Stassun2019}. The host star was modelled for the effective temperature, T$_{\text{eff}}$; surface gravity, $\log$\,$g$; metallicity, [M/H]; distance, d; color excess, \textit{E(B$-$V)}; the systemic radial velocity, v$_{0}$, using as priors either normal or uniform distributions as presented in Table~\ref{tab:priors}. The SED and best stellar model obtained with {\sc pastis} are displayed in Fig.~\ref{fig:sed} and the fitted and derived stellar parameters are listed in Table~\ref{tab:stellar_parameters}. The resulting stellar parameters are consistent within their uncertainties with the values obtained independently from the spectral analysis described in Sect.~\ref{sec:spectra_analysis}.

In order to corroborate the TOI-220's effective temperature obtained with {\sc pastis} and to discard any bias we could have introduced by using only NIR magnitudes, the observed and theoretical SEDs for TOI-220 were compared using the Virtual Observatory Spectral Analyzer \citep[VOSA,][]{Bayo2008}. The observed SED was constructed using the four IR bands {\it W1}-{\it W4} from allWISE, the {\it J}, {\it H}, and {\it Ks} bands from 2MASS \citep{2MASS}, and also including the 3 bands from Gaia \citep{Gaia2018}. As before, the theoretical fluxes were computed from the grids of theoretical stellar spectra BT-Settl \citep[AGSS2009,][]{Allard2012}. A best fit with the observed data was computed by $\chi^{2}$ minimization by testing effective temperature values ranging from 1000--7000\,K with steps of 100\,K. We also adopted the estimation of interstellar extinction provided in the \textit{TESS} input catalog \citep{Stassun2019}, then computing the reddening according to \cite{Fitzpatrick1999} and \cite{Schlafly2011}. The SED and the fitted stellar model are shown in Fig.~\ref{fig:SED2}. This independent analysis provides an effective temperature of 5300\,$\pm$\,50\,K, in very good agreement with the result from {\sc pastis}. This analysis also shows no IR-excess, suggesting the absence of circumstellar material around TOI-220, at least in the WISE infrared bands.

\begin{figure}
    \centering
    \includegraphics[width=\linewidth]{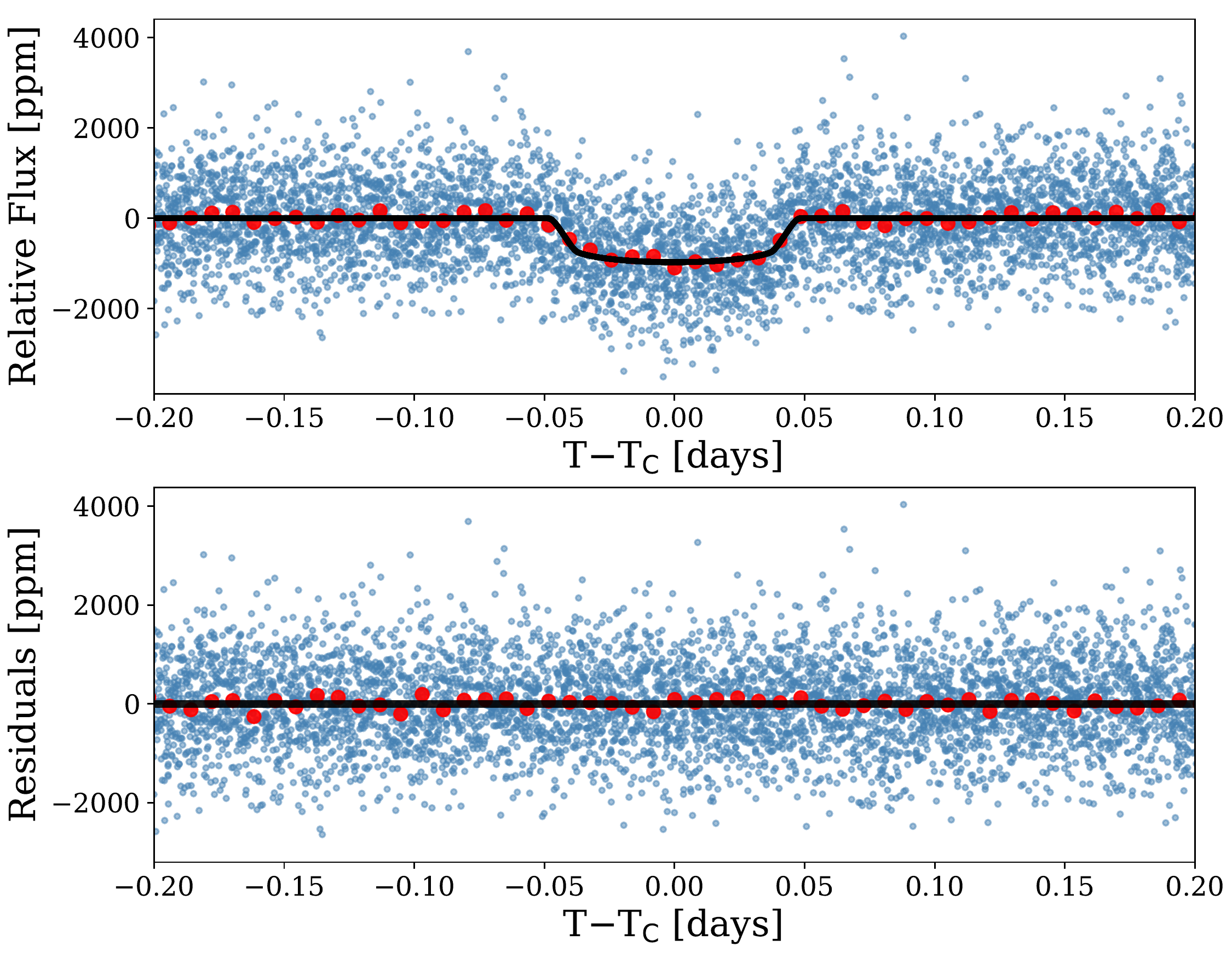} 
    \caption{Top: TOI-220\,$b$ transit light curve built by joining the 17 \textit{TESS} transits used in this work (blue points and 12\,min bins in red) together with the best model (solid curve) obtained with {\sc pastis} analysis. The residuals are shown in the bottom panel. }
    \label{fig:lc_all_joint}
\end{figure}

\subsubsection{Planetary analysis}
\label{sec:rv_1pl}

\begin{figure}
    \centering
    \includegraphics[width=\linewidth]{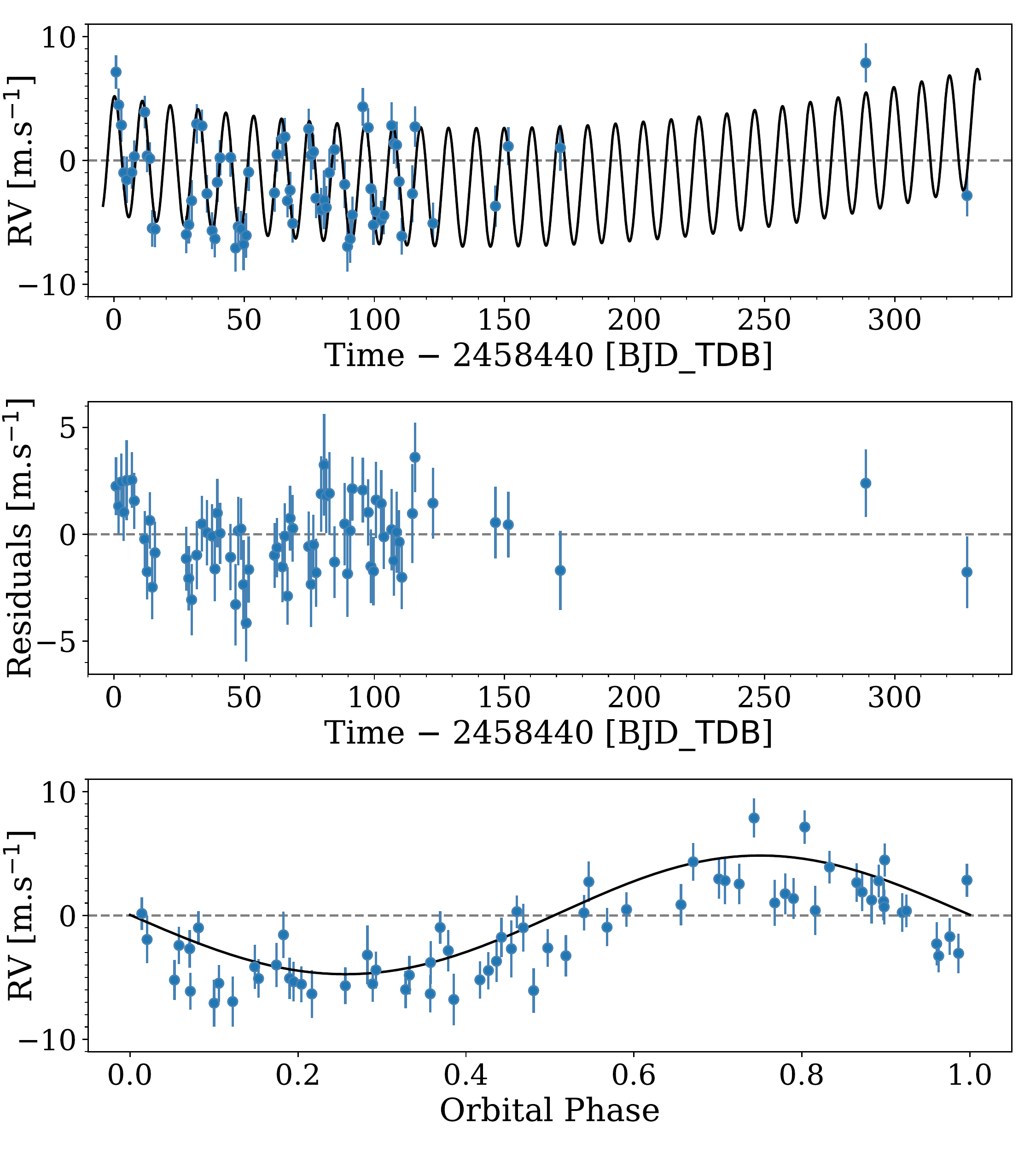} 
    \caption{Radial velocity time series with the best fitted model superimposed with the corresponding residuals are shown in the top and middle panel, respectively. The phased radial velocity measurements are shown in the bottom panel. The error bars include the fitted jitter for each case. Additional, a fitted quadratic long-term is represented by a solid red curve in the middle pannel (see Sect\,\ref{sec:rv_1pl} for details).}
    \label{fig:rvs_models}
\end{figure}

Based on the periodograms presented in Sect.~\ref{sec:periodograms}, the data analysis of the TOI-220 system was performed taking into account the presence of a single planet in the system to probe the 10.69\,d periodic signal. Additionally, we also included a quadratic, long-term RV trend in the modelling.

For the single planet scenario we modelled the planetary orbital period, P; the reference transit epoch, T$_{0}$; the star-planet radius ratio, $R_{\text{s}}/R_{\text{p}}$; the radial velocity semi-amplitude, K; and the inclination, eccentricity and argument of the periastron of the orbit: $i$, $e$ and $\omega$, respectively. 
The prior distribution used for each parameter is reported in Table \ref{tab:priors}. The posterior distribution of all the free parameters was drawn from the best 15 MCMC chains after removing the \textit{burn-in} (BI) time (7--22\,\% of the chain). The full list of fitted and derived parameters of the system together with its respective 68.3\,\% confidence interval are given in Table \ref{tab:stellar_parameters} and \ref{tab:planet_results} for the star and planet, respectively.
Despite the good fit of the keplerian model, it seems that a structure is still present in the RV residuals (see e.g. Fig.~\ref{fig:periodograms_activity}, second panel). To better assess the significance of this trend, we applied the Anderson-Darling (A-D) test \citep{Stephens1974} to the RV residuals (Table~\ref{tab:stats_results}). The A-D test is a modification of the K-S test to evaluate whether a data sample comes from a specific distribution; in our case we tested the RV residuals against a Normal distribution. The test's result is compared then to a set of critical values in a range of significance levels. In particular, the A-D test statistic (see Table~\ref{tab:stats_results}) of the RV residuals is greater than the critical value at 15\,\% significance, 0.580 > 0.547 ($\alpha=0.15$), therefore the normality hypothesis of the RV residuals is rejected evidencing that they may contain astrophysical information not included in our single planet model.

Therefore, we also explored the effect of including a radial velocity quadratic trend in the RV modelling. For this, we used the same initial setup as before but adding two extra terms to account for a quadratic drift in the stellar radial velocity. We set a wide uniform distribution as a prior for the amplitude of the drift coefficients (Table~\ref{tab:priors}). Here, we selected the best 20 MCMC chains after removing the BI (7--27\,\% of the chain) to draw the posterior distributions of the fitted parameters. The fitted drift coefficients are reported in Table~\ref{tab:stellar_parameters} and the RV drift together with the 1-planet model are shown in Fig.\,\ref{fig:rvs_models}. Despite the amplitude of the residuals being smaller when compared to the single planet scenario, the structure in the first epochs (t < 2458440.20 BJD) in the RV residuals seems to remain. But in this case, the normality of the residuals is not rejected by the A-D test. We show in Table~\ref{tab:stats_results} the result of this and other statistical tests which favour the inclusion of the RV drift in the modelling. Notably, the TOI-220\,$b$ parameters are not affected by including this RV long term drift (Table\,\ref{tab:planet_results}). In Appendix\,\ref{sec:RVdrift-mass-constraint} we show the mass constraints of a hypothetical second orbital body based on the estimated RV drift. 
Finally, we also checked the results when including a second planet in the modelling with a wide prior in the orbital period around the peak observed in the RV residual periodogram (300$\pm$100\,d). We found that the RMS of the RV residuals are reduced, however, the resulting parameters for the second planet are poorly constrained. Based on this together with the fact that all the obtained parameters of TOI-220\,$b$ remain fully compatible between models (see Table\,\ref{tab:planet_results}); and specially considering the baseline of the RV monitoring (only 327\,d compared with the periodogram peak of the RV residuals of $\sim$309\,d) and that the location of this peak strongly depends on the data points considered to calculate the periodogram; we adopt the values derived using the 1-planet with the quadratic RV long term drift shown in Tables~\ref{tab:stellar_parameters} and~\ref{tab:planet_results} for the star and planet, respectively. We obtained for TOI-220\,$b$ a planetary mass and radius of M$_{\text{p}}$=13.8$\pm$1.0\,M$_{\earth}$ and R$_{\text{p}}$=3.03$\pm$0.15\,R$_{\earth}$. The fitted orbital period and reference epoch are P=10.695264(86)\,d and T$_{0}$=2458335.9020(14) [BJD\_TDB], respectively. The best model for each individual transit is shown in Fig.~\ref{fig:lcs_all_vertical} and in Fig.~\ref{fig:lc_all_joint} for the complete folded light curve of the 17 \textit{TESS} transits. The best RV model including the RV quadratic drift and its residuals are displayed in Fig.~\ref{fig:rvs_models}.
As a sanity check, we performed an independent joint analysis of the transit photometry and HARPS Doppler measurements using the code \texttt{pyaneti} \citep{Barragan2019}. We adopted the same \texttt{PASTIS} RV models and found consistent results well within the nominal error bars, corroborating the results of our analysis.

\subsection{Transit Timing Variations}
\label{sec:ttv}

\begin{figure}
    \centering
    \includegraphics[width=\linewidth]{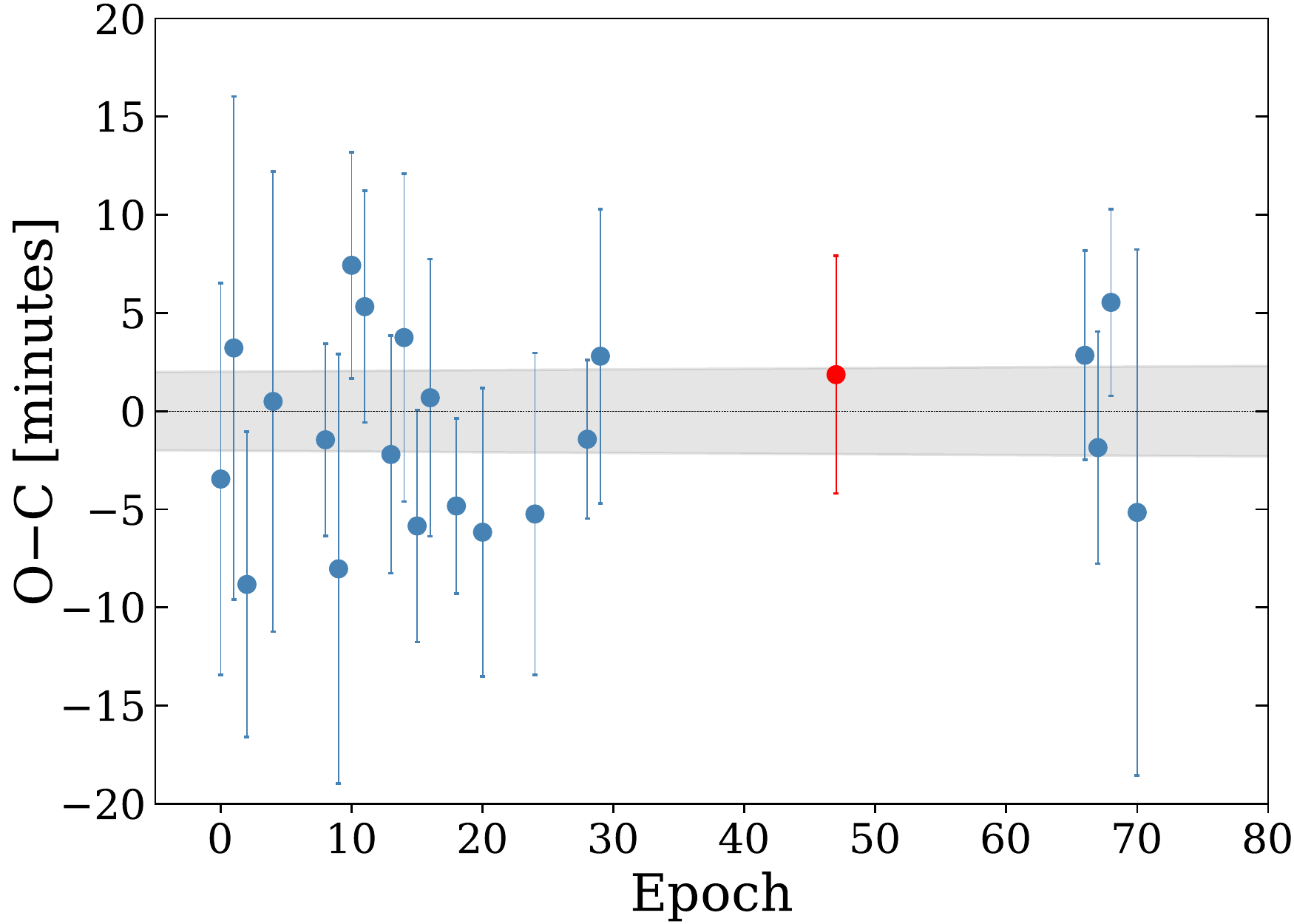} 
    \caption{\textit{Observed minus Calculated} diagram of the transit mid-times of TOI-220\,$b$. Blue symbols correspond to \textit{TESS} transits while the red symbol correspond to the second LCOGT transit observation. The shaded region represents the projected 1$\sigma$ uncertainties of ephemeris equation of TOI-220\,$b$ (Sect. \ref{sec:ttv}).}
    \label{fig:ttv}
\end{figure}

We performed the timing analysis of TOI-220\,$b$ using the 17 \textit{TESS} transits from sectors 1-12, the 4 transits of sectors 27-28; and the 2 transit observations obtained with the LCOGT telescopes. For this, we fixed all the transit parameters to the values obtained from {\sc pastis} analysis in Sect. \ref{sec:rv_1pl} except for the T$_{\text{c}}$, R$_{\text{p}}$/R$_{\text{s}}$ and the inclination which were let free to vary. As the two LCOGT transits have a relative lower quality in terms of photometric precision and sampling; and owing to the fact that they cover only the ingress of the transit (see Fig. \ref{fig:lco_transits}) we fixed also the inclination to i=87.88\,$\deg$ for the modelling of these light curves. For the mid-times of the \textit{TESS} transits, we used as priors a normal distribution with a 0.4\,d width centred in the values obtained from the BLS analysis (Table \ref{tab:tess_transits}). For each transit fit we used 20 MCMC chains of 10$^{5}$ iterations. The obtained values for R$_{\text{p}}$/R$_{\text{s}}$ and inclination are consistent within the uncertainties in all the transits. The resulting transit mid-times are reported in Table \ref{tab:o-c}. The average uncertainty of the transit mid-times is around 7\,min. The first LCOGT transit has an uncertainty larger than 30\,min in comparison with the 6\,min of second LCOGT transit. As this light curve has a very poor photometric quality and transit coverage, we removed it from the analysis. With the obtained mid-times and the ephemeris equation:
\begin{equation}
    \text{T}_{\text{c}}(n)=2458335.9021(14) [\text{BJD\_TBD}] +n\times10.695264(86),
    \label{eq:ephemeris}
\end{equation}
where P and T$_{0}$ are from  Sect. \ref{sec:rv_1pl}; we estimated the timing residuals of TOI-220\,$b$ as presented in Fig.~\ref{fig:ttv} and Table~\ref{tab:o-c}. All the transit times are consistent within the uncertainties with a constant orbital period; the RMS of the O$-$C points is around 4.7\,min. In addition, the statistical A-D test supports the hypothesis that the timing residuals are drawn from a normal distribution, discarding thus, orbital perturbations induced by a possible close companion.

    \begin{table}
        
        \caption{TOI-220 main IDs, celestial coordinates and magnitudes.}
        \label{tab:star-ids}
        \begin{tabular}{lll}
            \hline
            Parameter & value & reference \\
            \hline
            TIC & 150098860 & (1) \\
            TOI & 220 & (2) \\
            TYC & 8897-01263-1 & (1)\\
            APASS &29859752& (3)\\
            2MASS & J06071197-6159487 & (4) \\
            WISE &J060711.93-615949.5 & (5)\\
            Gaia DR2 & 5481210874877547904 & (6) \\
            RA (J2000) & 06:07:11.967 & (1) \\ 
            DEC (J2000)&-61:59:48.895 & (1) \\
            $\mu_{\text{RA}}$ [mas\,yr$^{-1}$] & -17.815 $\pm$ 0.046 &  (6)\\
            $\mu_{\text{DEC}}$ [mas\,yr$^{-1}$] & -68.361 $\pm$ 0.042 & (6)\\
            \hline
            Band & value & reference  \\
            \hline
            \textit{TESS} & $9.688 \pm 0.006$&   (1)    \\
            Johnson-{\it V} &	$10.416 \pm 0.263$ & (3) \\
            Johnson-{\it B} & $11.152 \pm 0.032$ & (3) \\
            2MASS-{\it J}	  & $9.030	\pm0.024$ &  (4) \\
            2MASS-{\it H}&	$8.631	\pm0.031$ & (4)  \\
            2MASS-{\it Ks}&	$8.542	\pm0.025$ & (4)  \\
            WISE-{\it W1}&	$8.476	\pm0.023$ & (5)  \\
            WISE-{\it W2}&	$8.525	\pm0.020$ & (5)  \\
            WISE-{\it W3} & $8.381	\pm0.021$ & (5)  \\
            WISE-{\it W4} & $8.526	\pm0.167$ & (5)  \\
            Gaia {\it G} & 10.1897 $\pm$ 0.003 &  (6) \\
            Gaia {\it B$_{p}$} & 10.6078 $\pm$ 0.0012 &  (6) \\
            Gaia {\it R$_{p}$} & 9.6408 $\pm$ 0.0007 &  (6) \\
            \hline
        \multicolumn{3}{l}{(1) TIC v8 \citep{Stassun2019}.} \\
        \multicolumn{3}{l}{(2) TOI Catalogue (Guerrero et al. Submitted).} \\
        \multicolumn{3}{l}{(3) AAVSO Photometric All-Sky Survey \citep{Henden2015}.} \\ 
        \multicolumn{3}{l}{(4) Two Micron All Sky survey \citep{2MASS}.}\\
        \multicolumn{3}{l}{(5) AllWISE Catalogue \citep{AllWISE2013}.}  \\ 
        \multicolumn{3}{l}{(6) Gaia DR2 \citep{Gaia2018}.}\\ 
        
        \end{tabular} \\
    \end{table}

\normalsize

    \begin{table}

        \caption{Stellar properties of TOI-220.}
        \label{tab:stellar_parameters}
        \begin{center}
        \begin{tabular}{lll}
            \hline
            Parameter & Value & Ref. \\
            \hline
            T$_{\text{eff}}$ [K] & 5273 $\pm$ 115 & (1)  \\ 
            T$_{\text{eff}}$ [K] ({\it adopted}) & 5298 $\pm$ 65 &  (2) \\ 
            T$_{\text{eff}}$ [K] &  $5384 \pm 77$ & (3) \\
            T$_{\text{eff}}$ [K] &  $5182 \pm 45$ & (4) \\
            $\log$\,$g$ [cgs] ({\it adopted}) & 4.37 $\pm$ 0.11 &  (2) \\
            $\log$\,$g$ [cgs] & 4.25 $\pm$ 0.07 &  (4) \\
            $\log$\,$g$ [cgs] & 4.49 $\pm$ 0.03 &  (3) \\
            V$_{\text{turb}}$ [km\,s$^{-1}$] & 0.694 $\pm$ 0.058 & (2) \\
            {}[Fe/H] [dex]& -0.217 $\pm$ 0.044 & (2) \\
            {}[M/H] [dex]& -0.19 $\pm$ 0.05 & (3)\\
            {}[Fe/H] [dex]& -0.2 $\pm$ 0.07 & (4) \\
            {}[Mg/H] [dex]& -0.10 $\pm$ 0.05 & (2) \\
            {}[Si/H] [dex]& -0.14 $\pm$ 0.04 & (2) \\
            {}[O/H]  [dex]& -0.031 $\pm$ 0.072 & (2) \\
            {}[C/H]  [dex]& -0.214 $\pm$ 0.045 & (2) \\
            {}[Cu/H] [dex]& -0.176 $\pm$ 0.044 & (2) \\
            {}[Zn/H] [dex]& -0.172 $\pm$ 0.027 & (2) \\
            {}[Sr/H] [dex]& -0.156 $\pm$ 0.077 & (2) \\
            {}[Y/H]  [dex]& -0.293 $\pm$ 0.094 & (2) \\
            {}[Zr/H] [dex]& -0.236 $\pm$ 0.071 & (2) \\
            {}[Ba/H] [dex]& -0.237 $\pm$ 0.025 & (2) \\
            {}[Ce/H] [dex]& -0.166 $\pm$ 0.045 & (2) \\
            {}[Nd/H] [dex]& -0.100 $\pm$ 0.039 & (2) \\
            $v$\,sin\,$i$ [km\,s$^{-1}$]& 2.90 $\pm$ 0.35 & (4) \\
            V$_{\text{mac}}$ [km\,s$^{-1}$]& 0.9 $\pm$ 0.4 & (4)\\
            V$_{\text{mic}}$ [km\,s$^{-1}$]& 0.85 $\pm$ 0.10& (4)\\
            Age [Gyr] ({\it adopted}) & 10.1 $\pm$ 1.4 & (2) \\
            \hline
            Age [Gyr] & 10.7$^{+2.1}_{-3.1}$& (3) \\
            Distance [pc]  & 89 $\pm$ 4 & (3)\\
            {}{\it E(B$-$V)} [mag] & 0.163 $\pm$ 0.087 &(3) \\
            Systemic RV, v$_{0}$ [m\,s$^{-1}$]  & 26.45827 $\pm$ 0.00037 & (3)\\
            RV drift lin. coeff. [m\,s$^{-1}$\,d$^{-1}$] & -0.0379(80) & (3) \\
            RV drift quad. coeff. [m\,s$^{-1}$\,d$^{-2}$]& 0.000149(31) & (3) \\
            Radius [R$_{\sun}$] & 0.858 $\pm$ 0.032 & (3)\\
            Mass [M$_{\sun}$] & 0.825 $\pm$ 0.028  & (3) \\
            Density [$\rho_{\sun}$] &1.31 $\pm$ 0.15  & (3) \\
            Limb darkening coeff. u$_{\text{a}}^{\text{\tiny{TESS}}}$ &0.352 $\pm$ 0.014 & (3)\\
            Limb darkening coeff. u$_{\text{b}}^{\text{\tiny{TESS}}}$ &0.2475 $\pm$ 0.0077& (3)\\
            \hline
            \multicolumn{3}{l}{(1) TIC v8 \citep{Stassun2019}.} \\ 
             \multicolumn{3}{l}{(2) From spectral analysis (Sect.\ref{sec:spectra_analysis}). Values correspond to weighted }\\
            \multicolumn{3}{l}{ average and the standard deviation.} \\ 
            \multicolumn{3}{l}{(3) From {\sc pastis} analysis (Sect. \ref{sec:pastis}). Derived values assume}\\
            \multicolumn{3}{l}{R$_{\sun}$=695\,508\,km and M$_{\sun}$= 1.98842 $\times$ 10$^{30}$\,kg.} \\
            \multicolumn{3}{l}{(4) From spectral analysis validation (Appendix \ref{sec:stellar_prop2})}.\\
            \end{tabular} \\
        \end{center}
    \end{table}

    \begin{table*}
        \caption{Results on the planetary parameters obtained from {\sc pastis} simultaneous analysis of the data assuming a single planet system and by including a long term quadratic drift in the RVs. Derived values assume M$_{\earth}$ = 5.9736$\times$10$^{24}$\,kg, R$_{\earth}$ = 6\,378\,137\,m, AU = 149\,597\,870.7\,km and zero albedo for the equilibrium temperature.}
        \label{tab:planet_results}
        \begin{center}
        
        \begin{tabular}{lcc}
            \hline
            Parameter (fitted) & 1 Planet & 1 Planet + RV drift ({\it adopted})  \\
            \hline
            Orbital period, P [d] & 10.695264(87) & 10.695264(86) \\
            Reference transit time,  T$_{0}$ [BJD$\_$TDB] & 2458335.9021(14) & 2458335.9020(14) \\
            Planet-to-star radius ratio, R$_{\text{p}}$/R$_{\text{s}}$& 0.03227 $\pm$ 0.00078 & 0.03235 $\pm$ 0.00076 \\
            Orbital inclination, $i$ [$\degr$] &$87.86 \pm 0.14 $&  $87.88 \pm 0.12$\\
            RV semi-amplitude, K [m\,s$^{-1}$] &  4.47 $\pm$ 0.36 & 4.56 $\pm$ 0.32\\
            Orbital eccentricity, $e$ &  $0.029^{\tiny{+0.034}}_{-0.021}$& $0.032^{+0.038}_{-0.023}$\\
            Argument of periastron, $\omega$ [$\degr$] & 159 $\pm$ 150 & $248^{+66}_{-190}$\\
            \hline
            Parameter (derived) && \\
            \hline
            Mass [M$_{\earth}$] & $13.6 \pm 1.2$ & $13.8 \pm 1.0$ \\
            Radius [R$_{\earth}$] & $3.02 \pm 0.17$  &  $3.03 \pm 0.15$ \\
            bulk density, $\rho_{\text{p}}$ [g\,cm$^{-3}$] & 2.69 $\pm$ 0.55 & 2.73 $\pm$ 0.47  \\
            System scale, a/R$_{\text{s}}$ & 22.3 $\pm$ 1.0 & 22.33 $\pm$ 0.80  \\
            Impact parameter, $b$ & $0.832 \pm 0.026$ & $0.836 \pm 0.022 $\\
            Transit duration, T$_{14}$ [h] & 2.231 $\pm$ 0.054 & 2.239 $\pm$ 0.055\\
            orbital semi-major axis, a [AU] &  0.08920 $\pm$ 0.0010 & 0.08911 $\pm$ 0.0010\\ 
            Mean equilibrium temperature, T$_{\text{eq}}$ [K]& 806 $\pm$ 25 & 805 $\pm$ 21\\
            \hline
            Instrument-related parameters &&\\
            \hline 
            HARPS jitter [m\,s$^{-1}$] & 1.71000 $\pm$ 0.00021 &1.35000 $\pm$ 0.00024 \\
            SED jitter [mag] & 0.048 $\pm$ 0.025 &0.049 $\pm$ 0.026 \\
            \textit{TESS} jitter [ppm] & 111$\pm$74 & 111$\pm$ 74 \\
            \textit{TESS} out-of-transit flux & 1.000093(30) & 1.000093(30)  \\
            \hline
        
        \end{tabular}\\
        \end{center}
    \end{table*}
\normalsize

\section{Planetary Structure}
\label{sec:composition}

\begin{figure}
    \centering
    \includegraphics[scale=0.4]{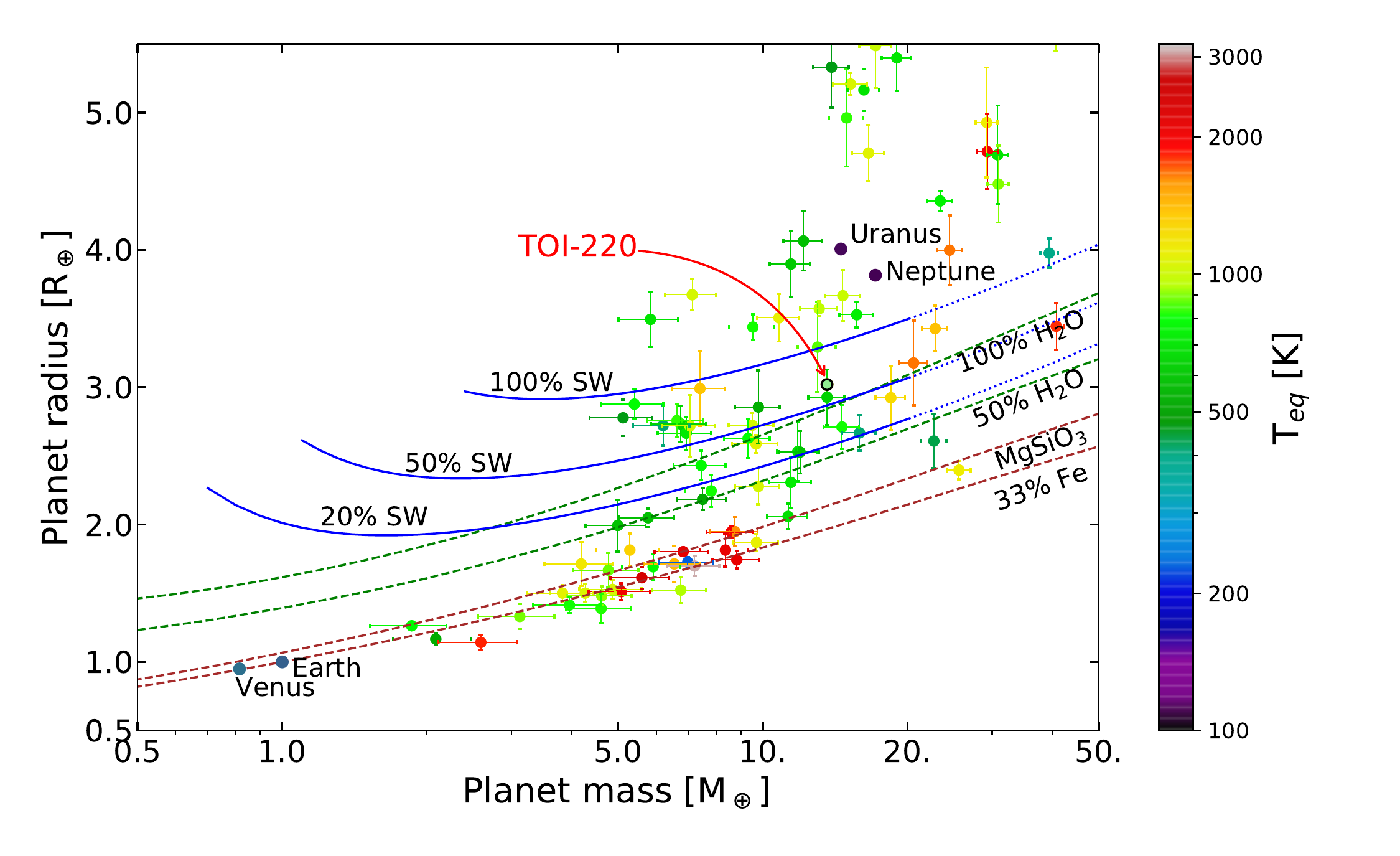} 
    \caption{Mass-Radius diagram of the confirmed exoplanets with mass and radius measurements better than 20\,\% and 10\,\%, respectively. The symbol's color scales with the equilibrium temperature of the planet. TOI-220\,$b$ location in the diagram is highlighted with the red arrow and black circle. Solar System planets are also shown for reference. Composition models from \citet{Zeng2019} are displayed with dash brown and green lines. Those for a planet at 800\,K with an Earth-like composition topped with various proportion of a steam and supercritical water layer \citep{mousis20} are displayed with blue solid lines.} 
    \label{fig:MRTemp}
\end{figure}

\begin{table}
        \centering
        \caption{Output parameters retrieved by the MCMC interior and atmosphere modeling.}
        \label{tab:internal_structure_results}
        \begin{tabular}{lc}
            \hline
            Parameter & Value \\
            \hline
            Total mass, M$_{\text{p}}$ [M$_{\oplus}$] & 13.8$\pm$0.7 \\
            Total radius, R$_{\text{p}}$ [R$_{\oplus}$] & 3.06$\pm$0.12 \\
            Total density, $\rho_{\text{p}}$ [g cm$^{-3}$] & 2.65$\pm$0.28 \\
            Core mass fraction, CMF & 0.08$\pm$0.03 \\
            Water mass fraction, WMF & 0.62$\pm$0.10 \\
            Fe/Si mole ratio & 0.64$\pm$0.11 \\
            Mg/Si mole ratio & 1.16$\pm$0.10 \\
            Mg number, \# Mg & 0.85-1.0 \\
            Temperature at 300 bar, T$_{\text{surf}}$ [K] & 3536$\pm$203 \\
            Planetary albedo, $a_{\text{p}}$ & 0.230$\pm$0.001 \\
            Atmospheric thickness, z [km] & 1111$\pm$99 \\
            Atmospheric mass, M$_{\text{atm}}$ [M$_{\oplus} \ 10^{-3}$] & 1.31$\pm$0.16 \\
            Core+Mantle radius, [R$_{\text{p}}$ units] & 0.47$\pm$0.05 \\
        
            \hline
        \end{tabular}
    \end{table}

\begin{figure}
    \centering
    \includegraphics[width=\linewidth]{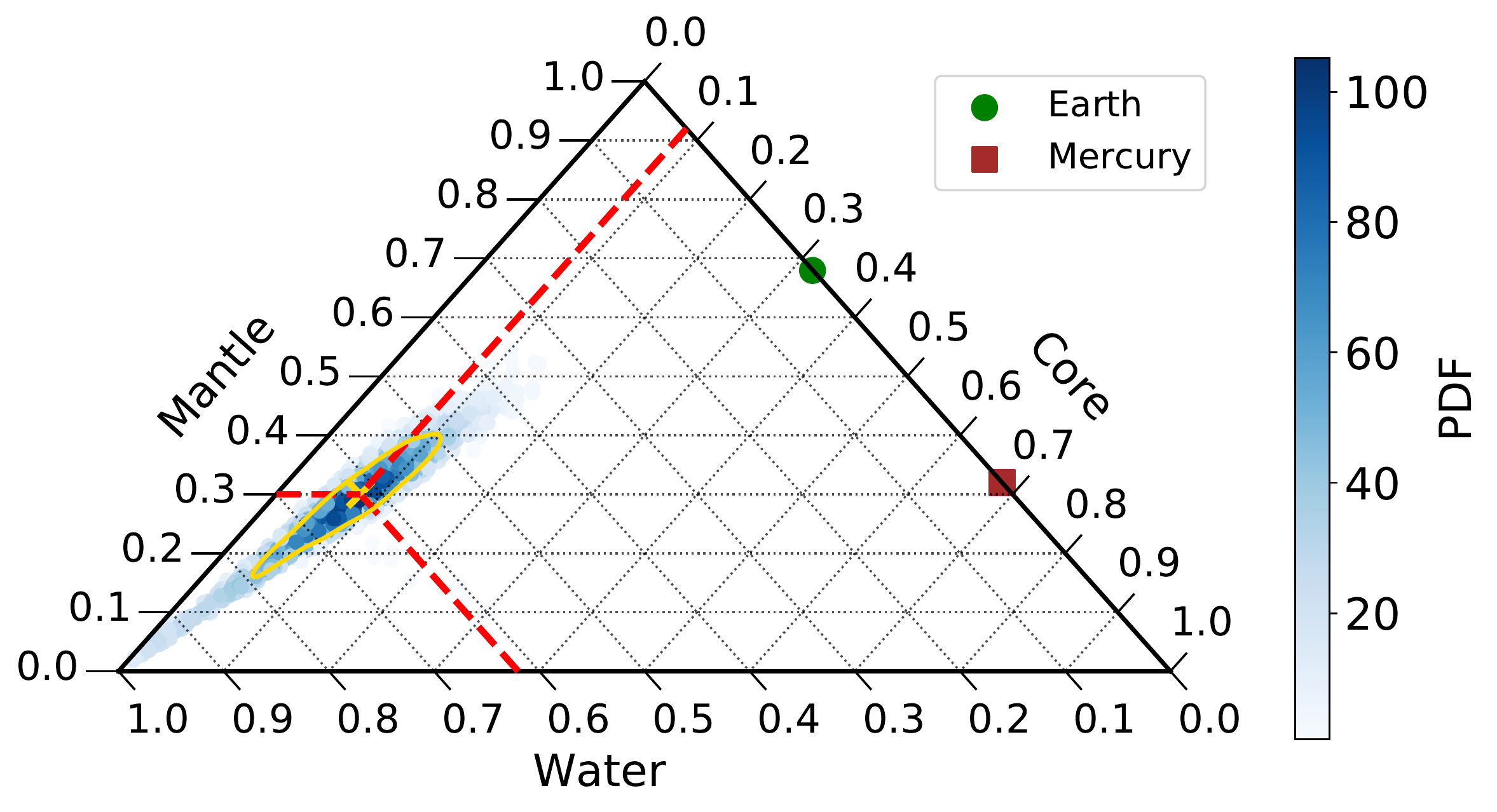}
    \caption{Sampled 2D marginal posterior distribution function for the CMF and WMF of TOI-220\,$b$ (blue region). The PDF mean and the 1$\sigma$ confidence interval is marked by the yellow cross and curve, respectively. TOI-220\,$b$ is consistent with a water rich planet (WMF=0.62$\pm$0.10) with a very small core (CMF=0.08$\pm$0.03). Earth and Mercury values are presented for comparison. } 
    \label{fig:ternary_diagram}
\end{figure}

\begin{figure}
    \centering
    \includegraphics[width=\linewidth]{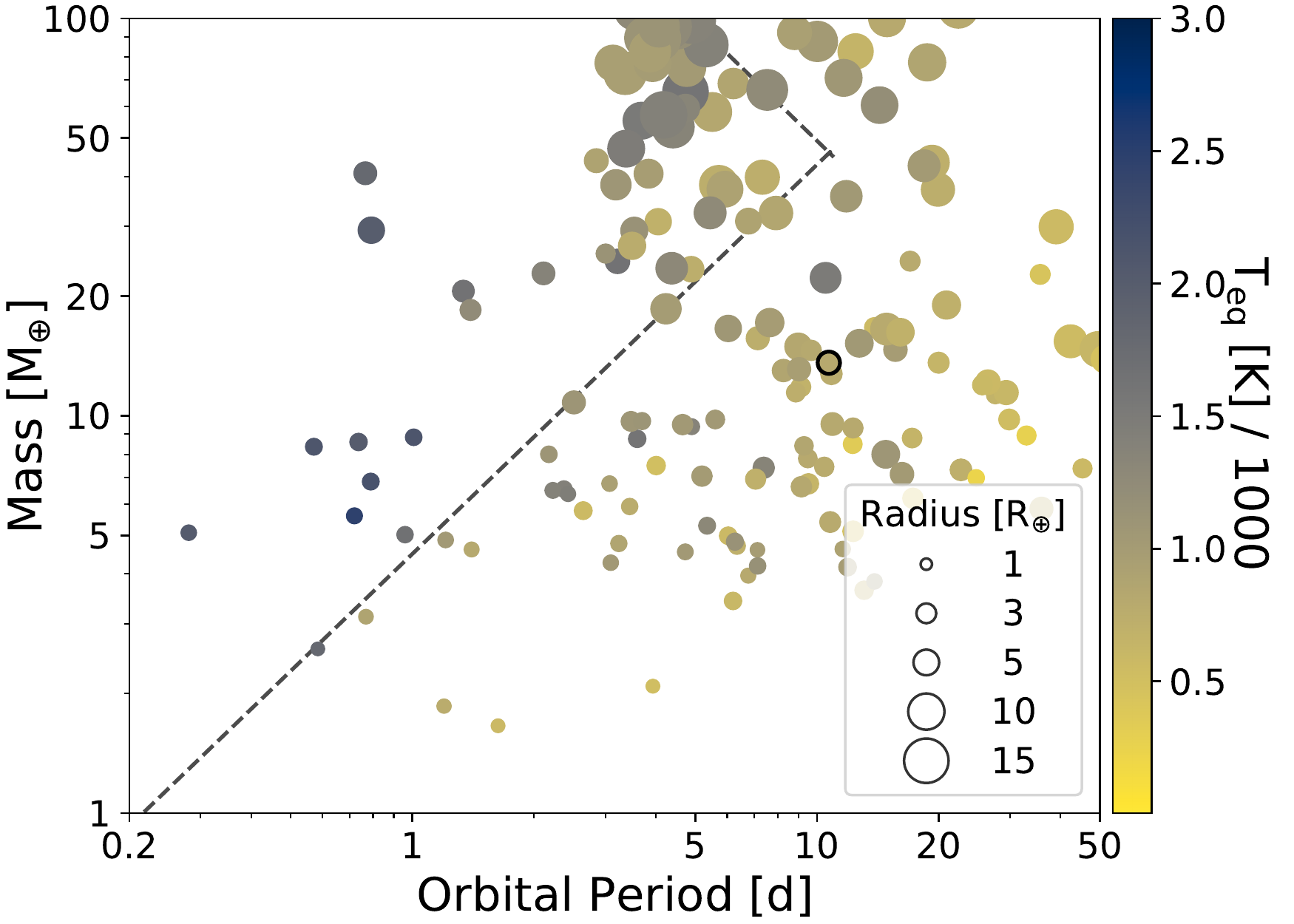}   
    \caption{Mass-Period diagram of the confirmed exoplanets to date with masses and periods below 100\,M$_{\earth}$ and 50\,d, respectively (same planets sample as Fig.~\ref{fig:MRTemp}). The symbols size represents the radius of the planet while its equilibrium temperature is coded by the color of the symbol. TOI-220\,$b$ is outlined with a black circle. The dashed lines mark the edges of the Neptune-desert from \citet{Mazeh2016}.}
    \label{fig:MPdiagram}
\end{figure}

Figure \ref{fig:MRTemp} shows the mass-radius diagram of all the confirmed exoplanets to date\footnote{from \url{http://exoplanet.eu/} as of 27 January 2021.}, with a precision better than 20\,\% and 10\,\% in their mass and radius, respectively. TOI-220\,$b$ is well located in a domain of the sub-Neptune size planets. We investigate the interior of TOI-220\,$b$ in order to get some insights into its composition. To that purpose we used the model of internal structure developed by \cite{brugger17} and recently updated by \cite{mousis20} to include a steam and supercritical water layer for the special case of highly-irradiated planets. When the pressure and temperature are above the critical point of water, water is in supercritical phase. To describe this water layer, we use the Equation of State (EOS) of \cite{Mazevet2019}, which includes experimental and theoretical data of water at pressures and temperatures higher than the critical point of water. TOI-220\,$b$ being strongly irradiated with an equilibrium temperature of 806~K, the model allows to address how its composition could differ from that of a Neptune-like planet with a large gaseous envelope overlying a rocky core. We followed the method described in \cite{Lillo-Box2020} and \cite{Acuna2021} exploring the core mass fraction (CMF, the ratio of the mass of the core and the total planetary mass) and the water mass fraction (WMF, the ratio of the mass of the water layer -steam and supercritical- and the total mass) of the planet as free parameters in an MCMC Bayesian framework adapted from \cite{dorn15}. 

We use as input parameters the planetary mass and radius, and the Fe/Si and Mg/Si mole ratios. The latter are calculated with the stellar abundances in Table \ref{tab:stellar_parameters} as Fe/Si = 0.65$\pm$0.09 and Mg/Si = 1.17$\pm$0.17. The code allows to explore the posterior distribution functions of three compositional parameters: CMF, WMF, and the Mg number ($\#$\,Mg), which is a proxy for the level of differentiation of the planet as described by \cite{sotin07} and \cite{brugger17}. If most of the Mg is located in the mantle and Fe in the core, the planet is highly differentiated and $\#$\,Mg is close to one (0.9 for the case of Earth). On the contrary, if Mg and Fe are mixed in the mantle, $\#$\,Mg will have lower values ($\simeq$ 0.6). In the MCMC algorithm, we assume uniform prior distributions of the CMF, WMF and the Mg number.

Table \ref{tab:internal_structure_results} shows the retrieved values of the three free compositional parameters, in addition to the observables (mass, radius and mole ratios), and the atmospheric variables. The posterior distributions of the CMF and WMF are also shown in Fig.~\ref{fig:ternary_diagram}. We used an MCMC algorithm to obtain samples of the posterior distribution of all our observable parameters: mass, radius, Fe/Si and Mg/Si. Convergence is achieved when the output parameters of the model reproduce the observed values within the 1$\sigma$ confidence intervals. Our results indicate that a planetary structure composed of a core rich in Fe, a silicate mantle and a supercritical water layer topped by a steam atmosphere in radiative-convective equilibrium is a likely scenario for TOI-220\,$b$.

Since its host star is less enriched in Fe than the Sun and Earth (Fe/Si = 0.93), its CMF is significantly lower ($0.08\pm0.03$) than the terrestrial value, 0.32. As can be seen in Table \ref{tab:internal_structure_results}, the large $\#$\,Mg obtained (0.85--1.0) is indicative of a highly-differentiated core and mantle configuration such as the Earth ($\#$\,Mg=0.9). Furthermore, as shown in Fig.~\ref{fig:ternary_diagram}, TOI-220\,$b$ could be a water-rich planet with a minimum WMF of 52\%, which is compatible with the composition derived for water-rich satellites in the Solar System, such as Titan, and a maximum WMF of 72\%, which is below the average water proportion found in comets \citep[see Fig. 12 of][]{McKay2019}. Moreover, most of the water mass would be under pressure and temperature conditions beyond the supercritical point, forming a supercritical and steam water layer that would constitute 47\% of the total radius.

\section{Discussion}
\label{sec:discussion}

\begin{figure}
    \centering
    \includegraphics[width=\linewidth]{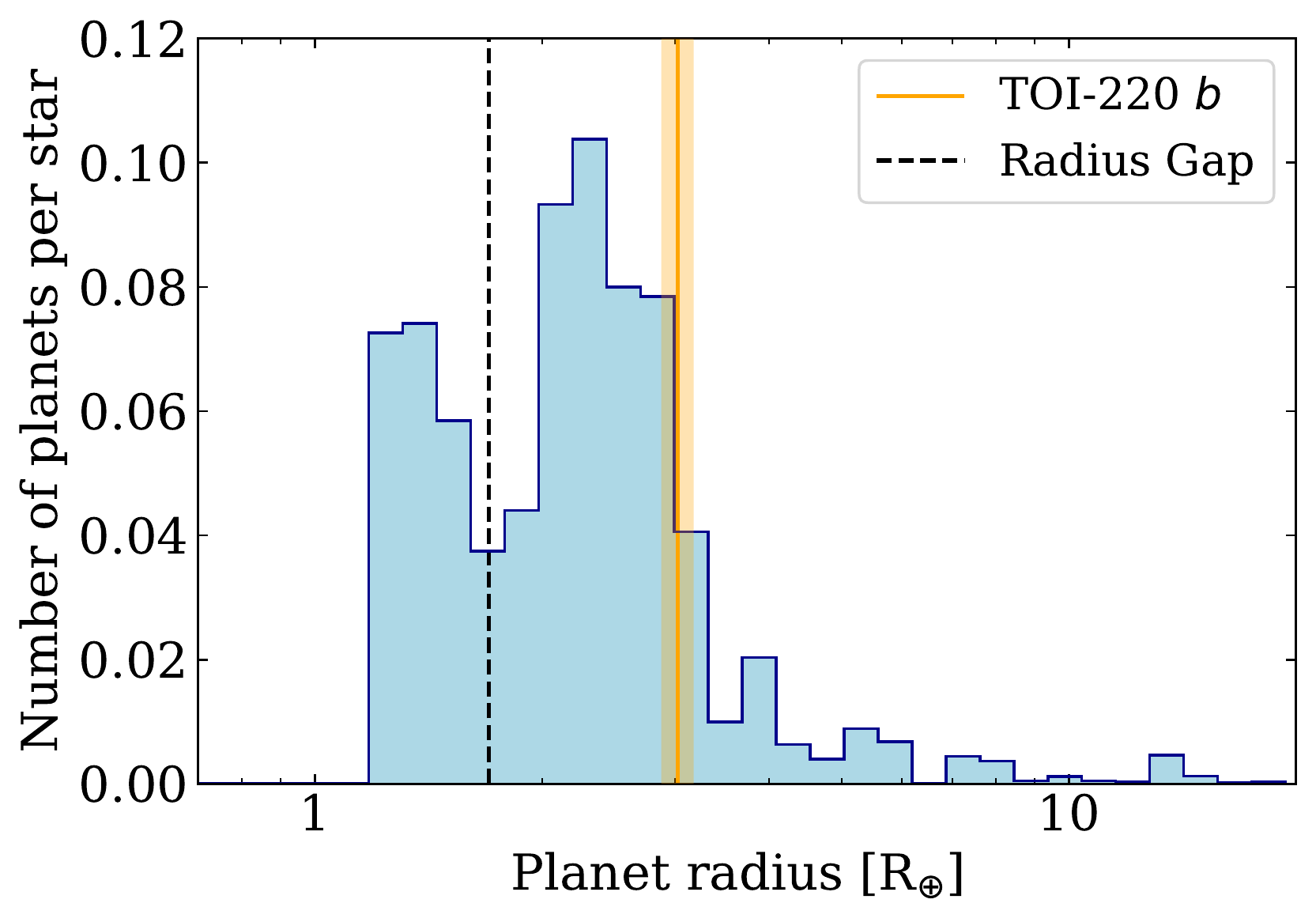}   
    \caption{Histogram of the planetary radius for planets with periods below 100\,d as presented in Fulton \& Petigura (2018).  The location of the radius gap and the size of TOI-220\,$b$ ($\pm$1\,$\sigma$) are shown with the black-dashed and orange-shaded lines, respectively.}
    \label{fig:radiusgap}
\end{figure}

The combined analysis of the photometric and spectroscopic data has fully confirmed the planetary nature of the TOI-220\,$b$ transiting candidate. 
We noted the presence of a structure in the RV residuals on the exact nature of which, instrumental or planetary, we could not conclude with the  data currently available. Therefore, any further investigation of the long-period signals would requires additional RVs. Thus, we adopt the values derived using a single planet model in the system and including a quadratic long term trend in the RVs.
In Fig.~\ref{fig:MPdiagram} we show the mass-period diagram of all the confirmed exoplanets including TOI-220\,$b$ (same sample as Fig.~\ref{fig:MRTemp}). The symbol size scales with the radii of planets while the color represents the planet's equilibrium temperature. TOI-220\,$b$ is located in the region of the \textit{warm sub-Neptunes} class of exoplanets (T$_{\text{eq}}$<1000\,K) with a relatively large size (3\,R$_{\earth}$).

A paucity of exoplanets has been reported in the planet size distribution around 1.5-2.0\,R$_{\earth}$ \citep[e.g.][]{Fulton2017, Fulton2018, VanEylen2018}, in particular for a M$_{\star} \sim$ 0.8\,M$_{\sun}$ the gap is around 1.6\,R$_{\earth}$. It is suggested that this gap is driven by photoevaporation mass-loss which acts shifting planets with gaseous envelopes towards a rocky Super-Earth population (R<1.7\,R$_{\earth}$). TOI-220\,$b$ lies above the gap in the {\it Fulton's diagram} (see Fig.\,\ref{fig:radiusgap}) in the not so populated high end of the size distribution of planets (R$\gtrsim$3\,R$_{\earth}$) orbiting low mass stars (M$_{\star}$<0.97\,M$_{\sun}$), as can be seen in Fig.\,8 and 9 in \cite{Fulton2018}. The atmospheric composition of TOI-220\,$b$, either with a thick H/He layer or water rich content, suggests a formation in the protoplanetary disk past the snow line followed by slow migration. The host's age of $\sim$10\,Gyr points to a very evolved system and the large planet’s radius suggests the planet was massive enough to keep its atmosphere despite the large irradiation received by its host star. 
To get a first estimate of the atmospheric escape of TOI-220\,$b$, we used equations of Aguichine et al. (Submitted). We found that the mass lost by Jeans' escape is negilible even for light species (H/He) due to the surface gravity of the planet. Only $\sim$0.1\,M$_{\earth}$ could be removed by XUV photo-evaporation along its 10.1\,Gyr of life. This picture is consistent with the idea that the fate of warm sub-Neptunes located above the gap in the {\it Fulton's diagram} (Fig.\,\ref{fig:radiusgap}) is not to end their evolution as naked cores \citep[see][and references therein]{Deleuil2020, Bean2021}. Precise measurements of physical parameters of planets and its host stars, like the ones presented in this work, are key to shape the knowledge of the formation and evolution processes that sculpt the planet population in this region of parameter space.

Complementarily, we have performed an interior structure and atmosphere analysis within a Bayesian framework to constrain the composition of TOI-220\,$b$ given its density and the stellar abundances of the host star. The low density of this strongly irradiated planet could be explained with a water-rich atmosphere that reaches the supercritical phase at its base, with a silicate mantle and an iron core as its bulk. As an alternative the planet might have a solid core surrounded by a thick H/He atmosphere. Future work should explore this possibility but also a composition with a well-mixed water and H/He atmosphere, since these are the most common volatiles that can form low-mass planets. The confirmation of the water-rich atmosphere scenario would also require conducting atmospheric characterisation to search for water spectral features. Following \cite{Kempton2018} we thus calculate TOI-220\,$b$ transmission spectroscopy metric to check whether the planet would be amenable to such a characterization by JWST or the ELT. We obtained a value of 45, which is well below the threshold of 84 for planets in this range of size, making of TOI-220\,$b$ atmosphere likely out of reach for the next decade.

\section*{Acknowledgements}
SH acknowledges CNES funding through the grant 837319. MD acknowledges the CNES support. V.A., E.D.M., S.G.S., N.C.S., S.C.C.B. and S.Hoj. acknowledge support by FCT - Funda\c{c}\~ao para a Ci\^encia e a Tecnologia (Portugal) through national funds and by FEDER through COMPETE2020 - Programa Operacional Competitividade e Internacionaliza\c{c}\~ao by these grants: UID/FIS/04434/2019; UIDB/04434/2020; UIDP/04434/2020; PTDC/FIS-AST/32113/2017 \& POCI-01-0145-FEDER-032113; PTDC/FIS-AST/28953/2017 \& POCI-01-0145-FEDER-028953. V.A., E.D.M., S.C.C.B., S.G.S. and O.D.S.D further acknowledge the support from FCT through Investigador FCT contracts IF/00650/2015/CP1273/CT0001, IF/00849/2015/CP1273/CT0003, IF/01312/2014/CP1215/CT0004, CEECIND/00826/2018--POPH/FSE(EC) and DL57/2016/CP1364/ CT0004, respectively. S.Hoj. further acknowledges the support by FCT through the fellowship PD/BD/128119/2016. DJA acknowledges support from the STFC via an Ernest Rutherford Fellowship (ST/R00384X/1). L.M.S. and D.G. gratefully acknowledge financial support from the CRT foundation under Grant No. 2018.2323 ``Gaseous or rocky? Unveiling the nature of small worlds". JRM, BLCM and ICL acknowledge continuous grants from the brazilian funding agencies CNPq, CAPES and FAPERN. This study was financed in part by the Coordena\c{c}\~{a}o de Aperfei\c{c}oamento de Pessoal de N{\'i}vel Superior-Brasil (CAPES)-Finance Code 001. PK and JS acknowledge the MSMT INTER-TRANSFER grant LTT20015. This paper includes data collected with the \textit{TESS} mission, which are publicly available from the Mikulski Archive for Space Telescopes (MAST). Funding for the {\it TESS} mission is provided by NASA's Science Mission directorate. We acknowledge the use of public \textit{TESS} Alert data from pipelines at the \textit{TESS} Science Office and at the \textit{TESS} Science Processing Operations Center. This research has made use of the Exoplanet Follow-up Observation Program website, which is operated by the California Institute of Technology, under contract with the National Aeronautics and Space Administration under the Exoplanet Exploration Program. Resources supporting this work were provided by the NASA High-End Computing (HEC) Program through the NASA Advanced Supercomputing (NAS) Division at Ames Research Center for the production of the SPOC data products. This research has made use of computing facilities operated by CeSAM data center at LAM, Marseille, France; and the Data \& Analysis Center for Exoplanets (DACE), which is a facility based at the University of Geneva (CH) dedicated to extrasolar planets data visualisation, exchange and analysis. DACE is a platform of the Swiss National Centre of Competence in Research (NCCR) PlanetS, federating the Swiss expertise in Exoplanet research. This research has made use of data obtained from the portal exoplanet.eu of The Extrasolar Planets Encyclopaedia. This research has made use of the NASA Exoplanet Archive, which is operated by the California Institute of Technology, under contract with the National Aeronautics and Space Administration under the Exoplanet Exploration Program. This work makes use of observations from the LCOGT network. D.G. warmly thanks Alessia Barbiero and Pietro Giordano for the inspiring conversations at the time the HARPS follow-up of TOI-220 was started in November 2018.\\

\textit{Software}: We gratefully acknowledge the open source software which made this work possible: \texttt{astropy} \citep{astropy2013, astropy2018}, \texttt{ipython} \citep{ipython}, \texttt{numpy} \citep{VanDerWalt2011}, \texttt{scipy} \citep{scipy}, \texttt{matplotlib} \citep{matplotlib}, \texttt{jupyter} \citep{jupyter2016}, \texttt{pandas} \citep{pandas-scipy-2010,pandas2020}.\\

\textit{Facilities}: \textit{TESS}, ESO:3.6m (HARPS spectrograph), Gemini:South (Zorro), LCOGT, Exoplanet Archive

\section{Data Availability}
The TESS photometric data used in this work is available via Mikulski Archive for Space Telescopes (MAST) archive and TFOP program.  Radial velocity measurements and the derived stellar activity indicators are provided in Tables \ref{tab:rv_data} and \ref{tab:serval_data}, respectively.




\bibliographystyle{mnras}
\bibliography{references}

\begin{thebibliography}{}
\makeatletter
\relax
\def\mn@urlcharsother{\let\do\@makeother \do\$\do\&\do\#\do\^\do\_\do\%\do\~}
\def\mn@doi{\begingroup\mn@urlcharsother \@ifnextchar [ {\mn@doi@}
  {\mn@doi@[]}}
\def\mn@doi@[#1]#2{\def\@tempa{#1}\ifx\@tempa\@empty \href
  {http://dx.doi.org/#2} {doi:#2}\else \href {http://dx.doi.org/#2} {#1}\fi
  \endgroup}
\def\mn@eprint#1#2{\mn@eprint@#1:#2::\@nil}
\def\mn@eprint@arXiv#1{\href {http://arxiv.org/abs/#1} {{\tt arXiv:#1}}}
\def\mn@eprint@dblp#1{\href {http://dblp.uni-trier.de/rec/bibtex/#1.xml}
  {dblp:#1}}
\def\mn@eprint@#1:#2:#3:#4\@nil{\def\@tempa {#1}\def\@tempb {#2}\def\@tempc
  {#3}\ifx \@tempc \@empty \let \@tempc \@tempb \let \@tempb \@tempa \fi \ifx
  \@tempb \@empty \def\@tempb {arXiv}\fi \@ifundefined
  {mn@eprint@\@tempb}{\@tempb:\@tempc}{\expandafter \expandafter \csname
  mn@eprint@\@tempb\endcsname \expandafter{\@tempc}}}

\bibitem[\protect\citeauthoryear{{Acu{\~n}a}, {Deleuil}, {Mousis}, {Marcq},
  {Levesque}  \& {Aguichine}}{{Acu{\~n}a} et~al.}{2021}]{Acuna2021}
{Acu{\~n}a} L.,  {Deleuil} M.,  {Mousis} O.,  {Marcq} E.,  {Levesque} M.,
  {Aguichine} A.,  2021, \mn@doi [\aap] {10.1051/0004-6361/202039885}, \href
  {https://ui.adsabs.harvard.edu/abs/2021A&A...647A..53A} {647, A53}

\bibitem[\protect\citeauthoryear{{Adibekyan}, {Sousa}, {Santos}, {Delgado
  Mena}, {Gonz{\'a}lez Hern{\'a}ndez}, {Israelian}, {Mayor}  \&
  {Khachatryan}}{{Adibekyan} et~al.}{2012}]{Adibekyan-12}
{Adibekyan} V.~Z.,  {Sousa} S.~G.,  {Santos} N.~C.,  {Delgado Mena} E.,
  {Gonz{\'a}lez Hern{\'a}ndez} J.~I.,  {Israelian} G.,  {Mayor} M.,
  {Khachatryan} G.,  2012, \mn@doi [\aap] {10.1051/0004-6361/201219401}, \href
  {http://adsabs.harvard.edu/abs/2012A%26A...545A..32A} {545, A32}

\bibitem[\protect\citeauthoryear{{Adibekyan} et~al.,}{{Adibekyan}
  et~al.}{2015}]{Adibekyan-15}
{Adibekyan} V.,  et~al., 2015, \mn@doi [\aap] {10.1051/0004-6361/201527120},
  \href {http://adsabs.harvard.edu/abs/2015A%26A...583A..94A} {583, A94}

\bibitem[\protect\citeauthoryear{{Akeson} et~al.,}{{Akeson}
  et~al.}{2013}]{Akeson2013}
{Akeson} R.~L.,  et~al., 2013, \mn@doi [\pasp] {10.1086/672273}, \href
  {https://ui.adsabs.harvard.edu/abs/2013PASP..125..989A} {125, 989}

\bibitem[\protect\citeauthoryear{{Allard}, {Homeier}  \& {Freytag}}{{Allard}
  et~al.}{2012}]{Allard2012}
{Allard} F.,  {Homeier} D.,   {Freytag} B.,  2012, \mn@doi [Philosophical
  Transactions of the Royal Society of London Series A]
  {10.1098/rsta.2011.0269}, \href
  {https://ui.adsabs.harvard.edu/abs/2012RSPTA.370.2765A} {370, 2765}

\bibitem[\protect\citeauthoryear{{Aller}, {Lillo-Box}, {Jones}, {Miranda}  \&
  {Barcel{\'o} Forteza}}{{Aller} et~al.}{2020}]{aller20}
{Aller} A.,  {Lillo-Box} J.,  {Jones} D.,  {Miranda} L.~F.,   {Barcel{\'o}
  Forteza} S.,  2020, \mn@doi [\aap] {10.1051/0004-6361/201937118}, \href
  {https://ui.adsabs.harvard.edu/abs/2020A&A...635A.128A} {635, A128}

\bibitem[\protect\citeauthoryear{{Armstrong} et~al.,}{{Armstrong}
  et~al.}{2020}]{Armstrong2020TOI849b}
{Armstrong} D.~J.,  et~al., 2020, \mn@doi [\nat] {10.1038/s41586-020-2421-7},
  \href {https://ui.adsabs.harvard.edu/abs/2020Natur.583...39A} {583, 39}

\bibitem[\protect\citeauthoryear{{Astropy Collaboration} et~al.,}{{Astropy
  Collaboration} et~al.}{2013}]{astropy2013}
{Astropy Collaboration} et~al., 2013, \mn@doi [\aap]
  {10.1051/0004-6361/201322068}, \href
  {http://adsabs.harvard.edu/abs/2013A%26A...558A..33A} {558, A33}

\bibitem[\protect\citeauthoryear{{Barrag{\'a}n}, {Gandolfi}  \&
  {Antoniciello}}{{Barrag{\'a}n} et~al.}{2019}]{Barragan2019}
{Barrag{\'a}n} O.,  {Gandolfi} D.,   {Antoniciello} G.,  2019, \mn@doi [\mnras]
  {10.1093/mnras/sty2472}, \href
  {https://ui.adsabs.harvard.edu/abs/2019MNRAS.482.1017B} {482, 1017}

\bibitem[\protect\citeauthoryear{{Bayo}, {Rodrigo}, {Barrado Y Navascu{\'e}s},
  {Solano}, {Guti{\'e}rrez}, {Morales-Calder{\'o}n}  \& {Allard}}{{Bayo}
  et~al.}{2008}]{Bayo2008}
{Bayo} A.,  {Rodrigo} C.,  {Barrado Y Navascu{\'e}s} D.,  {Solano} E.,
  {Guti{\'e}rrez} R.,  {Morales-Calder{\'o}n} M.,   {Allard} F.,  2008, \mn@doi
  [\aap] {10.1051/0004-6361:200810395}, \href
  {https://ui.adsabs.harvard.edu/abs/2008A&A...492..277B} {492, 277}

\bibitem[\protect\citeauthoryear{{Bean}, {Raymond}  \& {Owen}}{{Bean}
  et~al.}{2021}]{Bean2021}
{Bean} J.~L.,  {Raymond} S.~N.,   {Owen} J.~E.,  2021, \mn@doi [Journal of
  Geophysical Research (Planets)] {10.1029/2020JE006639}, \href
  {https://ui.adsabs.harvard.edu/abs/2021JGRE..12606639B} {126, e06639}

\bibitem[\protect\citeauthoryear{{Bensby}, {Feltzing}  \&
  {Lundstr{\"o}m}}{{Bensby} et~al.}{2003}]{Bensby-03}
{Bensby} T.,  {Feltzing} S.,   {Lundstr{\"o}m} I.,  2003, \mn@doi [\aap]
  {10.1051/0004-6361:20031213}, \href
  {https://ui.adsabs.harvard.edu/abs/2003A&A...410..527B} {410, 527}

\bibitem[\protect\citeauthoryear{{Bertran de Lis}, {Delgado Mena}, {Adibekyan},
  {Santos}  \& {Sousa}}{{Bertran de Lis} et~al.}{2015}]{Bertrandelis-15}
{Bertran de Lis} S.,  {Delgado Mena} E.,  {Adibekyan} V.~Z.,  {Santos} N.~C.,
  {Sousa} S.~G.,  2015, \mn@doi [\aap] {10.1051/0004-6361/201424633}, \href
  {http://adsabs.harvard.edu/abs/2015A%26A...576A..89B} {576, A89}

\bibitem[\protect\citeauthoryear{Bhatti, Bouma, Joshua, John  \&
  Price-Whelan}{Bhatti et~al.}{2020}]{astrobase}
Bhatti W.,  Bouma L.,  Joshua John  Price-Whelan A.,  2020,
  waqasbhatti/astrobase: astrobase v0.5.0, \mn@doi{10.5281/zenodo.3723832},
  \url {https://doi.org/10.5281/zenodo.3723832}

\bibitem[\protect\citeauthoryear{{Brown} et~al.,}{{Brown}
  et~al.}{2013}]{Brown:2013}
{Brown} T.~M.,  et~al., 2013, \mn@doi [Publications of the Astronomical Society
  of the Pacific] {10.1086/673168}, \href
  {https://ui.adsabs.harvard.edu/\#abs/2013PASP..125.1031B} {125, 1031}

\bibitem[\protect\citeauthoryear{{Brugger}, {Mousis}, {Deleuil}  \&
  {Deschamps}}{{Brugger} et~al.}{2017}]{brugger17}
{Brugger} B.,  {Mousis} O.,  {Deleuil} M.,   {Deschamps} F.,  2017, \mn@doi
  [\apj] {10.3847/1538-4357/aa965a}, \href
  {https://ui.adsabs.harvard.edu/abs/2017ApJ...850...93B} {850, 93}

\bibitem[\protect\citeauthoryear{{Bruntt} et~al.,}{{Bruntt}
  et~al.}{2010}]{Brunt2010}
{Bruntt} H.,  et~al., 2010, \mn@doi [\aap] {10.1051/0004-6361/201014143}, \href
  {https://ui.adsabs.harvard.edu/abs/2010A&A...519A..51B} {519, A51}

\bibitem[\protect\citeauthoryear{{Canto Martins} et~al.,}{{Canto Martins}
  et~al.}{2020}]{Martins2020a}
{Canto Martins} B.~L.,  et~al., 2020, arXiv e-prints, \href
  {https://ui.adsabs.harvard.edu/abs/2020arXiv200703079C} {p. arXiv:2007.03079}

\bibitem[\protect\citeauthoryear{{Carleo} et~al.,}{{Carleo}
  et~al.}{2020}]{Carleo2020}
{Carleo} I.,  et~al., 2020, arXiv e-prints, \href
  {https://ui.adsabs.harvard.edu/abs/2020arXiv200410095C} {p. arXiv:2004.10095}

\bibitem[\protect\citeauthoryear{{Chabrier}}{{Chabrier}}{2001}]{chabrier01}
{Chabrier} G.,  2001, \mn@doi [\apj] {10.1086/321401}, \href
  {http://adsabs.harvard.edu/abs/2001ApJ...554.1274C} {554, 1274}

\bibitem[\protect\citeauthoryear{{Claret} \& {Bloemen}}{{Claret} \&
  {Bloemen}}{2011}]{Claret2011}
{Claret} A.,  {Bloemen} S.,  2011, \mn@doi [\aap]
  {10.1051/0004-6361/201116451}, \href
  {https://ui.adsabs.harvard.edu/abs/2011A&A...529A..75C} {529, A75}

\bibitem[\protect\citeauthoryear{{Collier Cameron} et~al.,}{{Collier Cameron}
  et~al.}{2019}]{CollierCameron2019}
{Collier Cameron} A.,  et~al., 2019, \mn@doi [\mnras] {10.1093/mnras/stz1215},
  \href {https://ui.adsabs.harvard.edu/abs/2019MNRAS.487.1082C} {487, 1082}

\bibitem[\protect\citeauthoryear{{Collins}, {Kielkopf}, {Stassun}  \&
  {Hessman}}{{Collins} et~al.}{2017}]{Collins:2017}
{Collins} K.~A.,  {Kielkopf} J.~F.,  {Stassun} K.~G.,   {Hessman} F.~V.,  2017,
  \mn@doi [\aj] {10.3847/1538-3881/153/2/77}, \href
  {http://adsabs.harvard.edu/abs/2017AJ....153...77C} {153, 77}

\bibitem[\protect\citeauthoryear{{Crossfield} \& {Kreidberg}}{{Crossfield} \&
  {Kreidberg}}{2017}]{Crossfield2017}
{Crossfield} I. J.~M.,  {Kreidberg} L.,  2017, \mn@doi [\aj]
  {10.3847/1538-3881/aa9279}, \href
  {https://ui.adsabs.harvard.edu/abs/2017AJ....154..261C} {154, 261}

\bibitem[\protect\citeauthoryear{{Cutri} et~al.,}{{Cutri}
  et~al.}{2013}]{AllWISE2013}
{Cutri} R.~M.,  et~al., 2013, {Explanatory Supplement to the AllWISE Data
  Release Products}, Explanatory Supplement to the AllWISE Data Release
  Products

\bibitem[\protect\citeauthoryear{{Deleuil}, {Pollacco}, {Baruteau}, {Rauer}  \&
  {Blanc}}{{Deleuil} et~al.}{2020}]{Deleuil2020}
{Deleuil} M.,  {Pollacco} D.,  {Baruteau} C.,  {Rauer} H.,   {Blanc} M.,  2020,
  \mn@doi [\ssr] {10.1007/s11214-020-00726-2}, \href
  {https://ui.adsabs.harvard.edu/abs/2020SSRv..216..105D} {216, 105}

\bibitem[\protect\citeauthoryear{{Delgado Mena}, {Israelian}, {Gonz{\'a}lez
  Hern{\'a}ndez}, {Bond}, {Santos}, {Udry}  \& {Mayor}}{{Delgado Mena}
  et~al.}{2010}]{Delgado-10}
{Delgado Mena} E.,  {Israelian} G.,  {Gonz{\'a}lez Hern{\'a}ndez} J.~I.,
  {Bond} J.~C.,  {Santos} N.~C.,  {Udry} S.,   {Mayor} M.,  2010, \mn@doi
  [\apj] {10.1088/0004-637X/725/2/2349}, \href
  {http://adsabs.harvard.edu/abs/2010ApJ...725.2349D} {725, 2349}

\bibitem[\protect\citeauthoryear{{Delgado Mena} et~al.,}{{Delgado Mena}
  et~al.}{2014}]{Delgado-14}
{Delgado Mena} E.,  et~al., 2014, \mn@doi [\aap] {10.1051/0004-6361/201321493},
  \href {http://adsabs.harvard.edu/abs/2014A%26A...562A..92D} {562, A92}

\bibitem[\protect\citeauthoryear{{Delgado Mena}, {Tsantaki}, {Adibekyan},
  {Sousa}, {Santos}, {Gonz{\'a}lez Hern{\'a}ndez}  \& {Israelian}}{{Delgado
  Mena} et~al.}{2017}]{Delgado-17}
{Delgado Mena} E.,  {Tsantaki} M.,  {Adibekyan} V.~Z.,  {Sousa} S.~G.,
  {Santos} N.~C.,  {Gonz{\'a}lez Hern{\'a}ndez} J.~I.,   {Israelian} G.,  2017,
  \mn@doi [\aap] {10.1051/0004-6361/201730535}, \href
  {http://adsabs.harvard.edu/abs/2017A%26A...606A..94D} {606, A94}

\bibitem[\protect\citeauthoryear{{Delgado Mena} et~al.,}{{Delgado Mena}
  et~al.}{2019}]{Delgado-19}
{Delgado Mena} E.,  et~al., 2019, \mn@doi [\aap] {10.1051/0004-6361/201834783},
  \href {https://ui.adsabs.harvard.edu/abs/2019A&A...624A..78D} {624, A78}

\bibitem[\protect\citeauthoryear{{D{\'\i}az}, {Almenara}, {Santerne}, {Moutou},
  {Lethuillier}  \& {Deleuil}}{{D{\'\i}az} et~al.}{2014}]{Diaz2014}
{D{\'\i}az} R.~F.,  {Almenara} J.~M.,  {Santerne} A.,  {Moutou} C.,
  {Lethuillier} A.,   {Deleuil} M.,  2014, \mn@doi [\mnras]
  {10.1093/mnras/stu601}, \href
  {https://ui.adsabs.harvard.edu/abs/2014MNRAS.441..983D} {441, 983}

\bibitem[\protect\citeauthoryear{{Dorn}, {Khan}, {Heng}, {Connolly}, {Alibert},
  {Benz}  \& {Tackley}}{{Dorn} et~al.}{2015}]{dorn15}
{Dorn} C.,  {Khan} A.,  {Heng} K.,  {Connolly} J. A.~D.,  {Alibert} Y.,  {Benz}
  W.,   {Tackley} P.,  2015, \mn@doi [\aap] {10.1051/0004-6361/201424915},
  \href {https://ui.adsabs.harvard.edu/abs/2015A&A...577A..83D} {577, A83}

\bibitem[\protect\citeauthoryear{{Dotter}, {Chaboyer}, {Jevremovi{\'c}},
  {Kostov}, {Baron}  \& {Ferguson}}{{Dotter} et~al.}{2008}]{Dotter2008}
{Dotter} A.,  {Chaboyer} B.,  {Jevremovi{\'c}} D.,  {Kostov} V.,  {Baron} E.,
  {Ferguson} J.~W.,  2008, \mn@doi [\apjs] {10.1086/589654}, \href
  {https://ui.adsabs.harvard.edu/abs/2008ApJS..178...89D} {178, 89}

\bibitem[\protect\citeauthoryear{{Doyle}, {Davies}, {Smalley}, {Chaplin}  \&
  {Elsworth}}{{Doyle} et~al.}{2014}]{Doyle2014}
{Doyle} A.~P.,  {Davies} G.~R.,  {Smalley} B.,  {Chaplin} W.~J.,   {Elsworth}
  Y.,  2014, \mn@doi [\mnras] {10.1093/mnras/stu1692}, \href
  {https://ui.adsabs.harvard.edu/abs/2014MNRAS.444.3592D} {444, 3592}

\bibitem[\protect\citeauthoryear{{Dumusque} et~al.,}{{Dumusque}
  et~al.}{2015}]{Dumusque2015}
{Dumusque} X.,  et~al., 2015, \mn@doi [\apjl] {10.1088/2041-8205/814/2/L21},
  \href {https://ui.adsabs.harvard.edu/abs/2015ApJ...814L..21D} {814, L21}

\bibitem[\protect\citeauthoryear{{Fitzpatrick}}{{Fitzpatrick}}{1999}]{Fitzpatrick1999}
{Fitzpatrick} E.~L.,  1999, \mn@doi [\pasp] {10.1086/316293}, \href
  {https://ui.adsabs.harvard.edu/abs/1999PASP..111...63F} {111, 63}

\bibitem[\protect\citeauthoryear{{Fridlund} et~al.,}{{Fridlund}
  et~al.}{2020}]{Fridlund2020arXiv}
{Fridlund} M.,  et~al., 2020, arXiv e-prints, \href
  {https://ui.adsabs.harvard.edu/abs/2020arXiv200812535F} {p. arXiv:2008.12535}

\bibitem[\protect\citeauthoryear{{Fulton} \& {Petigura}}{{Fulton} \&
  {Petigura}}{2018}]{Fulton2018}
{Fulton} B.~J.,  {Petigura} E.~A.,  2018, \mn@doi [\aj]
  {10.3847/1538-3881/aae828}, \href
  {https://ui.adsabs.harvard.edu/abs/2018AJ....156..264F} {156, 264}

\bibitem[\protect\citeauthoryear{{Fulton} et~al.,}{{Fulton}
  et~al.}{2017}]{Fulton2017}
{Fulton} B.~J.,  et~al., 2017, \mn@doi [\aj] {10.3847/1538-3881/aa80eb}, \href
  {https://ui.adsabs.harvard.edu/abs/2017AJ....154..109F} {154, 109}

\bibitem[\protect\citeauthoryear{{Gaia Collaboration} et~al.,}{{Gaia
  Collaboration} et~al.}{2018}]{Gaia2018}
{Gaia Collaboration} et~al., 2018, \mn@doi [\aap]
  {10.1051/0004-6361/201833051}, \href
  {https://ui.adsabs.harvard.edu/abs/2018A&A...616A...1G} {616, A1}

\bibitem[\protect\citeauthoryear{{Gandolfi} et~al.,}{{Gandolfi}
  et~al.}{2018}]{Gandolfi2018}
{Gandolfi} D.,  et~al., 2018, \mn@doi [\aap] {10.1051/0004-6361/201834289},
  \href {https://ui.adsabs.harvard.edu/abs/2018A&A...619L..10G} {619, L10}

\bibitem[\protect\citeauthoryear{{Gao} et~al.,}{{Gao}
  et~al.}{2020}]{Gao2020Nat}
{Gao} P.,  et~al., 2020, \mn@doi [Nature Astronomy]
  {10.1038/s41550-020-1114-3}, \href
  {https://ui.adsabs.harvard.edu/abs/2020NatAs...4..951G} {4, 951}

\bibitem[\protect\citeauthoryear{{Ginzburg}, {Schlichting}  \&
  {Sari}}{{Ginzburg} et~al.}{2018}]{Ginzburg2018}
{Ginzburg} S.,  {Schlichting} H.~E.,   {Sari} R.,  2018, \mn@doi [\mnras]
  {10.1093/mnras/sty290}, \href
  {https://ui.adsabs.harvard.edu/abs/2018MNRAS.476..759G} {476, 759}

\bibitem[\protect\citeauthoryear{{Girardi} et~al.,}{{Girardi}
  et~al.}{2012}]{girardi12}
{Girardi} L.,  et~al., 2012, \mn@doi [Astrophysics and Space Science
  Proceedings] {10.1007/978-3-642-18418-5_17}, \href
  {https://ui.adsabs.harvard.edu/abs/2012ASSP...26..165G} {26, 165}

\bibitem[\protect\citeauthoryear{{Gupta} \& {Schlichting}}{{Gupta} \&
  {Schlichting}}{2019}]{Gupta2019}
{Gupta} A.,  {Schlichting} H.~E.,  2019, \mn@doi [\mnras]
  {10.1093/mnras/stz1230}, \href
  {https://ui.adsabs.harvard.edu/abs/2019MNRAS.487...24G} {487, 24}

\bibitem[\protect\citeauthoryear{{Hall}, {Lockwood}  \& {Skiff}}{{Hall}
  et~al.}{2007}]{Hall2007}
{Hall} J.~C.,  {Lockwood} G.~W.,   {Skiff} B.~A.,  2007, \mn@doi [\aj]
  {10.1086/510356}, \href
  {https://ui.adsabs.harvard.edu/abs/2007AJ....133..862H} {133, 862}

\bibitem[\protect\citeauthoryear{{Henden}, {Levine}, {Terrell}  \&
  {Welch}}{{Henden} et~al.}{2015}]{Henden2015}
{Henden} A.~A.,  {Levine} S.,  {Terrell} D.,   {Welch} D.~L.,  2015, in
  American Astronomical Society Meeting Abstracts \#225. p. 336.16

\bibitem[\protect\citeauthoryear{{Howell}, {Everett}, {Sherry}, {Horch}  \&
  {Ciardi}}{{Howell} et~al.}{2011}]{howell2011}
{Howell} S.~B.,  {Everett} M.~E.,  {Sherry} W.,  {Horch} E.,   {Ciardi} D.~R.,
  2011, \mn@doi [\aj] {10.1088/0004-6256/142/1/19}, \href
  {https://ui.adsabs.harvard.edu/abs/2011AJ....142...19H} {142, 19}

\bibitem[\protect\citeauthoryear{{Hunter}}{{Hunter}}{2007}]{matplotlib}
{Hunter} J.~D.,  2007, \mn@doi [Computing in Science and Engineering]
  {10.1109/MCSE.2007.55}, \href
  {http://adsabs.harvard.edu/abs/2007CSE.....9...90H} {9, 90}

\bibitem[\protect\citeauthoryear{{Jenkins}}{{Jenkins}}{2002}]{Jenkins2002}
{Jenkins} J.~M.,  2002, \mn@doi [\apj] {10.1086/341136}, \href
  {https://ui.adsabs.harvard.edu/abs/2002ApJ...575..493J} {575, 493}

\bibitem[\protect\citeauthoryear{{Jenkins} et~al.,}{{Jenkins}
  et~al.}{2010}]{Jenkins2010}
{Jenkins} J.~M.,  et~al., 2010, in Software and Cyberinfrastructure for
  Astronomy. p. 77400D, \mn@doi{10.1117/12.856764}

\bibitem[\protect\citeauthoryear{{Jenkins} et~al.,}{{Jenkins}
  et~al.}{2016}]{Jenkins2016}
{Jenkins} J.~M.,  et~al., 2016, in \procspie. p. 99133E,
  \mn@doi{10.1117/12.2233418}

\bibitem[\protect\citeauthoryear{{Jenkins} et~al.,}{{Jenkins}
  et~al.}{2020}]{Jenkins2020LTT9779b}
{Jenkins} J.~S.,  et~al., 2020, \mn@doi [Nature Astronomy]
  {10.1038/s41550-020-1142-z}, \href
  {https://ui.adsabs.harvard.edu/abs/2020NatAs...4.1148J} {4, 1148}

\bibitem[\protect\citeauthoryear{{Jensen}}{{Jensen}}{2013}]{Jensen:2013}
{Jensen} E.,  2013, {Tapir: A web interface for transit/eclipse observability},
  Astrophysics Source Code Library (\mn@eprint {ascl} {1306.007})

\bibitem[\protect\citeauthoryear{{Johnson} \& {Soderblom}}{{Johnson} \&
  {Soderblom}}{1987}]{Johnson-87}
{Johnson} D. R.~H.,  {Soderblom} D.~R.,  1987, \mn@doi [\aj] {10.1086/114370},
  \href {https://ui.adsabs.harvard.edu/abs/1987AJ.....93..864J} {93, 864}

\bibitem[\protect\citeauthoryear{Jones, Oliphant, Peterson  et~al.}{Jones
  et~al.}{2001}]{scipy}
Jones E.,  Oliphant T.,  Peterson P.,   et~al., 2001, {SciPy}: Open source
  scientific tools for {Python}, \url {http://www.scipy.org/}

\bibitem[\protect\citeauthoryear{{Kempton} et~al.,}{{Kempton}
  et~al.}{2018}]{Kempton2018}
{Kempton} E. M.~R.,  et~al., 2018, \mn@doi [\pasp] {10.1088/1538-3873/aadf6f},
  \href {https://ui.adsabs.harvard.edu/abs/2018PASP..130k4401K} {130, 114401}

\bibitem[\protect\citeauthoryear{Kluyver et~al.,}{Kluyver
  et~al.}{2016}]{jupyter2016}
Kluyver T.,  et~al., 2016, in Loizides F.,  Schmidt B.,  eds, Positioning and
  Power in Academic Publishing: Players, Agents and Agendas. pp 87 -- 90

\bibitem[\protect\citeauthoryear{{Kurucz}}{{Kurucz}}{1993}]{Kurucz-93}
{Kurucz} R.~L.,  1993, VizieR Online Data Catalog, \href
  {https://ui.adsabs.harvard.edu/abs/1993yCat.6039....0K} {p. VI/39}

\bibitem[\protect\citeauthoryear{{Li}, {Tenenbaum}, {Twicken}, {Burke},
  {Jenkins}, {Quintana}, {Rowe}  \& {Seader}}{{Li} et~al.}{2019}]{Li2019}
{Li} J.,  {Tenenbaum} P.,  {Twicken} J.~D.,  {Burke} C.~J.,  {Jenkins} J.~M.,
  {Quintana} E.~V.,  {Rowe} J.~F.,   {Seader} S.~E.,  2019, \mn@doi [\pasp]
  {10.1088/1538-3873/aaf44d}, \href
  {https://ui.adsabs.harvard.edu/abs/2019PASP..131b4506L} {131, 024506}

\bibitem[\protect\citeauthoryear{{Lillo-Box}, {Barrado}  \& {Bouy}}{{Lillo-Box}
  et~al.}{2014}]{lillo-box14b}
{Lillo-Box} J.,  {Barrado} D.,   {Bouy} H.,  2014, \mn@doi [\aap]
  {10.1051/0004-6361/201423497}, \href
  {http://adsabs.harvard.edu/abs/2014A%26A...566A.103L} {566, A103}

\bibitem[\protect\citeauthoryear{{Lillo-Box} et~al.,}{{Lillo-Box}
  et~al.}{2020}]{Lillo-Box2020}
{Lillo-Box} J.,  et~al., 2020, \mn@doi [\aap] {10.1051/0004-6361/202037896},
  \href {https://ui.adsabs.harvard.edu/abs/2020A&A...640A..48L} {640, A48}

\bibitem[\protect\citeauthoryear{{Lopez} et~al.,}{{Lopez}
  et~al.}{2019}]{Lopez2019}
{Lopez} T.~A.,  et~al., 2019, \mn@doi [\aap] {10.1051/0004-6361/201936267},
  \href {https://ui.adsabs.harvard.edu/abs/2019A&A...631A..90L} {631, A90}

\bibitem[\protect\citeauthoryear{{Lovis} \& {Pepe}}{{Lovis} \&
  {Pepe}}{2007}]{Lovis2007}
{Lovis} C.,  {Pepe} F.,  2007, \mn@doi [\aap] {10.1051/0004-6361:20077249},
  \href {http://adsabs.harvard.edu/abs/2007A%26A...468.1115L} {468, 1115}

\bibitem[\protect\citeauthoryear{{Mayor} et~al.,}{{Mayor}
  et~al.}{2003}]{Mayor2003}
{Mayor} M.,  et~al., 2003, The Messenger, \href
  {https://ui.adsabs.harvard.edu/abs/2003Msngr.114...20M} {114, 20}

\bibitem[\protect\citeauthoryear{{Mazeh}, {Holczer}  \& {Faigler}}{{Mazeh}
  et~al.}{2016}]{Mazeh2016}
{Mazeh} T.,  {Holczer} T.,   {Faigler} S.,  2016, \mn@doi [\aap]
  {10.1051/0004-6361/201528065}, \href
  {https://ui.adsabs.harvard.edu/abs/2016A&A...589A..75M} {589, A75}

\bibitem[\protect\citeauthoryear{{Mazevet}, {Licari}, {Chabrier}  \&
  {Potekhin}}{{Mazevet} et~al.}{2019}]{Mazevet2019}
{Mazevet} S.,  {Licari} A.,  {Chabrier} G.,   {Potekhin} A.~Y.,  2019, \mn@doi
  [\aap] {10.1051/0004-6361/201833963}, \href
  {https://ui.adsabs.harvard.edu/abs/2019A&A...621A.128M} {621, A128}

\bibitem[\protect\citeauthoryear{{McKay} et~al.,}{{McKay}
  et~al.}{2019}]{McKay2019}
{McKay} A.~J.,  et~al., 2019, \mn@doi [\aj] {10.3847/1538-3881/ab32e4}, \href
  {https://ui.adsabs.harvard.edu/abs/2019AJ....158..128M} {158, 128}

\bibitem[\protect\citeauthoryear{McKinney}{McKinney}{2010}]{pandas-scipy-2010}
McKinney W.,  2010, in van~der Walt S.,  Millman J.,  eds, Proceedings of the
  9th Python in Science Conference. pp 51 -- 56

\bibitem[\protect\citeauthoryear{{Morley}, {Fortney}, {Marley}, {Zahnle},
  {Line}, {Kempton}, {Lewis}  \& {Cahoy}}{{Morley} et~al.}{2015}]{Morley2015}
{Morley} C.~V.,  {Fortney} J.~J.,  {Marley} M.~S.,  {Zahnle} K.,  {Line} M.,
  {Kempton} E.,  {Lewis} N.,   {Cahoy} K.,  2015, \mn@doi [\apj]
  {10.1088/0004-637X/815/2/110}, \href
  {https://ui.adsabs.harvard.edu/abs/2015ApJ...815..110M} {815, 110}

\bibitem[\protect\citeauthoryear{Mousis, Deleuil, Aguichine, Marcq, Naar,
  Aguirre, Brugger  \& Gon{\c{c}}alves}{Mousis et~al.}{2020}]{mousis20}
Mousis O.,  Deleuil M.,  Aguichine A.,  Marcq E.,  Naar J.,  Aguirre L.~A.,
  Brugger B.,   Gon{\c{c}}alves T.,  2020, \mn@doi [The Astrophysical Journal]
  {10.3847/2041-8213/ab9530}, 896, L22

\bibitem[\protect\citeauthoryear{{Murdoch}, {Hearnshaw}  \& {Clark}}{{Murdoch}
  et~al.}{1993}]{Murdoch1993}
{Murdoch} K.~A.,  {Hearnshaw} J.~B.,   {Clark} M.,  1993, \mn@doi [\apj]
  {10.1086/173003}, \href
  {https://ui.adsabs.harvard.edu/abs/1993ApJ...413..349M} {413, 349}

\bibitem[\protect\citeauthoryear{{Nielsen} et~al.,}{{Nielsen}
  et~al.}{2020}]{Nielsen2020}
{Nielsen} L.~D.,  et~al., 2020, \mn@doi [\mnras] {10.1093/mnras/staa197}, \href
  {https://ui.adsabs.harvard.edu/abs/2020MNRAS.492.5399N} {492, 5399}

\bibitem[\protect\citeauthoryear{{Owen} \& {Lai}}{{Owen} \&
  {Lai}}{2018}]{OwenLai2018}
{Owen} J.~E.,  {Lai} D.,  2018, \mn@doi [\mnras] {10.1093/mnras/sty1760}, \href
  {https://ui.adsabs.harvard.edu/abs/2018MNRAS.479.5012O} {479, 5012}

\bibitem[\protect\citeauthoryear{{Pandas development team.}}{{Pandas
  development team.}}{2020}]{pandas2020}
{Pandas development team.} 2020, pandas-dev/pandas: Pandas,
  \mn@doi{10.5281/zenodo.3509134}, \url
  {https://doi.org/10.5281/zenodo.3509134}

\bibitem[\protect\citeauthoryear{{Pepe} et~al.,}{{Pepe}
  et~al.}{2002}]{Pepe2002}
{Pepe} F.,  et~al., 2002, The Messenger, \href
  {https://ui.adsabs.harvard.edu/abs/2002Msngr.110....9P} {110, 9}

\bibitem[\protect\citeauthoryear{P\'erez \& Granger}{P\'erez \&
  Granger}{2007}]{ipython}
P\'erez F.,  Granger B.~E.,  2007, \mn@doi [Computing in Science and
  Engineering] {10.1109/MCSE.2007.53}, 9, 21

\bibitem[\protect\citeauthoryear{{Piskunov} \& {Valenti}}{{Piskunov} \&
  {Valenti}}{2017}]{PiskunovValenti2017}
{Piskunov} N.,  {Valenti} J.~A.,  2017, \mn@doi [\aap]
  {10.1051/0004-6361/201629124}, \href
  {https://ui.adsabs.harvard.edu/abs/2017A&A...597A..16P} {597, A16}

\bibitem[\protect\citeauthoryear{{Price-Whelan} et~al.,}{{Price-Whelan}
  et~al.}{2018}]{astropy2018}
{Price-Whelan} A.~M.,  et~al., 2018, \mn@doi [\aj] {10.3847/1538-3881/aabc4f},
  \href {https://ui.adsabs.harvard.edu/#abs/2018AJ....156..123T} {156, 123}

\bibitem[\protect\citeauthoryear{{Ricker} et~al.,}{{Ricker}
  et~al.}{2015}]{Ricker2015}
{Ricker} G.~R.,  et~al., 2015, \mn@doi [Journal of Astronomical Telescopes,
  Instruments, and Systems] {10.1117/1.JATIS.1.1.014003}, \href
  {https://ui.adsabs.harvard.edu/abs/2015JATIS...1a4003R} {1, 014003}

\bibitem[\protect\citeauthoryear{{Santerne} et~al.,}{{Santerne}
  et~al.}{2015}]{Santerne2015}
{Santerne} A.,  et~al., 2015, \mn@doi [\mnras] {10.1093/mnras/stv1080}, \href
  {https://ui.adsabs.harvard.edu/abs/2015MNRAS.451.2337S} {451, 2337}

\bibitem[\protect\citeauthoryear{{Santerne} et~al.,}{{Santerne}
  et~al.}{2019}]{Santerne2019}
{Santerne} A.,  et~al., 2019, arXiv e-prints, \href
  {https://ui.adsabs.harvard.edu/abs/2019arXiv191107355S} {p. arXiv:1911.07355}

\bibitem[\protect\citeauthoryear{{Santos} et~al.,}{{Santos}
  et~al.}{2013}]{Santos-13}
{Santos} N.~C.,  et~al., 2013, \mn@doi [\aap] {10.1051/0004-6361/201321286},
  \href {http://adsabs.harvard.edu/abs/2013A%26A...556A.150S} {556, A150}

\bibitem[\protect\citeauthoryear{{Schlafly} \& {Finkbeiner}}{{Schlafly} \&
  {Finkbeiner}}{2011}]{Schlafly2011}
{Schlafly} E.~F.,  {Finkbeiner} D.~P.,  2011, \mn@doi [\apj]
  {10.1088/0004-637X/737/2/103}, \href
  {https://ui.adsabs.harvard.edu/abs/2011ApJ...737..103S} {737, 103}

\bibitem[\protect\citeauthoryear{{Sch{\"o}nrich}, {Binney}  \&
  {Dehnen}}{{Sch{\"o}nrich} et~al.}{2010}]{Schonrich-10}
{Sch{\"o}nrich} R.,  {Binney} J.,   {Dehnen} W.,  2010, \mn@doi [\mnras]
  {10.1111/j.1365-2966.2010.16253.x}, \href
  {https://ui.adsabs.harvard.edu/abs/2010MNRAS.403.1829S} {403, 1829}

\bibitem[\protect\citeauthoryear{{Skrutskie} et~al.,}{{Skrutskie}
  et~al.}{2006}]{2MASS}
{Skrutskie} M.~F.,  et~al., 2006, \mn@doi [\aj] {10.1086/498708}, \href
  {https://ui.adsabs.harvard.edu/abs/2006AJ....131.1163S} {131, 1163}

\bibitem[\protect\citeauthoryear{{Smith} et~al.,}{{Smith}
  et~al.}{2012}]{Smith2012}
{Smith} J.~C.,  et~al., 2012, \mn@doi [\pasp] {10.1086/667697}, \href
  {https://ui.adsabs.harvard.edu/abs/2012PASP..124.1000S} {124, 1000}

\bibitem[\protect\citeauthoryear{{Sneden}}{{Sneden}}{1973}]{Sneden-73}
{Sneden} C.~A.,  1973, PhD thesis, THE UNIVERSITY OF TEXAS AT AUSTIN.

\bibitem[\protect\citeauthoryear{{Sotin}, {Grasset}  \& {Mocquet}}{{Sotin}
  et~al.}{2007}]{sotin07}
{Sotin} C.,  {Grasset} O.,   {Mocquet} A.,  2007, \mn@doi [\icarus]
  {10.1016/j.icarus.2007.04.006}, \href
  {https://ui.adsabs.harvard.edu/abs/2007Icar..191..337S} {191, 337}

\bibitem[\protect\citeauthoryear{{Sousa}}{{Sousa}}{2014}]{Sousa-14}
{Sousa} S.~G.,  2014, [arXiv:1407.5817], \href
  {http://adsabs.harvard.edu/abs/2014arXiv1407.5817S} {}

\bibitem[\protect\citeauthoryear{{Sousa}, {Santos}, {Adibekyan}, {Delgado-Mena}
   \& {Israelian}}{{Sousa} et~al.}{2015}]{Sousa-15}
{Sousa} S.~G.,  {Santos} N.~C.,  {Adibekyan} V.,  {Delgado-Mena} E.,
  {Israelian} G.,  2015, \mn@doi [\aap] {10.1051/0004-6361/201425463}, \href
  {http://adsabs.harvard.edu/abs/2015A%26A...577A..67S} {577, A67}

\bibitem[\protect\citeauthoryear{{Southworth}}{{Southworth}}{2008}]{Southworth2008}
{Southworth} J.,  2008, \mn@doi [\mnras] {10.1111/j.1365-2966.2008.13145.x},
  \href {https://ui.adsabs.harvard.edu/abs/2008MNRAS.386.1644S} {386, 1644}

\bibitem[\protect\citeauthoryear{{Stassun} et~al.,}{{Stassun}
  et~al.}{2019}]{Stassun2019}
{Stassun} K.~G.,  et~al., 2019, \mn@doi [\aj] {10.3847/1538-3881/ab3467}, \href
  {https://ui.adsabs.harvard.edu/abs/2019AJ....158..138S} {158, 138}

\bibitem[\protect\citeauthoryear{Stephens}{Stephens}{1974}]{Stephens1974}
Stephens M.~A.,  1974, \mn@doi [Journal of the American Statistical
  Association] {10.1080/01621459.1974.10480196}, 69, 730

\bibitem[\protect\citeauthoryear{{Stumpe} et~al.,}{{Stumpe}
  et~al.}{2012}]{Stumpe2012}
{Stumpe} M.~C.,  et~al., 2012, \mn@doi [\pasp] {10.1086/667698}, \href
  {https://ui.adsabs.harvard.edu/abs/2012PASP..124..985S} {124, 985}

\bibitem[\protect\citeauthoryear{{Stumpe}, {Smith}, {Catanzarite}, {Van Cleve},
  {Jenkins}, {Twicken}  \& {Girouard}}{{Stumpe} et~al.}{2014}]{Stumpe2014}
{Stumpe} M.~C.,  {Smith} J.~C.,  {Catanzarite} J.~H.,  {Van Cleve} J.~E.,
  {Jenkins} J.~M.,  {Twicken} J.~D.,   {Girouard} F.~R.,  2014, \mn@doi [\pasp]
  {10.1086/674989}, \href
  {https://ui.adsabs.harvard.edu/abs/2014PASP..126..100S} {126, 100}

\bibitem[\protect\citeauthoryear{{Szab{\'o}} \& {Kiss}}{{Szab{\'o}} \&
  {Kiss}}{2011}]{SzaboKiss2011}
{Szab{\'o}} G.~M.,  {Kiss} L.~L.,  2011, \mn@doi [\apjl]
  {10.1088/2041-8205/727/2/L44}, \href
  {https://ui.adsabs.harvard.edu/abs/2011ApJ...727L..44S} {727, L44}

\bibitem[\protect\citeauthoryear{{Twicken} et~al.,}{{Twicken}
  et~al.}{2018}]{Twicken2018}
{Twicken} J.~D.,  et~al., 2018, \mn@doi [\pasp] {10.1088/1538-3873/aab694},
  \href {https://ui.adsabs.harvard.edu/abs/2018PASP..130f4502T} {130, 064502}

\bibitem[\protect\citeauthoryear{{Valenti} \& {Piskunov}}{{Valenti} \&
  {Piskunov}}{1996}]{ValentiPiskunov1996}
{Valenti} J.~A.,  {Piskunov} N.,  1996, \aaps, \href
  {https://ui.adsabs.harvard.edu/abs/1996A&AS..118..595V} {118, 595}

\bibitem[\protect\citeauthoryear{{Van Der Walt}, {Colbert}  \&
  {Varoquaux}}{{Van Der Walt} et~al.}{2011}]{VanDerWalt2011}
{Van Der Walt} S.,  {Colbert} S.~C.,   {Varoquaux} G.,  2011, preprint, \href
  {http://adsabs.harvard.edu/abs/2011arXiv1102.1523V} {} (\mn@eprint {arXiv}
  {1102.1523})

\bibitem[\protect\citeauthoryear{{Van Eylen}, {Agentoft}, {Lundkvist},
  {Kjeldsen}, {Owen}, {Fulton}, {Petigura}  \& {Snellen}}{{Van Eylen}
  et~al.}{2018}]{VanEylen2018}
{Van Eylen} V.,  {Agentoft} C.,  {Lundkvist} M.~S.,  {Kjeldsen} H.,  {Owen}
  J.~E.,  {Fulton} B.~J.,  {Petigura} E.,   {Snellen} I.,  2018, \mn@doi
  [\mnras] {10.1093/mnras/sty1783}, \href
  {https://ui.adsabs.harvard.edu/abs/2018MNRAS.479.4786V} {479, 4786}

\bibitem[\protect\citeauthoryear{{West} et~al.,}{{West}
  et~al.}{2019}]{West2019NGTS4b}
{West} R.~G.,  et~al., 2019, \mn@doi [\mnras] {10.1093/mnras/stz1084}, \href
  {https://ui.adsabs.harvard.edu/abs/2019MNRAS.486.5094W} {486, 5094}

\bibitem[\protect\citeauthoryear{{Zechmeister} \& {K{\"u}rster}}{{Zechmeister}
  \& {K{\"u}rster}}{2009}]{Zechmeister2009}
{Zechmeister} M.,  {K{\"u}rster} M.,  2009, \mn@doi [\aap]
  {10.1051/0004-6361:200811296}, \href
  {https://ui.adsabs.harvard.edu/abs/2009A&A...496..577Z} {496, 577}

\bibitem[\protect\citeauthoryear{{Zechmeister} et~al.,}{{Zechmeister}
  et~al.}{2018}]{Zechmeister2018}
{Zechmeister} M.,  et~al., 2018, \mn@doi [\aap] {10.1051/0004-6361/201731483},
  \href {https://ui.adsabs.harvard.edu/abs/2018A&A...609A..12Z} {609, A12}

\bibitem[\protect\citeauthoryear{{Zeng} et~al.,}{{Zeng}
  et~al.}{2019}]{Zeng2019}
{Zeng} L.,  et~al., 2019, \mn@doi [Proceedings of the National Academy of
  Science] {10.1073/pnas.1812905116}, \href
  {https://ui.adsabs.harvard.edu/abs/2019PNAS..116.9723Z} {116, 9723}

\makeatother
\end{thebibliography}



\appendix

\section{Validation of stellar parameters}
\label{sec:stellar_prop2}
\begin{figure}
     \centering
     \includegraphics[scale=0.23]{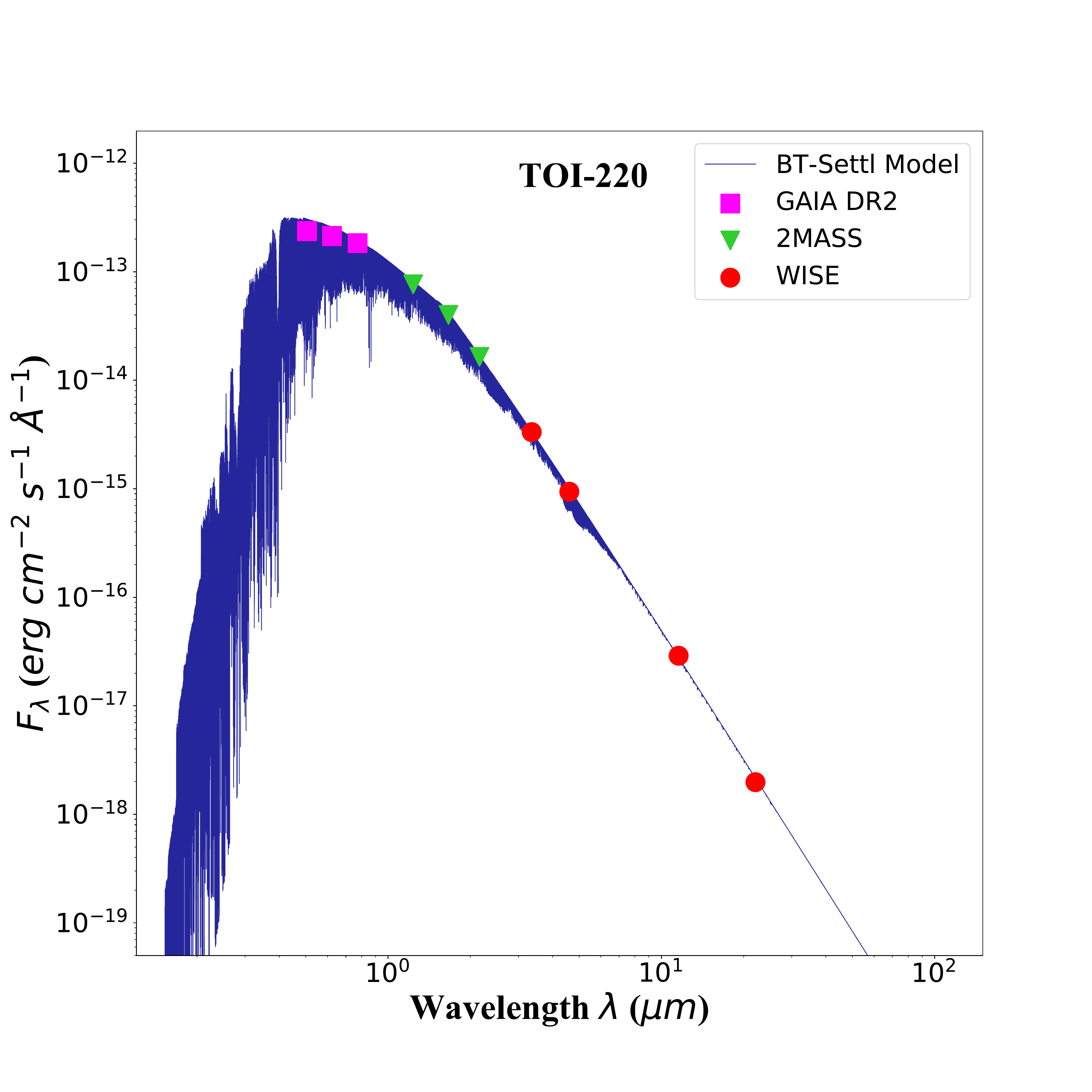}
     \caption{Independent SED fitting of TOI-220 using Gaia (magenta squares), 2MASS (green triangles) and WISE magnitudes (red circles). The blue solid line represents the fitted theoretical stellar model. See Sect.\ref{sec:pastis} for details.}
     \label{fig:SED2}
 \end{figure}

In addition to what has been presented in Sect.~\ref{sec:spectra_analysis}, as an independent validation, we also derived the main stellar parameters by applying the Spectroscopy Made Easy package \citep[SME,][]{ValentiPiskunov1996, PiskunovValenti2017} to the co-added high signal-to-noise HARPS spectrum. SME uses grids of 1-D LTE stellar models to iteratively calculate synthetic spectra where by following a minimizing strategy each synthetic spectrum is compared to the normalized co-added spectrum observed data. The process is repeated by changing input parameters until we obtain a match for the line profiles of individual diagnostic lines, leading to values for T$_\mathrm{eff}$, $\log$\,$g$, metallicity, and $v$\,sin\,$i$. The turbulent velocities are calculated according to the empirical relations of \cite{Brunt2010} and \cite{Doyle2014} and are held fixed in the determination of the other parameters. We refer to \cite{Fridlund2020arXiv} for more details and references. With this, we found a slightly cooler T$_{\text{eff}}$\,=\,5182\,$\pm$\,45\,K, while the rest of the calculated values $\log$\,$g$\,=\,4.25\,$\pm$\,0.07, Fe/H\,=\,$-$0.2\,$\pm$\,0.07, $v$\,sin\,$i$\,=\,2.9\,$\pm$\,0.35\,km\,s$^{-1}$,  V$_{mic}$\,=\,0.85\,$\pm$\,0.1 km\,s$^{-1}$ (fixed) and V$_{mac}$ = 0.9\,$\pm$\,0.4\,km\,s$^{-1}$, are in good agreement with the respective results of Sect.~\ref{sec:spectra_analysis} presented in Table~\ref{tab:stellar_parameters}. Finally, the result of the independent SED fitting of TOI-220 described in Sect. \ref{sec:pastis_sed} is shown in Fig.~\ref{fig:SED2}. 

\section{Priors of the Bayesian analysis}

In Table \ref{tab:priors} we present the list of the prior distributions used for each parameter fitted with {\sc pastis} code in Sect. \ref{sec:pastis}. 

\begin{table}
    \centering
    \caption{List of the prior distributions used in {\sc pastis} analysis: $\mathcal{N}(\mu,\sigma^{2})$: Normal with mean $\mu$ and width $\sigma$; $\mathcal{U}(a,b)$: Uniform dist. between $a$ and $b$; $\mathcal{T}(\mu,\sigma^{2},a,b)$: Truncated normal distribution with mean $\mu$, width $\sigma$ between $a$ and $b$; and $\mathcal{S}(a,b)$: Sine dist. between $a$ and $b$. (*): parameters included in the RV drift modelling.}
    \label{tab:priors}
    \begin{tabular}{l|c}
    \hline
    \hline
    Stellar parameters & distribution \\
    \hline
    Effective Temperature, T$_{\text{eff}}$ [K]  &  $\mathcal{N}(5300,100)$   \\
    Surface gravity, $\log$\,$g$ [cgs]  &  $\mathcal{N}(4.5,0.1)$   \\
    Metallicity, [M/H] [dex] & $\mathcal{N}(-0.2,0.05)$ \\
    Distance, d [pc] & $\mathcal{N}(90.54,5)$ \\
    \textit{E(B$-$V)} [mag] & $\mathcal{U}(0,1)$ \\
    Systemic RV, v$_{0}$ [m\,s$^{-1}$] & $\mathcal{U}(20,30)$ \\
    Reference Time RV [BJD\_TDB] (*)  &  2458450 \\
    RV drift lin. coeff. [m\,s$^{-1}$\,d$^{-1}$] (*)&  $\mathcal{U}(-50,50)$ \\
    RV drift quad. coeff. [m\,s$^{-1}$\,d$^{-2}$] (*)& $\mathcal{U}(-50,50)$ \\
    \hline
    Planetary parameters &\\
    \hline
    Orbital period, P [d] & $\mathcal{N}(10.6,0.5)$ \\
    Reference transit time, T$_{0}$ [BJD\_TDB] & $\mathcal{N}(2458335.9026,0.1)$ \\
    Planet-to-star radius ratio, R$_{\text{p}}$/R$_{\text{s}}$ &$\mathcal{U}(0.029,0.04)$ \\
    RV semi-amplitude, K [m\,s$^{-1}$] & $\mathcal{U}(0,10)$  \\
    Orbital inclination, $i$ [\degr] & $\mathcal{S}(85,90)$  \\
    Orbital eccentricity, $e$ & $\mathcal{T}(0,0.32,0,1)$ \\
    Argument of periastron, $\omega$ [$\degr$] & $\mathcal{U}(0,360)$ \\
    \hline
    Instrument parameters&\\
    \hline
    \textit{TESS} contamination [\%] & $\mathcal{T}(0,0.005, 0,1)$ \\
    \textit{TESS} jitter [ppm] &  $\mathcal{U}(0.0, 0.1)$\\ 
    \textit{TESS} out-of-trasit flux & $\mathcal{U}(0.99, 1.01)$\\
    HARPS jitter [km\,s$^{-1}$] & $\mathcal{U}(0.0, 0.1)$ \\
    SED jitter [mag] & $\mathcal{U}(0,1)$ \\
    \hline
    \end{tabular}
 
\end{table}
 
 \section{Search for transits of planet-c in TESS light curve}
 \label{sec:bls_residuals}
 
 As mentioned in Sect. \ref{sec:obs}, by the end of the writing of this manuscript two new sectors (27 and 28) from the second pass of {\it TESS} in the Southern hemisphere became available. These new data extend the photometric baseline for additional 43 days to the initial 311 days of the first sectors (after a gap of 395 days), allowing for a search of evidence of transits of additional planet in longer orbital periods. For this, we used the Box Least Squares (BLS) periodogram method over the TOI-220 full {\it TESS} light curve, i.e. the photometric time series containing the data of all the available sectors. We identified and masked the transits of TOI-220\,$b$ at the orbital period of 10.69\,d before conducting a wide search of transit-like features in a wide range of periods and durations adopting the minimun number of events in the light curve as two. It is worth mentioning that a photometric feature at BTJD$\sim$1417 was consistently identified as the first event by the BLS method.
 We confirmed that this {\it event} correspond to a known systematic spike reported in the {\it TESS} data release notes of the sector 4. After removing it, we found no significant peaks in the BLS. All the {\it TESS} light curves were also visually inspected searching for individual transits but only very low SNR features were identified. Therefore, there is no significant evidence of transits of an additional planet in the {\it TESS} data.

  \section{Statistics and RV residuals of the 2 fitted models}

In Table \ref{tab:stats_results} we reported different statistical metrics we used to compare the models analyzed in Sect. \ref{sec:rv_1pl}: the single planet model with and without a radial velocity drift. We show the likelihood of each model and the $\Delta$AIC (which favours the inclusion of the RV drift) as well as the RMS of the radial velocity residuals and the Anderson-Darling test to check for the normality of the residuals. See detailed discussion in Sect.\,\ref{sec:rv_1pl}.  
 
    \begin{table}
        
        \caption{Statistical comparison of the 2 models explored in this work. }
        \label{tab:stats_results}
        \begin{center}
            
        \begin{tabular}{lcc}
            \hline
            Parameter & 1 planet & 1 planet + RV drift \\
            \hline
            $\log$\,\textit{L}  & 55474 & 55484\\
            $\Delta$AIC   &  15 & 0 \\
            RV residuals RMS [m\,s$^{-1}$] & 1.93 & 1.69 \\ 
            A-D test & 0.580 & 0.294 \\
            
            \hline
            \multicolumn{3}{l}{$\log$\,{\it L}: log of the model likelihood.}\\
            \multicolumn{3}{l}{AIC: \textit{Akaike} information criterion.} \\
            \multicolumn{3}{l}{ A-D: Anderson-Darling test, to be compared with a critical value of} \\
            \multicolumn{3}{l}{0.547, corresponding to a 15\,\% significance for normality rejection.}
        \end{tabular}
        \end{center}
    \end{table}       

 \section{Mass constraints of a hypothetical companion based on the RV drift}
\label{sec:RVdrift-mass-constraint}
 We note that the RV drift defined by the linear and quadratic coefficients of Table \ref{tab:stellar_parameters}, which had been defined against the epoch of the first RV point (2458440.752469 BJD), can be replaced by a purely quadratic expression, given by $RV_{dr} = c_2 ( t - T_{0,dr})^2+ k$, where $c_2$ is the quadratic coefficient of Table \ref{tab:stellar_parameters}, $k$ is an offset without physical implications, and $T_{0,dr}$ defines the 'vertex' of the parabolic function. With $c_1$ being the linear coefficient of the RV drift, the time of the vertex relative to the first RV-point is given by: $t_v = -c_1 / (2 c_2) =  127.18$\,d, which results in an absolute value of $T_{0,dr} = 2458567.933677$\,BJD. This conversion permits us to describe the RV drift as a consequence of a hypothetical circular Keplerian orbit, with the drift given by:
 \begin{equation}
 y= -K \cos{\frac{2\pi t}{P}},    
 \end{equation}
  where $K$, $P$ and $t$ are the RV-amplitude, the period, and the time relative to $T_{0,dr}$. Considering that the second order Taylor expansion of $y= a\  \cos{bx}$ is given by $ y= a - \frac{1}{2} x^{2} (a b^2) + O(x^4)$, the RVs of the Keplerian orbit can be approximated to second order by:
  \begin{equation}
    y' = -K + [\frac{ 2\pi^{2} K}{P^{2}}] t^{2},      
  \end{equation}
  where the term in brackets corresponds now to the RV-drift's quadratic coefficient of Table~\ref{tab:stellar_parameters}. Hence:
  \begin{equation}
      c_2 = \frac{2 \pi^{2} K}{P^{2} }= 0.000149(31) \,[\text{m}\,\text{s}^{-1}\,\text{d}^{-2}]
      \label{eq:hans}
  \end{equation}
  For a given value of $P$, we may now solve this for $K$, and -- using the stellar mass indicated in Table~\ref{tab:stellar_parameters} -- derive the mass of the hypothetical orbiter. This leads to the mass limits shown in Table~\ref{tab:mass-rv-drift}. A lower limit of 750 days has been set, which is given by the validity of the Taylor approximation, which gets worse for shorter orbital periods. We note that only periods of less then $\approx$10000\,d or 30 years lead to realistic masses, since longer periods correspond to stellar mass components that would have been detected by other means; e.g. through the SED of the target.
  
   \begin{table}
        \caption{Mass constraints based on the quadratic RV drift. $K$, $Kmin$ and $Kmax$ are the RV amplitudes corresponding to equation \ref{eq:hans} solved for $K$, with the lower and upper errors set by the errors of $c_2$. M$_{pl}$ is the orbiting object's mass in units of Jupiter masses, based on the central $K$ value and a stellar mass of 0.828\,M$_{\sun}$.}
        \label{tab:mass-rv-drift}
        \begin{center}
            
        \begin{tabular}{llllc}
        \hline
        P [d] &	$K_{min}$ [m\,s$^{-1}$] &	$K$ [m\,s$^{-1}$] &	$K_{max}$ [m\,s$^{-1}$]	& M$_{pl}$ [M$_{Jup}$]\\
        \hline
        750	& 3.36	&4.25&	5.13&	0.17 \\
        1000 &	5.98	&7.55&	9.12 &	0.33 \\
        3000 &	53.80	&67.94&	82.07&	4.25 \\
        10000 &	597.80	&754.84&	911.90 & 70.5 \\
        30000 &	5380.16	&6793.59 &	8207.02& 916 \\
        \hline
       \end{tabular}
        \end{center}
    \end{table}

\section{Timing analysis of TOI-220\,\lowercase{$b$}}

\begin{figure}
    \centering
    \includegraphics[width=\linewidth]{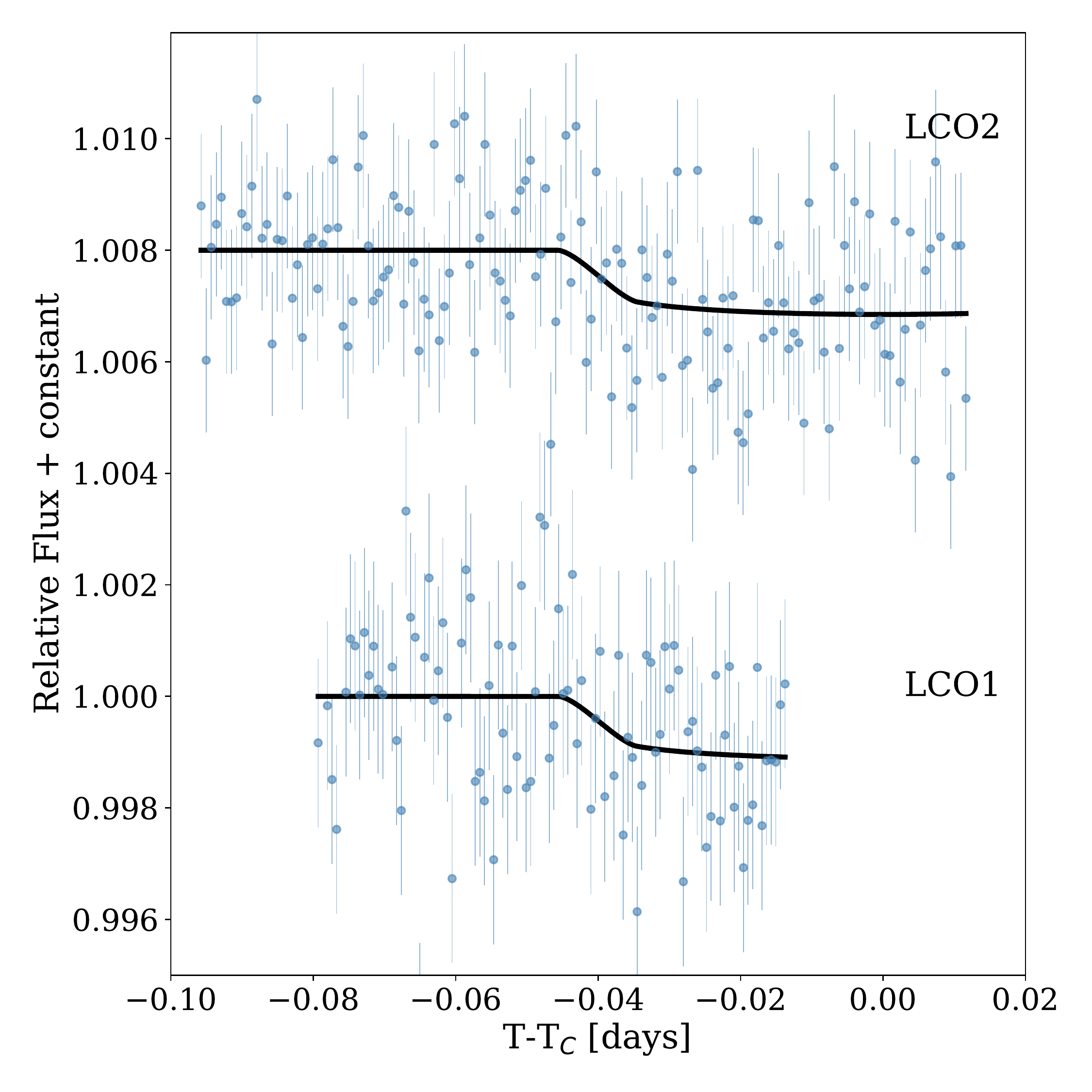} 
    \caption{Two transits of TOI-220\,$b$ obtained with telescopes of the LCOGT network. The solid curves correspond to the fitted model of the timing analysis of the system (Sect \ref{sec:ttv}) and are labelled with the corresponding ID of Table \ref{tab:o-c}.}
    \label{fig:lco_transits}
\end{figure}

In Table\,\ref{tab:o-c} we reported the fitted transit time for each of the transits analyzed in Sect. \ref{sec:ttv}.  The transit ID is the same used in Table \ref{tab:tess_transits}. The results of the two LCO transits (Fig.\,\ref{fig:lco_transits}) are also shown. We reported the difference between these transit times and the estimated TOI-220\,$b$ ephemeris equation (Eq.\,\ref{eq:ephemeris}) which are presented in Fig.\,\ref{fig:ttv}.

\begin{table}
    \centering
    \begin{tabular}{l|c|l|l|c}
    \hline
    Transit ID & T$_{\text{C}}$ [BJD\_TBD]   & +$\sigma$& -$\sigma$&O$-$C [min] \\
    \hline
    S01a&2458335.8997&0.0052&0.0083&-3.46\\
    S01b&2458346.5996&0.0089&0.0089&3.22\\
    S02a&2458357.2865&0.0054&0.0054&-8.82\\
    S02b&2458378.6835&0.0057&0.0100&0.50\\
    S04a&2458421.4632&0.0034&0.0034&-1.46\\
    S04b&2458432.1539&0.0076&0.0076&-8.03\\
    S05a&2458442.8599&0.0040&0.0040&7.43\\
    S05b&2458453.5537&0.0041&0.0041&5.32\\
    S06a&2458474.9390&0.0042&0.0042&-2.21\\
    S06b&2458485.6384&0.0058&0.0058&3.75\\
    S07a&2458496.3270&0.0041&0.0041&-5.85\\
    S07b&2458507.0268&0.0049&0.0049&0.69\\
    S08&2458528.4135&0.0031&0.0031&-4.83\\
    S09&2458549.8031&0.0051&0.0051&-6.16\\
    S10&2458592.5848&0.0066&0.0046&-5.24\\
    S12a&2458635.3685&0.0028&0.0028&-1.43\\
    S12b&2458646.0667&0.0052&0.0052&2.80\\
    LCO1&2458528.420&0.024&0.024&4.53\\
    LCO2&2458838.5808&0.0042&0.0042&1.86\\
    S27a&2459041.7915&0.0037&0.0037&2.85\\
    S27b&2459052.4835&0.0041&0.0041&-1.86\\
    S28a&2459063.1839&0.0033&0.0033&5.54\\
    S28b&2459084.5670&0.0093&0.0093&-5.16\\
    \hline
    \end{tabular}
    \caption{Transit mid-times obtained from individual fitting  and their respective timing residuals (Sect. \ref{sec:ttv}) using P=10.695264(87)\,d and T$_{0}$=2458335.9021(14)\,[BJD\_TBD] (Sect.\ref{sec:rv_1pl}).}
    \label{tab:o-c}
\end{table}

\section{Radial velocity data}

In Tables \ref{tab:rv_data} and \ref{tab:serval_data} we report the radial velocity measurements together with the stellar activity indicators used in these work derived from the HARPS {\sc DRS} and {\sc SERVAL}, respectively (see Sect. \ref{subsec:harps}). 

\begin{table*}
\caption{HARPS {\sc DRS} radial velocity measurements and activity indicators of TOI-220. Exposure time and signal-to-noise ratio per pixel at 550\,nm are given in the last two columns.} 
\label{tab:rv_data}
\begin{tabular}{cccccccccc}
\hline
BJD\_TDB       & RV        & $\sigma_{\text{RV}}$   & BIS  & FWHM & CCF & $\log$\,R$^{\prime}_\mathrm{HK}$ & $\sigma_{\text{$\log$\,R$^{\prime}_\mathrm{HK}$}}$ & EXPTIME    & SNR (550\,nm)  \\
$[$d] & [km\,s$^{-1}$] & [km\,s$^{-1}$] & [km\,s$^{-1}$] & [km\,s$^{-1}$] & CONTRAST&&& [s] & per pixel\\
\hline
2458440.708866 & 26.4650 & 0.0010 & -0.0273 & 6.1504 & 40.3215 &  -5.0453 &   0.0102 &  1800 & 74.8 \\
2458440.800232 & 26.4658 & 0.0010 & -0.0233 & 6.1550 & 40.3239 &  -5.0352 &   0.0114 &  1800 & 73.4 \\
2458441.758120 & 26.4637 & 0.0010 & -0.0211 & 6.1550 & 40.2942 &  -5.0432 &   0.0111 &  1800 & 73.8 \\
2458441.778259 & 26.4620 & 0.0009 & -0.0253 & 6.1559 & 40.2737 &  -5.0481 &   0.0080 &  1800 & 83.0 \\
2458442.807269 & 26.4601 & 0.0009 & -0.0254 & 6.1487 & 40.3025 &  -5.0262 &   0.0104 &  1800 & 80.2 \\
2458442.828449 & 26.4621 & 0.0010 & -0.0260 & 6.1543 & 40.3186 &  -5.0557 &   0.0122 &  1800 & 79.9 \\
2458443.686568 & 26.4573 & 0.0009 & -0.0256 & 6.1541 & 40.2811 &  -5.0555 &   0.0101 &  1800 & 78.1 \\
2458443.775781 & 26.4572 & 0.0010 & -0.0267 & 6.1510 & 40.2879 &  -5.0614 &   0.0119 &  1800 & 71.4 \\
2458444.809444 & 26.4567 & 0.0015 & -0.0265 & 6.1551 & 40.2335 &  -5.0456 &   0.0181 &  1800 & 50.3 \\
2458446.782081 & 26.4577 & 0.0008 & -0.0211 & 6.1489 & 40.2909 &  -5.0626 &   0.0092 &  1800 & 90.4 \\
2458446.830311 & 26.4568 & 0.0009 & -0.0252 & 6.1547 & 40.2613 &  -5.0637 &   0.0111 &  1800 & 89.9 \\
2458447.738845 & 26.4586 & 0.0008 & -0.0237 & 6.1512 & 40.2769 &  -5.0318 &   0.0082 &  1800 & 87.4 \\
2458447.834355 & 26.4585 & 0.0009 & -0.0277 & 6.1535 & 40.2766 &  -5.0801 &   0.0114 &  1800 & 81.7 \\
2458451.730670 & 26.4629 & 0.0008 & -0.0270 & 6.1515 & 40.3034 &  -5.0619 &   0.0086 &  1800 & 92.5 \\
2458451.805994 & 26.4611 & 0.0009 & -0.0250 & 6.1557 & 40.2822 &  -5.0528 &   0.0118 &  1800 & 83.3 \\
2458452.702375 & 26.4586 & 0.0008 & -0.0249 & 6.1487 & 40.3235 &  -5.0401 &   0.0081 &  1500 & 88.2 \\
2458452.782399 & 26.4587 & 0.0008 & -0.0268 & 6.1491 & 40.3037 &  -5.0383 &   0.0098 &  1800 & 90.7 \\
2458453.631164 & 26.4596 & 0.0009 & -0.0255 & 6.1596 & 40.2811 &  -5.0500 &   0.0090 &  1800 & 83.5 \\
2458453.773665 & 26.4572 & 0.0009 & -0.0216 & 6.1523 & 40.3094 &  -5.0261 &   0.0093 &  1500 & 83.5 \\
2458454.627488 & 26.4527 & 0.0012 & -0.0228 & 6.1487 & 40.3376 &  -5.0787 &   0.0169 &  1500 & 61.2 \\
2458454.750359 & 26.4528 & 0.0014 & -0.0269 & 6.1542 & 40.2969 &  -5.0989 &   0.0200 &  1800 & 53.3 \\
2458455.673973 & 26.4538 & 0.0012 & -0.0219 & 6.1500 & 40.3438 &  -5.0740 &   0.0158 &  1500 & 60.4 \\
2458455.790907 & 26.4516 & 0.0012 & -0.0247 & 6.1566 & 40.3110 &  -5.0154 &   0.0146 &  1800 & 62.2 \\
2458467.755644 & 26.4523 & 0.0009 & -0.0251 & 6.1536 & 40.3024 &  -5.0391 &   0.0109 &  1500 & 80.6 \\
2458468.701619 & 26.4530 & 0.0010 & -0.0242 & 6.1563 & 40.3013 &  -5.0400 &   0.0101 &  1500 & 76.3 \\
2458469.796343 & 26.4550 & 0.0012 & -0.0271 & 6.1548 & 40.2970 &  -5.0806 &   0.0184 &  1500 & 65.4 \\
2458471.743546 & 26.4612 & 0.0011 & -0.0286 & 6.1537 & 40.3077 &  -5.0479 &   0.0144 &  1500 & 69.9 \\
2458473.769591 & 26.4618 & 0.0008 & -0.0210 & 6.1545 & 40.2550 &  -5.0396 &   0.0090 &  1800 & 91.0 \\
2458473.790575 & 26.4602 & 0.0008 & -0.0206 & 6.1549 & 40.2674 &  -5.0433 &   0.0103 &  1800 & 90.8 \\
2458475.700867 & 26.4556 & 0.0010 & -0.0263 & 6.1564 & 40.2620 &  -5.0359 &   0.0112 &  1500 & 75.3 \\
2458477.683346 & 26.4526 & 0.0010 & -0.0214 & 6.1553 & 40.2887 &  -5.0420 &   0.0107 &  1800 & 78.4 \\
2458478.764436 & 26.4519 & 0.0010 & -0.0224 & 6.1545 & 40.2691 &  -5.0313 &   0.0123 &  1800 & 81.5 \\
2458479.671230 & 26.4565 & 0.0011 & -0.0252 & 6.1502 & 40.2764 &  -5.0191 &   0.0115 &  1800 & 66.7 \\
2458480.721428 & 26.4585 & 0.0008 & -0.0259 & 6.1560 & 40.2723 &  -5.0374 &   0.0088 &  1800 & 90.2 \\
2458484.774635 & 26.4585 & 0.0010 & -0.0226 & 6.1556 & 40.2878 &  -4.9952 &   0.0122 &  1500 & 74.8 \\
2458486.707656 & 26.4512 & 0.0015 & -0.0257 & 6.1485 & 40.3665 &  -5.0866 &   0.0265 &  1500 & 52.9 \\
2458487.719322 & 26.4529 & 0.0011 & -0.0192 & 6.1580 & 40.2910 &  -5.0331 &   0.0144 &  1500 & 69.9 \\
2458488.654032 & 26.4539 & 0.0012 & -0.0229 & 6.1566 & 40.3144 &  -5.0545 &   0.0145 &  1500 & 60.1 \\
2458488.794703 & 26.4516 & 0.0012 & -0.0229 & 6.1561 & 40.3274 &  -5.0834 &   0.0222 &  1500 & 68.0 \\
2458489.695779 & 26.4506 & 0.0025 & -0.0164 & 6.1595 & 40.3359 &  -5.1050 &   0.0572 &  1500 & 34.6 \\
2458489.817157 & 26.4522 & 0.0024 & -0.0165 & 6.1478 & 40.3351 &  -5.2218 &   0.0715 &  1500 & 36.1 \\
2458490.775790 & 26.4522 & 0.0014 & -0.0237 & 6.1567 & 40.3240 &  -5.1718 &   0.0303 &  1500 & 58.0 \\
2458491.667815 & 26.4571 & 0.0013 & -0.0252 & 6.1522 & 40.2983 &  -4.9942 &   0.0147 &  1500 & 60.4 \\
2458491.787386 & 26.4577 & 0.0017 & -0.0199 & 6.1502 & 40.2760 &  -5.0854 &   0.0313 &  1500 & 48.0 \\
2458501.597810 & 26.4573 & 0.0013 & -0.0258 & 6.1493 & 40.2993 &  -5.0926 &   0.0183 &  1500 & 56.1 \\
2458501.714141 & 26.4537 & 0.0014 & -0.0271 & 6.1552 & 40.2865 &  -5.0426 &   0.0208 &  1500 & 54.5 \\
2458502.585805 & 26.4593 & 0.0011 & -0.0274 & 6.1552 & 40.2717 &  -5.0338 &   0.0130 &  1500 & 65.8 \\
2458502.714208 & 26.4582 & 0.0011 & -0.0206 & 6.1536 & 40.3083 &  -5.0889 &   0.0159 &  1500 & 73.2 \\
2458504.675128 & 26.4600 & 0.0012 & -0.0258 & 6.1602 & 40.2833 &  -5.0751 &   0.0170 &  1500 & 65.3 \\
2458505.654443 & 26.4601 & 0.0010 & -0.0212 & 6.1560 & 40.2635 &  -5.0220 &   0.0109 &  1500 & 72.6 \\
2458506.564996 & 26.4558 & 0.0009 & -0.0261 & 6.1547 & 40.2964 &  -5.0245 &   0.0083 &  1500 & 83.5 \\
2458506.726419 & 26.4538 & 0.0011 & -0.0222 & 6.1567 & 40.2421 &  -5.1025 &   0.0181 &  1500 & 75.1 \\
2458507.570571 & 26.4557 & 0.0014 & -0.0259 & 6.1586 & 40.3063 &  -5.0686 &   0.0203 &  1500 & 55.0 \\
2458507.729911 & 26.4560 & 0.0014 & -0.0248 & 6.1560 & 40.2585 &  -5.1037 &   0.0263 &  1500 & 58.5 \\
2458508.653091 & 26.4533 & 0.0015 & -0.0202 & 6.1525 & 40.3208 &  -5.0814 &   0.0274 &  1800 & 54.4 \\
2458508.673450 & 26.4530 & 0.0015 & -0.0162 & 6.1554 & 40.2769 &  -5.0783 &   0.0275 &  1800 & 55.3 \\
2458514.784388 & 26.4608 & 0.0012 & -0.0229 & 6.1629 & 40.2459 &  -5.0436 &   0.0185 &  1800 & 69.8 \\
2458515.753041 & 26.4586 & 0.0016 & -0.0213 & 6.1576 & 40.2386 &  -5.0440 &   0.0302 &  1240 & 50.8 \\
2458516.577968 & 26.4592 & 0.0010 & -0.0270 & 6.1551 & 40.3106 &  -5.0600 &   0.0126 &  1500 & 79.0 \\
2458516.748836 & 26.4584 & 0.0015 & -0.0263 & 6.1649 & 40.2796 &  -5.1390 &   0.0314 &  1500 & 55.9 \\
2458517.578103 & 26.4552 & 0.0011 & -0.0230 & 6.1566 & 40.3041 &  -5.0307 &   0.0140 &  1500 & 69.1 \\
Continued &&&&&&&&&\\
\hline
\end{tabular}
\end{table*}

\begin{table*}
 \contcaption{HARPS {\sc DRS} radial velocity measurements and activity indicators of TOI-220. Exposure time and signal-to-noise ratio per pixel at 550\,nm are given in the last two columns.}
 \label{tab:rv_continuacion}
\begin{tabular}{cccccccccc}
\hline
BJD\_TDB       & RV        & $\sigma_{\text{RV}}$   & BIS  & FWHM & CCF & $\log$\,R$^{\prime}_\mathrm{HK}$ & $\sigma_{\text{$\log$\,R$^{\prime}_\mathrm{HK}$}}$ & EXPTIME    & SNR (550\,nm)  \\
$[$d] & [km\,s$^{-1}$] & [km\,s$^{-1}$] & [km\,s$^{-1}$] & [km\,s$^{-1}$] & CONTRAST&&& [s] & per pixel \\
\hline
2458519.545756 & 26.4548 & 0.0015 & -0.0236 & 6.1538 & 40.3182 &  -5.0835 &   0.0255 &  1500 & 52.5 \\
2458519.730269 & 26.4523 & 0.0029 & -0.0186 & 6.1508 & 40.2656 &  -5.2730 &   0.1117 &  1500 & 31.6 \\
2458520.745206 & 26.4550 & 0.0021 & -0.0245 & 6.1680 & 40.2821 &  -5.1150 &   0.0508 &  1500 & 41.5 \\
2458521.549624 & 26.4544 & 0.0013 & -0.0215 & 6.1534 & 40.3222 &  -5.0607 &   0.0192 &  1500 & 59.9 \\
2458522.730498 & 26.4573 & 0.0016 & -0.0215 & 6.1522 & 40.3176 &  -5.1530 &   0.0509 &  1500 & 55.9 \\
2458524.738624 & 26.4591 & 0.0012 & -0.0232 & 6.1538 & 40.3246 &  -5.0988 &   0.0283 &  1500 & 69.5 \\
2458528.634749 & 26.4563 & 0.0015 & -0.0292 & 6.1511 & 40.3093 &  -5.1221 &   0.0343 &  1500 & 53.7 \\
2458529.725206 & 26.4513 & 0.0017 & -0.0294 & 6.1658 & 40.1514 &  -4.9232 &   0.0266 &  1500 & 51.2 \\
2458530.732839 & 26.4519 & 0.0016 & -0.0271 & 6.1521 & 40.3033 &  -5.2255 &   0.0526 &  1500 & 54.9 \\
2458531.549651 & 26.4538 & 0.0009 & -0.0238 & 6.1536 & 40.3071 &  -5.0648 &   0.0120 &  1800 & 80.7 \\
2458535.589291 & 26.4626 & 0.0010 & -0.0252 & 6.1534 & 40.2066 &  -4.9657 &   0.0097 &  1800 & 79.7 \\
2458537.672021 & 26.4609 & 0.0010 & -0.0262 & 6.1585 & 40.2762 &  -5.0910 &   0.0222 &  1800 & 82.0 \\
2458538.689271 & 26.4560 & 0.0013 & -0.0312 & 6.1611 & 40.2935 &  -5.1449 &   0.0319 &  1800 & 66.9 \\
2458539.678801 & 26.4530 & 0.0011 & -0.0244 & 6.1564 & 40.2975 &  -5.0551 &   0.0220 &  1800 & 76.0 \\
2458540.700230 & 26.4541 & 0.0014 & -0.0254 & 6.1546 & 40.3270 &  -5.1352 &   0.0364 &  1800 & 63.8 \\
2458542.669892 & 26.4534 & 0.0011 & -0.0263 & 6.1511 & 40.3212 &  -5.1218 &   0.0237 &  1800 & 80.4 \\
2458543.677293 & 26.4538 & 0.0010 & -0.0270 & 6.1592 & 40.3075 &  -5.1119 &   0.0218 &  1800 & 89.6 \\
2458546.690397 & 26.4611 & 0.0015 & -0.0279 & 6.1612 & 40.3362 &  -5.1225 &   0.0379 &  1800 & 57.7 \\
2458547.561096 & 26.4596 & 0.0012 & -0.0283 & 6.1577 & 40.3057 &  -5.0859 &   0.0208 &  1800 & 70.5 \\
2458548.556934 & 26.4595 & 0.0015 & -0.0190 & 6.1614 & 40.2781 &  -5.0799 &   0.0259 &  1800 & 55.3 \\
2458549.552540 & 26.4565 & 0.0009 & -0.0240 & 6.1554 & 40.3031 &  -5.0554 &   0.0135 &  1800 & 88.0 \\
2458550.576618 & 26.4521 & 0.0009 & -0.0246 & 6.1621 & 40.2942 &  -5.0500 &   0.0152 &  1800 & 88.9 \\
2458554.664806 & 26.4555 & 0.0020 & -0.0280 & 6.1579 & 40.3187 &  -5.0843 &   0.0476 &  1800 & 44.5 \\
2458555.650955 & 26.4610 & 0.0011 & -0.0202 & 6.1623 & 40.3268 &  -5.1202 &   0.0256 &  1800 & 75.4 \\
2458562.535601 & 26.4532 & 0.0012 & -0.0251 & 6.1541 & 40.1392 &  -5.0067 &   0.0142 &  1800 & 69.5 \\
2458586.560780 & 26.4545 & 0.0012 & -0.0288 & 6.1560 & 40.3136 &  -5.0919 &   0.0252 &  1800 & 72.1 \\
2458591.490221 & 26.4594 & 0.0010 & -0.0267 & 6.1569 & 40.2479 &  -5.0704 &   0.0163 &  1800 & 84.4 \\
2458611.494962 & 26.4593 & 0.0015 & -0.0273 & 6.1592 & 40.2856 &  -5.0685 &   0.0292 &  1800 & 60.8 \\
2458728.879893 & 26.4661 & 0.0011 & -0.0235 & 6.1538 & 40.2937 &  -5.0360 &   0.0131 &  1800 & 76.9 \\
2458767.767044 & 26.4554 & 0.0012 & -0.0270 & 6.1482 & 40.1839 &  -5.0234 &   0.0118 &  2100 & 65.9 \\
\hline
\end{tabular}
\end{table*}
\newpage

\begin{table*}
\caption{HARPS {\sc SERVAL} activity indicators of TOI-220.} 
\label{tab:serval_data}
\begin{tabular}{crrrrcccc}
\hline
BJD\_TDB &          dLW  & $\sigma_{\text{dLW}}$ & CRX & $\sigma_{\text{CRX}}$ & H$\alpha$ & $\sigma_{\text{H$\alpha$}}$ &  Na\,D & $\sigma_{\text{Na D}}$  \\
 $[$d]     & & & [m\,s$^{-1}$] & [m\,s$^{-1}$] &  &  &  &   \\
\hline
2458440.708866 &  -0.2623 &   1.4290 &  -5.8768 &  11.5422 &   0.4656 &   0.0010 &   0.2665 &   0.0012 \\
2458440.800232 &  -4.9625 &   1.5490 &   6.7761 &  11.0373 &   0.4668 &   0.0010 &   0.2670 &   0.0012 \\
2458441.758120 &  -0.3565 &   1.3821 &  -1.7498 &  10.5452 &   0.4678 &   0.0011 &   0.2648 &   0.0012 \\
2458441.778259 &   1.3307 &   1.1907 &   7.7788 &   9.8335 &   0.4675 &   0.0010 &   0.2669 &   0.0011 \\
2458442.807269 &  -0.7192 &   1.3148 &  13.7782 &   9.0388 &   0.4691 &   0.0010 &   0.2604 &   0.0011 \\
2458442.828449 &  -2.5532 &   1.2666 &  20.1963 &   9.1391 &   0.4644 &   0.0009 &   0.2596 &   0.0011 \\
2458443.686568 &   2.2460 &   1.3634 &  -7.9982 &  10.3545 &   0.4654 &   0.0010 &   0.2678 &   0.0011 \\
2458443.775781 &   2.9989 &   1.3117 &  -4.4235 &  11.2185 &   0.4667 &   0.0011 &   0.2664 &   0.0012 \\
2458444.809444 &   3.6119 &   2.0043 &   2.4141 &  13.7423 &   0.4705 &   0.0016 &   0.2609 &   0.0018 \\
2458446.782081 &   0.0413 &   1.0744 &   0.8902 &   9.4182 &   0.4691 &   0.0008 &   0.2610 &   0.0010 \\
2458446.830311 &   3.3628 &   1.2715 &  -7.4150 &  10.0903 &   0.4681 &   0.0008 &   0.2610 &   0.0010 \\
2458447.738845 &   0.6403 &   1.1213 &   0.1112 &  11.5124 &   0.4648 &   0.0009 &   0.2603 &   0.0010 \\
2458447.834355 &   0.6680 &   1.4356 &  -2.1942 &  11.0709 &   0.4663 &   0.0009 &   0.2604 &   0.0011 \\
2458451.730670 &  -0.0467 &   1.1565 &   6.0690 &   8.4936 &   0.4657 &   0.0008 &   0.2607 &   0.0009 \\
2458451.805994 &   1.0959 &   1.3075 &  10.4416 &   9.3948 &   0.4649 &   0.0009 &   0.2611 &   0.0010 \\
2458452.702375 &  -4.0832 &   1.2065 & -13.0851 &   7.1234 &   0.4647 &   0.0009 &   0.2615 &   0.0010 \\
2458452.782399 &  -2.7432 &   1.2199 &  -4.6426 &   6.7290 &   0.4642 &   0.0008 &   0.2597 &   0.0009 \\
2458453.631164 &  -0.7047 &   1.0873 &  -2.9315 &   8.2828 &   0.4656 &   0.0009 &   0.2620 &   0.0011 \\
2458453.773665 &  -1.1784 &   1.1849 & -11.5157 &   8.6985 &   0.4646 &   0.0009 &   0.2622 &   0.0010 \\
2458454.627488 &  -6.1300 &   1.7282 &  13.0497 &   9.5861 &   0.4664 &   0.0012 &   0.2596 &   0.0014 \\
2458454.750359 &  -1.1140 &   2.5133 &  10.3401 &  13.2155 &   0.4679 &   0.0014 &   0.2617 &   0.0017 \\
2458455.673973 &  -4.3860 &   1.6552 &   0.0656 &   9.5256 &   0.4649 &   0.0013 &   0.2591 &   0.0015 \\
2458455.790907 &  -0.4042 &   1.7392 &   8.1142 &  10.3294 &   0.4678 &   0.0012 &   0.2686 &   0.0014 \\
2458467.755644 &   0.0746 &   1.2988 &   0.2065 &   8.0561 &   0.4688 &   0.0009 &   0.2618 &   0.0011 \\
2458468.701619 &  -1.4562 &   1.3661 &  11.2744 &   9.5247 &   0.4670 &   0.0010 &   0.2618 &   0.0012 \\
2458469.796343 &  -1.7262 &   1.4104 &  -5.0228 &  10.2855 &   0.4640 &   0.0011 &   0.2618 &   0.0013 \\
2458471.743546 &  -0.8041 &   1.4308 &  -4.8355 &  10.5433 &   0.4666 &   0.0011 &   0.2586 &   0.0012 \\
2458473.769591 &   2.2120 &   0.9652 & -11.8209 &   8.3775 &   0.4665 &   0.0008 &   0.2613 &   0.0010 \\
2458473.790575 &   1.0648 &   1.2348 & -21.6860 &   8.0858 &   0.4656 &   0.0008 &   0.2668 &   0.0009 \\
2458475.700867 &   4.6992 &   1.3896 &   8.9426 &   9.7111 &   0.4637 &   0.0010 &   0.2592 &   0.0012 \\
2458477.683346 &   1.4648 &   1.2915 &   7.2851 &   7.8859 &   0.4664 &   0.0010 &   0.2667 &   0.0011 \\
2458478.764436 &  -0.6453 &   1.4899 &  -4.9377 &   8.2332 &   0.4655 &   0.0009 &   0.2686 &   0.0011 \\
2458479.671230 &  -0.8130 &   1.7958 &   4.7311 &  10.6195 &   0.4703 &   0.0012 &   0.2661 &   0.0013 \\
2458480.721428 &   2.7656 &   1.0797 & -13.2663 &   8.3028 &   0.4663 &   0.0008 &   0.2659 &   0.0010 \\
2458484.774635 &   0.3673 &   1.5322 &   0.8046 &   9.6450 &   0.4645 &   0.0010 &   0.2612 &   0.0012 \\
2458486.707656 &  -3.1221 &   1.8534 & -14.3388 &  11.9351 &   0.4633 &   0.0014 &   0.2587 &   0.0017 \\
2458487.719322 &   1.1626 &   1.6061 &   8.9141 &   9.9225 &   0.4673 &   0.0011 &   0.2610 &   0.0013 \\
2458488.654032 &  -2.4478 &   1.6747 &   5.4204 &   9.9891 &   0.4653 &   0.0013 &   0.2635 &   0.0015 \\
2458488.794703 &  -3.6027 &   1.6175 &  -3.4279 &   8.9609 &   0.4658 &   0.0011 &   0.2627 &   0.0013 \\
2458489.695779 &  -4.0812 &   3.2346 &   8.1978 &  20.4807 &   0.4661 &   0.0021 &   0.2585 &   0.0027 \\
2458489.817157 &   1.4793 &   3.1791 &  35.1790 &  16.5534 &   0.4573 &   0.0020 &   0.2648 &   0.0026 \\
2458490.775790 &  -1.9411 &   1.9919 &   7.9932 &  11.3483 &   0.4641 &   0.0012 &   0.2607 &   0.0015 \\
2458491.667815 &  -1.8741 &   1.9072 & -27.1674 &  10.1383 &   0.4719 &   0.0012 &   0.2612 &   0.0015 \\
2458491.787386 &  -1.3321 &   2.4345 &  -0.3721 &  13.7161 &   0.4666 &   0.0015 &   0.2677 &   0.0018 \\
2458501.597810 &  -2.3896 &   1.7019 & -19.7716 &  12.1489 &   0.4692 &   0.0014 &   0.2673 &   0.0016 \\
2458501.714141 &   0.0764 &   1.8841 &  -3.7394 &  10.5970 &   0.4663 &   0.0014 &   0.2596 &   0.0016 \\
2458502.585805 &  -0.1945 &   1.7044 &  -6.8271 &   9.7077 &   0.4652 &   0.0012 &   0.2662 &   0.0013 \\
2458502.714208 &  -1.9466 &   1.5951 &  -1.6920 &   8.8068 &   0.4656 &   0.0010 &   0.2603 &   0.0012 \\
2458504.675128 &  -3.1907 &   1.5990 &  -9.1703 &  10.9906 &   0.4645 &   0.0012 &   0.2599 &   0.0014 \\
2458505.654443 &   1.9881 &   1.5946 &  -6.0949 &   8.9858 &   0.4649 &   0.0011 &   0.2605 &   0.0012 \\
2458506.564996 &  -0.4999 &   1.2270 &  -4.2423 &   8.2120 &   0.4657 &   0.0009 &   0.2584 &   0.0011 \\
2458506.726419 &   2.2123 &   1.6614 & -19.0888 &   9.1581 &   0.4646 &   0.0010 &   0.2610 &   0.0012 \\
2458507.570571 &  -0.8163 &   2.1560 &  10.4649 &  15.1414 &   0.4643 &   0.0015 &   0.2610 &   0.0017 \\
2458507.729911 &   4.2407 &   2.0254 &  -7.7070 &  10.3599 &   0.4634 &   0.0013 &   0.2596 &   0.0015 \\
2458508.653091 &   0.6389 &   1.9808 &  35.3264 &  12.6350 &   0.4602 &   0.0014 &   0.2590 &   0.0017 \\
2458508.673450 &   1.6420 &   1.9984 &  22.2776 &  14.2334 &   0.4580 &   0.0014 &   0.2585 &   0.0016 \\
2458514.784388 &   7.0675 &   1.4771 &  -0.6879 &  10.9105 &   0.4560 &   0.0010 &   0.2667 &   0.0013 \\
2458515.753041 &   8.4969 &   2.3699 & -13.9389 &  13.3166 &   0.4573 &   0.0015 &   0.2661 &   0.0018 \\
2458516.577968 &  -2.3914 &   1.5421 & -24.5056 &   9.4269 &   0.4591 &   0.0010 &   0.2653 &   0.0011 \\
2458516.748836 &   3.8234 &   2.0109 &  -4.3809 &  12.7091 &   0.4593 &   0.0014 &   0.2638 &   0.0016 \\
2458517.578103 &  -2.6623 &   1.2879 & -18.1996 &   9.1335 &   0.4581 &   0.0011 &   0.2682 &   0.0013 \\
Continued &&&&&&&\\
\hline
\end{tabular}
\end{table*}

\begin{table*}
 \contcaption{HARPS {\sc SERVAL} activity indicators of TOI-220.}
 \label{tab:serval_continuacion}
\begin{tabular}{crrrrcccc}
\hline
BJD\_TDB &          dLW  & $\sigma_{\text{dLW}}$ & CRX & $\sigma_{\text{CRX}}$ & H$\alpha$ & $\sigma_{\text{H$\alpha$}}$ &  Na\,D & $\sigma_{\text{Na D}}$  \\
 $[$d]     & [m\,s$^{-1}$] & [m\,s$^{-1}$] &  &  &  &  &  &   \\
\hline
2458519.545756 &  -2.7058 &   1.7916 &  -0.4067 &  11.4625 &   0.4669 &   0.0015 &   0.2652 &   0.0017 \\
2458519.730269 &   2.1747 &   4.0675 & -18.2779 &  19.2282 &   0.4634 &   0.0024 &   0.2696 &   0.0031 \\
2458520.745206 &   1.2080 &   2.5333 & -13.1655 &  15.8645 &   0.4589 &   0.0017 &   0.2636 &   0.0022 \\
2458521.549624 &  -0.3015 &   1.6485 &  -5.2641 &  10.4197 &   0.4608 &   0.0013 &   0.2682 &   0.0015 \\
2458522.730498 &  -0.9532 &   2.2842 &  27.7059 &  13.1662 &   0.4589 &   0.0013 &   0.2598 &   0.0016 \\
2458524.738624 &  -1.2092 &   1.7960 & -10.3374 &  11.4734 &   0.4593 &   0.0011 &   0.2622 &   0.0013 \\
2458528.634749 &  -1.0433 &   1.9821 &  -8.1356 &  10.8765 &   0.4671 &   0.0014 &   0.2639 &   0.0017 \\
2458529.725206 &   5.8613 &   2.6058 &  -5.5268 &  13.0208 &   0.4623 &   0.0015 &   0.2629 &   0.0018 \\
2458530.732839 &   3.2712 &   2.0613 & -26.2329 &  13.9351 &   0.4638 &   0.0014 &   0.2629 &   0.0016 \\
2458531.549651 &   1.3428 &   1.2768 &  -6.5268 &   9.4249 &   0.4632 &   0.0010 &   0.2630 &   0.0011 \\
2458535.589291 &   9.7591 &   1.3480 &  -3.8071 &   8.7267 &   0.4649 &   0.0010 &   0.2609 &   0.0011 \\
2458537.672021 &   1.5254 &   1.2464 &  -3.3699 &   9.7638 &   0.4674 &   0.0009 &   0.2659 &   0.0011 \\
2458538.689271 &  -0.6371 &   1.9865 &  15.3358 &  10.8076 &   0.4588 &   0.0012 &   0.2667 &   0.0013 \\
2458539.678801 &  -0.8648 &   1.6380 &  -8.4122 &  10.2376 &   0.4589 &   0.0010 &   0.2644 &   0.0012 \\
2458540.700230 &  -3.6376 &   1.8298 & -11.8585 &  11.4789 &   0.4622 &   0.0012 &   0.2636 &   0.0014 \\
2458542.669892 &  -2.4210 &   1.4654 &   6.2483 &   9.7577 &   0.4602 &   0.0010 &   0.2644 &   0.0011 \\
2458543.677293 &  -1.3194 &   1.3547 & -10.0779 &  10.2875 &   0.4590 &   0.0009 &   0.2672 &   0.0010 \\
2458546.690397 &  -3.4933 &   2.0798 & -16.0068 &  13.8284 &   0.4601 &   0.0013 &   0.2658 &   0.0016 \\
2458547.561096 &  -2.6385 &   1.7150 &  18.2028 &  12.7411 &   0.4627 &   0.0011 &   0.2661 &   0.0013 \\
2458548.556934 &  -1.0987 &   1.8158 &   6.1270 &  12.1296 &   0.4673 &   0.0015 &   0.2676 &   0.0017 \\
2458549.552540 &  -0.2272 &   1.2355 &  -9.1661 &   8.6772 &   0.4652 &   0.0009 &   0.2668 &   0.0010 \\
2458550.576618 &  -0.8848 &   1.3658 &  -3.7954 &   9.2115 &   0.4648 &   0.0009 &   0.2675 &   0.0010 \\
2458554.664806 &  -1.0631 &   2.8411 &  26.3565 &  16.8388 &   0.4660 &   0.0018 &   0.2629 &   0.0021 \\
2458555.650955 &  -2.6379 &   1.6708 &  -3.0134 &  12.1884 &   0.4660 &   0.0010 &   0.2617 &   0.0012 \\
2458562.535601 &  11.6500 &   1.7232 &   2.8796 &   8.4236 &   0.4655 &   0.0012 &   0.2603 &   0.0013 \\
2458586.560780 &  -2.4692 &   1.5497 &   1.8631 &  10.8953 &   0.4670 &   0.0011 &   0.2604 &   0.0012 \\
2458591.490221 &   2.5875 &   1.5793 &  -6.1143 &   8.8415 &   0.4663 &   0.0010 &   0.2578 &   0.0010 \\
2458611.494962 &   0.6165 &   2.0178 &  -8.6563 &  11.9310 &   0.4648 &   0.0013 &   0.2675 &   0.0015 \\
2458728.879893 &   0.9451 &   1.7133 &  -8.4924 &  10.5144 &   0.4646 &   0.0011 &   0.2620 &   0.0012 \\
2458767.767044 &  10.1703 &   1.7910 &  -0.4932 &  11.3817 &   0.4690 &   0.0013 &   0.2663 &   0.0014 \\
\hline
\end{tabular}
\end{table*}
\newpage

\section{Affiliations}
\label{sec:affiliations}
$^{1}$ Aix Marseille Univ, CNRS, CNES, LAM, Marseille, France.\\
$^{2}$ Dipartimento di Fisica, Universita degli Studi di Torino, via Pietro Giuria 1, I-10125, Torino, Italy.\\
$^{3}$ Centre for Exoplanets and Habitability, University of Warwick, Gibbet Hill Road, Coventry, CV4 7AL, UK.\\
$^{4}$ Department of Physics, University of Warwick, Gibbet Hill Road, Coventry CV4 7AL, UK. \\
$^{5}$ Departamento de Física Teórica e Experimental, Universidade Federal do Rio Grande do Norte, Campus Universitário, Natal, RN, 59072-970, Brazil.\\
$^{6}$ Centro de Astrobiolog{\'i}a (CAB, CSIC-INTA), Depto. de Astrof{\'i}sica, ESAC campus 28692 Villanueva de la Ca{\~n}ada (Madrid).\\
$^{7}$ Instituto de Astrof\'isica e Ci\^encias do Espa\c{c}o, Universidade do Porto, CAUP, Rua das Estrelas, 4150-762 Porto, Portugal. \\
$^{8}$ Departamento de F\'isica e Astronomia, Faculdade de Ci\^encias, Universidade do Porto, Rua do Campo Alegre, 4169-007 Porto, Portugal.\\
$^{9}$ Leiden Observatory, Leiden University, 2333CA Leiden, The Netherlands.\\
$^{10}$ Department of Space, Earth and Environment, Chalmers University of Technology, Onsala Space Observatory, 439 92 Onsala, Sweden.\\
$^{11}$ Center for Astrophysics \textbar \ Harvard \& Smithsonian, 60 Garden Street, Cambridge, MA 02138, USA.\\
$^{12}$ NASA Ames Research Center, Moffett Field, CA 94035, USA.\\
$^{13}$ Instituto de Astrof{\'i}sica de Canarias, 38205 La Laguna, Tenerife, Spain.\\
$^{14}$ Departamento de Astrof{\'i}sica, Universidad de La Laguna, 38206 La Laguna, Tenerife, Spain.\\
$^{15}$ Sub-department of Astrophysics, Department of Physics, University of Oxford, Oxford, OX1 3RH, UK.\\
$^{16}$ George Mason University, 4400 University Drive, Fairfax, VA, 22030 USA.\\
$^{17}$ International Center for Advanced Studies (ICAS) and ICIFI (CONICET), ECyT-UNSAM, Campus Miguelete, 25 de Mayo y Francia, (1650) Buenos Aires, Argentina.\\
$^{18}$ Th\"uringer Landessternwarte Tautenburg, Sternwarte 5, 07778 Tautenburg, Germany.\\
$^{19}$ NASA Exoplanet Science Institute, Caltech/IPAC, Mail Code 100-22, 1200 E. California Blvd., Pasadena, CA 91125, USA.\\
$^{20}$ Dept.\ of Physics \& Astronomy, Swarthmore College, Swarthmore PA 19081, USA.\\
$^{21}$ U.S. Naval Observatory, 3450 Massachusetts Avenue NW, Washington, D.C. 20392, USA.\\
$^{22}$ Observatoire de l'Universit{\'e} de Gen{\`e}ve, Chemin des Maillettes 51, 1290 Versoix, Switzerland.\\
$^{23}$ University of Z{\"u}rich, Institute for Computational Science, Winterthurerstrasse 190, 8057 Z{\"u}rich, Switzerland\\
$^{24}$ European Southern Observatory, Alonso de Cordova 3107,Vitacura, Santiago, Chile.\\
$^{25}$ Department of Physics and Kavli Institute for Astrophysics and Space Research, Massachusetts Institute of Technology, Cambridge, MA 02139, USA \\
$^{26}$Department of Earth, Atmospheric and Planetary Sciences, Massachusetts Institute of Technology, Cambridge, MA 02139, USA \\
$^{27}$Department of Aeronautics and Astronautics, MIT, 77 Massachusetts Avenue, Cambridge, MA 02139, USA \\
$^{28}$ Hazelwood Observatory, Australia.\\
$^{29}$ Mullard Space Science Laboratory, University College London, Holmbury St. Mary, Dorking, Surrey, RH5 6NT, UK.\\
$^{30}$ Department of Astrophysical Sciences, Princeton University, 4 Ivy Lane, Princeton, NJ 08544, USA. \\
$^{31}$ Univ. Grenoble Alpes, CNRS, IPAG, F-38000 Grenoble, France\\
$^{32}$ Astrophysics Science Division, NASA Goddard Space Flight Center, Greenbelt, MD 20771, USA\\
$^{33}$ Deutsches Zentrum f\"ur Luft- und Raumfahrt, Institut f\"ur Planetenforschung, 12489 Berlin, Rutherfordstrasse 2., Germany.\\
$^{34}$ Center for Planetary Systems Habitability and McDonald Observatory, The University of Texas at Austin, Austin, TX 78730, USA.\\
$^{35}$ Noqsi Aerospace Ltd., 15 Blanchard Avenue, Billerica, MA 01821, USA \\
$^{36}$ Rheinisches Institut f\"ur Umweltforschung an der Universit\"at zu K\"oln, Aachener Strasse 209, 50931 K\"oln, Germany.\\
$^{37}$ Astronomical Institute of the Czech Academy of Sciences, Fri\v{c}ova 298, Ond\v{r}ejov 25168, Czech Republic\\
$^{38}$ Stellar Astrophysics Centre, Department of Physics and Astronomy, Aarhus University, Ny Munkegade 120,DK-8000 Aarhus C, Denmark \\
$^{39}$ Department of Astronomy, University of Tokyo, 7-3-1 Hongo, Bunkyo-ku, Tokyo 113-0033, Japan.\\
$^{40}$ Mikulski Archive for Space Telescopes.\\
$^{41}$ NCCR/PlanetS, Centre for Space \& Habitability, University of Bern, Bern, Switzerland. \\
$^{42}$ Astronomy Department and Van Vleck Observatory, Wesleyan University, Middletown, CT 06459, USA \\
$^{43}$ SETI Institute, Mountain View, CA 94043, USA. \\
$^{44}$ Department of Astronomy, Columbia University, 550 W 120th Street, New York NY 10027, USA \\


\bsp	
\label{lastpage}
\end{document}